\documentclass[article,showpacs,preprintnumbers,amsmath,amssymb]{revtex4}
\usepackage[dvips]{color}
\usepackage{rotating}
\usepackage{graphicx}
\usepackage{dcolumn}
\usepackage{bm}
\usepackage{epsfig}
\usepackage{hyperref}
\usepackage{ulem}

\newcommand{\lton}{\mathrel{\lower.9ex
\hbox{$\stackrel{\displaystyle <}{\sim}$}}}
\newcommand{\gton}{\mathrel{\lower.9ex
\hbox{$\stackrel{\displaystyle >}{\sim}$}}}

\newcommand{\shk}[1]{{k}\!\!\!/}
\newcommand{\shq}[1]{{q}\!\!\!/}

\newcommand{\barQ}{{\bar{Q}}}
\newcommand{\barc}{{\bar{c}}}
\newcommand{\barb}{{\bar{b}}}

\newcommand{\bfk}{{\bf{k}}}

\newcommand{\bfr}{{\bf{r}}}
\newcommand{\hatr}{{\hat{r}}}
\newcommand{\hatk}{{\hat{k}}}
\newcommand{\hatt}{\hat{t}}

\newcommand{\hats}{\hat{s}}
\newcommand{\tilR}{{\tilde{R}}}
\newcommand{\tilpsi}{{\tilde{\psi}}}

\newcommand{\calO}{{\cal{O}}}
\newcommand{\rGeV}{{\rm GeV}}
\newcommand{\rdof}{{\rm dof}}
\newcommand{\rf}{{\rm form}}
\newcommand{\rd}{{\rm diss}}
\newcommand{\rIS}{{\rm IS}}
\newcommand{\charm}{{\rm{charm}}}
\newcommand{\bottom}{{\rm{bottom}}}

\begin{document}

\begin{flushright}
\vskip .5cm
\end{flushright} \vspace{1cm}

\title{High transverse momentum quarkonium production and dissociation \\ in heavy ion
collisions}

\author{Rishi Sharma$^1$}%
\email{rishi@triumf.ca}

\author{Ivan Vitev$^2$}
\email{ivitev@lanl.gov}

\affiliation{$^1$ TRIUMF, Theory Group, 4004 Wesbrook Mall, Vancouver, Canada}
\affiliation{$^2$ Los Alamos National Laboratory, Theoretical Division, Los Alamos, NM 87545, USA}

\vspace*{1cm}

\begin{abstract}
We calculate the yields of quarkonia in heavy ion collisions at RHIC and the
LHC as a function of their transverse momentum.  Based upon non-relativistic
quantum chromodynamics, our results include both color-singlet and color-octet contributions
and feed-down effects from excited states. In reactions with ultra-relativistic
nuclei, we focus on the consistent implementation of dynamically calculated
nuclear matter effects, such as coherent power corrections, cold nuclear matter
energy loss, and the Cronin effect in the initial state. In the final state, we
consider  radiative energy loss for the color-octet state and collisional
dissociation of quarkonia  as they traverse through the QGP.
Theoretical results are presented for $J/\psi$  and $\Upsilon$ and compared to
experimental data where applicable. At RHIC, a good description of the
high-$p_T$ $J/\psi$  modification observed in central Cu$+$Cu and Au$+$Au
collisions can be achieved within the model uncertainties. We find that
measurements of $J/\psi$ yields in proton-nucleus reactions are needed to
constrain the magnitude of cold nuclear matter effects.  At the LHC, a  good
description of the experimental data can be achieved only in mid-central and
peripheral Pb+Pb collisions. The large five-fold suppression of prompt $J/\psi$
in the most central nuclear reactions  may indicate for the first time possible
thermal effects at the level of the quarkonium  wavefunction at  large
transverse momenta.     
\end{abstract}


\maketitle
\section{Introduction~\label{section:Introduction}}
Melting of heavy quarkonium states, like the $J/\psi$ and the
$\Upsilon$, due to color screening in a deconfined quark-gluon plasma 
(QGP)~\cite{Matsui:1986} has been proposed as one of the principal 
signatures for its  formation. An expected experimental consequence of this 
melting  in the thermal medium created in heavy ion collisions (HIC) is a
suppression of the yields of heavy mesons, when compared to their yields in
nucleon-nucleon (NN) collisions scaled with the number of binary 
interactions.  

In a simplified picture, one can think of a quarkonium as a $Q\barQ$ pair
in a color-singlet bound state, where the heavy quark ($Q$) and the
anti-quark ($\bar{Q}$) are separated by distances $\sim 1/(m_Qv)$,
smaller than $1/\Lambda_{QCD}$. Here, $v\sim \alpha_s(m_Q v)$ is the relative
velocity between $Q$ and $\barQ$, and the state is held together by an
effective potential interaction. The sizes of the higher excited states are
larger, and the sizes of the bottomonia are smaller than
the corresponding charmonia. The presence of a thermal medium screens the
interactions between the $Q$ and $\barQ$ and leads to melting at some
characteristic temperature~\cite{Mocsy:2007jz} that depends on
the meson. This picture suggests that it may be possible to observe sequential
melting of narrower quarkonia as we explore thermal media of increasing
temperatures~\cite{Karsch:1991,Karsch:2006}. Several studies
(see~\cite{ConesadelValle:2011fw} for a recent review) have calculated
the modification of quarkonium yields in collisions at SPS, RHIC, and the
LHC~\cite{Zhao:2008vu,Lourenco:2009,Rapp:2009my,
Strickland:LongAndShort,Margotta:2011ta,Song:2011nu}.

A more sophisticated description of quarkonia is provided by non-relativistic
quantum chromodynamics (NRQCD)~\cite{Bodwin:1994jh}. In this picture, the
$Q\barQ$ state in the color-singlet combination is the lowest order Fock space
component of the full quarkonium wavefunction. Higher Fock components in the
wavefunction have additional partons and are suppressed by positive powers of
the small parameter $v$. In our calculation we will take into account both
color-singlet and color-octet $Q\barQ$ contributions in the production. The
NRQCD formalism has been used to calculate differential yields of heavy mesons
as a function of their transverse momentum, $p_T$, in p$+$p
collisions~\cite{Bodwin:1994jh,Cho:1995ce,Cho:1995vh,Braaten:2000cm,Cooper:LongAndShort}.
The picture of the production process in NRQCD is as follows. The initial hard
collision produces a ``proto-quarkonium'', a short distance ($\sim1/m_Q$)
$Q\barQ$ pair  where the $Q$ and the $\barQ$ can be in the color-singlet or the
octet state. Since this process is short-distance, it can be calculated using
perturbative QCD~\cite{Kang:2011mg}. This ``proto-quarkonium'' then evolves
into the quarkonium state with probabilities that are given by non-perturbative
matrix elements.  For color-octet states, this evolution process also involves
the emission of soft partons to form a net color-singlet object. The final
state thus formed has a hierarchy of Fock space components that are given by
NRQCD scaling rules. In this paper, we combine the NRQCD formalism with  cold
nuclear matter (CNM) effects and the effects of propagation of quarkonia
through a hot QGP to calculate the final yields in HIC. We then obtain the
nuclear modification ratio, 
\begin{equation}
R_{AB} (p_T; N_{\rm{part}}) = \frac{ d\sigma^{ Q \bar{Q}}_{AB}/dydp_T  } {  N^{\rm
coll}_{AB}  d\sigma^{ Q \bar{Q}}_{pp}/dydp_T  },\;\label{eq:RAAdef}
\end{equation}
in reactions with heavy nuclei. In Eq.~\ref{eq:RAAdef}, $N_{\rm{part}}$ refers
to the number of participants in the collision, which depends on the nuclei $A$
and $B$, and the impact parameter. $N^{coll}_{AB}$ is the number of binary
collisions in the process. We present results for Au$+$Au and Cu$+$Cu
collisions at RHIC at $\sqrt{S}=0.2$~TeV per nucleon pair and Pb$+$Pb
collisions at LHC at $\sqrt{S}=2.76$~TeV per nucleon pair.

We emphasize up front that our goal is not to fit data in heavy ion reactions
but to investigate systematically the relative significance of CNM and QGP effects
and discuss ways in which these effects can be better constrained. Our
formalism is limited to high transverse momentum and not applicable to
$p_T$-integrated yields. The CNM~\cite{Vogt:2010aa} effects we consider include
nuclear shadowing (here implemented as coherent power corrections)  and initial
state energy loss. These are calculated in the same framework as previously used
for light hadrons~\cite{Vitev:2003xu,Gyulassy:2003mc,
Qiu:2003vd,Qiu:2004da,Vitev:2006,Vitev:2007ve,Neufeld:2011,Kang:2011bp} and open heavy
flavor~\cite{Vitev:2008vk,Vitev:2006bi,Kang:2011rt}. We also consider transverse momentum
broadening~\cite{Accardi:2002ik,Vitev:2003xu,Vitev:2006bi,Kang:2011rt} (as a model for 
the Cronin effect).  We give results for the p(d)$+$A collisions at RHIC and the
LHC with and without considering the Cronin effect. Since the CNM effects are
generated on a short time scale, $~\sim (1{\rm{fm}}/c) A^{1/3}/\gamma$ (where
$A$ is the atomic mass number of the nucleus and $\gamma=1/\sqrt{1-v^2}$ is the
Lorentz factor, and $A^{1/3}/\gamma$ is roughly the Lorentz contracted size of
the nucleus in the lab frame in fm), compared to the formation time of the
quarkonia, we take the $Q\bar{Q}$ pair in the final state in the CNM
calculation as the ``proto-quarkonium'' state.

We include two effects of $Q\barQ$ propagation through the thermal medium.
First, on time scales shorter than the formation time of quarkonia, the
color-octet component of the proto-quarkonium state will be quenched. This
effect is included by calculating the quenching factor for a massive
color-octet state passing through the medium for the formation time. Second, on
time scales longer than the formation time, the quarkonia can be dissociated by
the medium.  We include this by solving the rate
equations~\cite{hep-ph/0611109,Sharma:2009hn} describing the change in the
yields of quarkonia as a function of time. The microscopic input
for the equations are the time scales for the formation and the dissociation of
the quarkonia. Since the size of proto-quarkonium is small, it is typically
assumed that the formation time $t_\rf$ for the ``proto-quarkonium'' to
``expand'' into the quarkonium wavefunction, is not significantly modified by
the presence of the QGP~\cite{Strickland:LongAndShort}. Hence, $t_\rf$ in the
rest frame of the meson is $\sim 1/(m_Q v^2)$. We estimate $t_\rf$ by dividing
the radial size of the wavefunction by the radial velocity and assign a factor
of 2 uncertainty to it. (Some equilibrium approaches~\cite{Zhao:2008vu} include
effects of a finite time of formation by suppressing the decay rate.)

For dissociation times, most studies use results for thermally equilibrated
quarkonia~\cite{Zhao:2008vu,Strickland:LongAndShort,Rapp:2009my}.  For a
quarkonium which is stationary or slowly moving through the thermal medium it might be
safe to assume that it reaches thermal equilibrium with the medium. In this
picture, the wavefunction of the quarkonium (the lowest Fock component) is
better described by the solution of the Schr\"{o}dinger equation with a
modified potential that depends on the
temperature~\cite{Mocsy:2007jz,Rapp:2009my,Strickland:LongAndShort,Margotta:2011ta}.
The potentials used are based on finite temperature lattice calculations of the
potential between two static charges~\cite{Kaczmarek:2005ui}. The dissociation
rates have been calculated in several approaches.
In~\cite{Laine:2006ns,Brambilla:2008cx} the imaginary part of the potentials
--- which is related to dissociation --- has been calculated in the hard
thermal loop framework. In~\cite{Park:2007zza}  the thermal width has been
calculated using the calculation of collisional decay rates~\cite{Peskin:1979}.
In~\cite{Rapp:2009my} a  $T-$matrix approach was used to calculate decay rates
of thermal states. For a review of thermal properties of quarkonia using 
AdS/CFT techniques see~\cite{CasalderreySolana:2011us}.

If a quarkonium state is produced on a short formation time scale and is moving 
rapidly through the QGP, it  may not have sufficient time to thermally equilibrate with 
the medium.  As it propagates through the hot and dense matter, collisions with
thermal gluons can dissociate the quarkonium on a time scale
$t_\rd$~\cite{hep-ph/0611109,Sharma:2009hn,Dominguez:2008be,Dominguez:2008aa}.
In this paper, we take this high-$p_T$ limit and explore the consequence of
assuming  that  the wavefunction of the meson is the same as in the vacuum, and
that the main impact of the medium is to dissociate the meson as it propagates
through it.  This dissociation mechanism has been phenomenologically successful
in describing the modification of the yields of open heavy flavor mesons as a
function of the transverse momentum $p_T$. At RHIC, this mechanism is consistent with the 
large suppression of non-photonic electrons~\cite{Adare:2006nq,Abelev:2006db,Knospe:2012qf}.  
The predictions~\cite{Sharma:2009hn} for $D$-meson suppression at the LHC are
compatible with preliminary experimental results~\cite{Dainese:2011vb}. The
formation times for quarkonia are also small. Therefore, in-medium formation and
dissociation can modify the quarkonia yields. Before the formation of the
quarkonium state, the production of $Q\bar{Q}$ in the color-octet state is the
dominant mechanism for quarkonia production. As mentioned above this colored object loses
energy, which quenches the yield of quarkonia. After formation, an overall
color neutral object with a dominant color-singlet wavefunction propagates
through the medium and is dominantly lost via dissociation. We find the
dissociation time $t_\rd$ for the color-singlet Fock component of the
quarkonium wavefunction by calculating the rate of broadening of the
wavefunction due to in-medium interactions, as described for open heavy flavor
in~\cite{hep-ph/0611109}. One important difference with the heavy-light mesons
is that the probability of reformation after dissociation is small since the
probability of refragmentation to quarkonia is small. In principle,
recombination~\cite{Bass:2006vu} with thermal $\barQ$ and $Q$ from the medium
can re-generate quarkonia. Since the population of thermal heavy quarks is
exponentially suppressed ($\sim e^{-m_T/T}$), especially at large $p_T$, we
ignore this process.

Our results for the $R_{AA}$ of $J/\psi$ mesons, obtained using this formalism,
are consistent with PHENIX and STAR data for central Au$+$Au and Cu$+$Cu
collisions.  p$+$A quarkonium yields at high $p_T$ are necessary to
accurately constrain the magnitude of CNM effects. Results from ATLAS and CMS
at the LHC for the $J/\psi$ suppression are well reproduced by the calculation
only in peripheral and mid-central collisions.  It is difficult (with or
without transverse momentum broadening) to obtain a five-fold suppression  of
prompt $J/\psi$, as reported by the CMS experiment at the LHC, without assuming
thermal effects at the level of the quarkonium wavefunction. For the
$\Upsilon$s, this approach predicts that the suppression for the $\Upsilon(2S)$
and $\Upsilon(3S)$ at high $p_T$ is comparable to the $\Upsilon(1S)$
suppression. (This statement is true for the direct production of $\Upsilon(1S)$
and $\Upsilon(2S)$. The feed-down contributions to $\Upsilon(1S)$, $\Upsilon(2S)$,
and $\Upsilon(3S)$ from the $p$-wave states are suppressed differently, but we
find their $R_{AA}$ to be comparable even when we include feed-down from $p-$
wave states.) This prediction is also different from the expectation of the
equilibrium models, which predict stronger suppression (smaller $R_{AA}$) for
the excited states.  Differential high transverse momentum data is also
necessary to shed light on the possible thermal modification of boosted
$\Upsilon$s at the LHC.

Our paper is organized as follows. In Section~\ref{section:ppProduction} we
present  the calculation of the p$+$p baseline yields of quarkonia. 
In Section~\ref{section:CNM} we
discuss how CNM effects modify the production yields of quarkonia in nuclear
collisions.  In Section~\ref{section:Wavefunction} we evaluate the
wavefunctions needed for the evaluation of the formation and dissociation
time scales.  In Section~\ref{section:QGPdynamics} we calculate the dissociation
time scale in the QGP and discuss the dynamics of quarkonia as they propagate
through the strongly-interacting medium. Results for the nuclear modification
factor  in p$+$A and A$+$A collisions  are given in
Section~\ref{section:Results}. We present our conclusions in
Section~\ref{section:Conclusions}. We discuss the details of fitting the
color-octet matrix elements for $J/\psi$ ($\Upsilon$) production in
Appendix~\ref{appendix:LHCbaseJ} (Appendix~\ref{appendix:LHCbaseU}). The
contribution of feed-down from $B-$decay is given in
Appendix~\ref{appendix:Bfeed} and results for $R_{AA}$ at forward rapidity at LHC
energies is shown in Appendix~\ref{appendix:y3}.

\section{Quarkonium production in p$+$p collisions~\label{section:ppProduction}}

In this section we describe the production of quarkonia at high transverse
momenta in p$+$p collisions. This provides a baseline for the calculation of
the nuclear modification factor $R_{AB} (p_T)$ defined above in
Eq.~\ref{eq:RAAdef}.  It also gives the initial unquenched spectrum of
``proto-quarkonia'' in the heavy ion collision, excluding CNM effects and
effects of the propagation of the quarkonium states through the QGP  medium.

The dominant processes in evaluating the differential yields of heavy mesons 
as a function of $p_T$ are the $2\rightarrow 2$
processes of the kind $g+q\rightarrow H+q$, $q+\bar{q}\rightarrow H+g$ and
$g+g\rightarrow H+g$, where $H$ refers to the heavy meson. We label the process
generically as $a+b\rightarrow c+d$, where $a$ and $b$ are light incident
partons, $c$ refers to $H$ and $d$ is a light final-state parton. Given the 
scattering matrix for the process, ${\cal{M}}_{ab\rightarrow cd}$, the cross section has
the form
\begin{equation}
\begin{split}
\frac{d\sigma}{dp_Tdy} &= \int dx_a \phi_{a}(x_a, \mu_F)\phi_{b}(x_b, \mu_F) 
\frac{2p_T}{x_a-\frac{m_T}{\sqrt{S}}e^{y}} x_a x_b
\frac{d\sigma}{d\hatt}(ab\rightarrow cd)\;,~\label{eq:dsigmabydpT}
\end{split}
\end{equation}
where $\phi_a$ ($\phi_b$) is the distribution function of  parton $a$ ($b$) in
the incident hadron traveling in the $+z$ ($-z$) direction. (In our
calculations we use leading order (LO) 2008 MSTW distribution 
functions~\cite{MSTW}.) We denote by $x_a$ ($x_b$)  the fraction of the large light-cone  momentum of the hadron 
carried by the parton. In Eq.~\ref{eq:dsigmabydpT}, $\sqrt{S}$ is the  center-of-mass energy of the
incident hadrons. Momentum-energy conservation fixes
\begin{equation}
x_b = \frac{1}{\sqrt{S}}
\frac{x_a\sqrt{S}m_Te^{-y}-m_H^2}{x_a\sqrt{S}-m_T e^{y}}.
\end{equation}
We take the factorization and renormalization scales $\mu_F$, $\mu_R$ to be
$m_T=\sqrt{p_T^2+m_H^2}$, where $m_H\sim2m_Q$ is the meson mass. We will
analyze the uncertainty associated with the scale by varying the scale from
$m_T/2$ to $2m_T$. The invariant cross section is given by 
\begin{equation}
\frac{d\sigma}{d\hatt} = \frac{|{\cal{M}}|^2}{16\pi\hats^2}\;,
~\label{eq:dsigmavsMsq}
\end{equation}
where $\hat{s}$, $\hat{t}$, and $\hat{u}$ are the parton level 
Mandelstam variables.

\subsection{NRQCD calculation of $d\sigma/d\hatt$~\label{section:Formalism}}
We use LO NRQCD~\cite{Bodwin:1994jh} results to calculate
the production of quarkonia in p$+$p collisions. NRQCD provides a systematic
procedure to compute any quantity as an expansion in the relative velocity $v$
of the heavy quarks in the meson. For example, the wavefunction of the $J/\psi$
meson (analogous expressions hold for the $\psi(2S)$, $\Upsilon(1S)$,
$\Upsilon(2S)$ and $\Upsilon(3S)$) is written as
\begin{equation}
\begin{split}
|J/\psi\rangle = &|Q\barQ([^3S_1]_{1})\rangle 
  + \calO(v)|Q\barQ([^1S_0]_{8}g)\rangle 
  + \calO(v^2)|Q\barQ([^3S_1]_{8}gg)\rangle\\
  &+ \calO(v^1)|Q\barQ([^3P_0]_{8}g)\rangle
  + \calO(v^1)|Q\barQ([^3P_1]_{8}g)\rangle
  + \calO(v^1)|Q\barQ([^3P_2]_{8}g)\rangle
  +\cdot\cdot\cdot
  \label{eq:Jfock}
\end{split}
\end{equation}
The differential cross section for the prompt (as opposed to inclusive, which
includes contributions from $B-$hadron decay) and direct (as opposed to
indirect, from the decay of heavier charmed mesons) production of $J/\psi$ can
also be calculated in NRQCD. It can be written as the sum of the contributions,
\begin{equation}
\begin{split}
d\sigma(J/\psi) &= d\sigma(Q\barQ([^3S_1]_{1}))
                  \langle \calO(Q\barQ([^3S_1]_{1})\rightarrow J/\psi)\rangle 
                +  d\sigma(Q\barQ([^1S_0]_{8}))
                  \langle \calO(Q\barQ([^1S_0]_{8})\rightarrow J/\psi)\rangle\\ 
                &+  d\sigma(Q\barQ([^3S_1]_{8}))
                  \langle \calO(Q\barQ([^3S_1]_{8})\rightarrow J/\psi)\rangle 
                +  d\sigma(Q\barQ([^3P_0]_{8}))
                  \langle \calO(Q\barQ([^3P_0]_{8})\rightarrow J/\psi)\rangle\\ 
                &+  d\sigma(Q\barQ([^3P_1]_{8}))
                  \langle \calO(Q\barQ([^3P_1]_{8})\rightarrow J/\psi)\rangle
                +  d\sigma(Q\barQ([^3P_2]_{8}))
                  \langle \calO(Q\barQ([^3P_2]_{8})\rightarrow J/\psi)\rangle
                + \cdot\cdot\cdot  \;,
\label{eq:dsigmaJ}
\end{split}
\end{equation}
where the quantity in the brackets $[\;]$ represents the angular momentum
quantum numbers of the $Q\barQ$ pair in the Fock expansion. The subscript on
$[\;]$ refers to the color structure of the $Q\barQ$ pair, $1$ being 
the color-singlet and $8$ being the color-octet. The dots represent terms which contribute at
higher powers of $v$. The short distance cross sections $d\sigma(Q\barQ)$
correspond to the production of a $Q\barQ$ pair in a particular color and spin
configuration, while the long distance matrix element
$\langle\calO(Q\barQ)\rightarrow J/\psi\rangle$ corresponds to the probability
of the $Q\barQ$ state to convert to the quarkonium wavefunction. This
probability includes any necessary prompt emission of soft gluons to prepare a
color neutral system that matches onto the corresponding Fock component of the
quarkonium wavefunction.

Power counting rules tell us that contributions from the color-octet matrix
elements in Eq.~\ref{eq:dsigmaJ} are suppressed by $v^4$ compared to the color
singlet matrix elements. More specifically,
\begin{equation}
\begin{split}
\langle \calO(Q\barQ([^3S_1]_{1})\rightarrow J/\psi)\rangle &= \calO(m_Q^3 v^3)\;, \\
\langle \calO(Q\barQ([^3S_1]_{8})\rightarrow J/\psi)\rangle &= \calO(m_Q^3 v^7)\;, \\
\langle \calO(Q\barQ([^1S_0]_{8})\rightarrow J/\psi)\rangle &= \calO(m_Q^3 v^7)\;, \\
\langle \calO(Q\barQ([^3P_J]_{8})\rightarrow J/\psi)\rangle &= 
\calO(m_Q^5 v^7) \;. ~\label{eq:scalingsJ}
\end{split}
\end{equation}
These operators are multiplied by the short distance differential cross sections,
which are related to the probability to create $Q\barQ$ pairs in specific
quantum states. Since these are short distance operators, they can be
calculated in perturbation theory. We use the expressions for the short
distance color-singlet cross sections given in~\cite{Baier:1983, Humpert:1987} 
and the color-octet cross sections given in~\cite{Cho:1995ce,Cho:1995vh,Braaten:2000cm}.

The case of the $p$-wave bound states ($\chi_{c0}$, $\chi_{c1}$, and
$\chi_{c2}$, sometimes collectively referred to as $\chi_{cJ}$, and the
corresponding states of the $b$ quark) is slightly
different. The wavefunction of $\chi_c$ states can be written as
\begin{equation}
\begin{split}
|\chi_{cJ}\rangle = & |Q\barQ([^3P_J]_{1})\rangle 
  + \calO(v)|Q\barQ([^3S_1]_{8}g)\rangle
  + \calO(v^2)|Q\barQ([^1S_0]_{8}g)\rangle
  + \calO(v)|Q\barQ([^3D_J]_{8}g)\rangle\\
  &+ \calO(v^2)|Q\barQ([^1P_1]_{8}g)\rangle
  + \calO(v^2)|Q\barQ([^3P_J]_{8}gg)\rangle
  +\cdot\cdot\cdot
  \label{eq:chifock}
\end{split}
\end{equation}
The color-singlet state $Q\barQ[^3P_J]_{1}$ and the color-octet state
$Q\barQ[^3S_1]_{8}$ contribute to the same order in $v$ because of the angular
momentum barrier for $p-$wave states, and hence both need to be included for a consistent
calculation in $v$.  For the calculation of the production cross section, we
consistently take the contributions to the lowest order in $v$.
For example the $\chi_{c}$ contribution is
\begin{equation}
\begin{split}
d\sigma(\chi_{cJ}) &= d\sigma(Q\barQ([^3P_J]_{1}))
                  \langle \calO(Q\barQ([^3P_J]_{1})\rightarrow \chi_{cJ})\rangle 
                +  d\sigma(Q\barQ([^3S_1]_{8}))
                  \langle \calO(Q\barQ([^3S_1]_{8})\rightarrow \chi_{cJ})\rangle
                + \cdot\cdot\cdot  
\label{eq:dsigmachi}
\end{split}
\end{equation}
Similar expressions hold for the $\chi_b(1P), \chi_b(2P)$ and $\chi_b(3P)$ 
mesons. The expressions for the short distance coefficients are given
in~\cite{Cho:1995vh}. The scaling of the matrix elements is given as
\begin{equation}
\begin{split}
\langle \calO(Q\barQ([^3P_J]_{1})\rightarrow \chi_{cJ})\rangle &= 
\calO(m_Q^5 v^5) \;,\\
\langle \calO(Q\barQ([^3S_1]_{8})\rightarrow \chi_{cJ})\rangle &= 
\calO(m_Q^5 v^5)\;.~\label{eq:scalingschi}
\end{split}
\end{equation}

Therefore we need the color-singlet and color-octet matrix elements
to obtain theoretical results for the production of quarkonia at RHIC for
$\sqrt{S}=0.2$~TeV and at the LHC for $\sqrt{S}=2.76$~TeV. We estimate the matrix
elements by fitting to yields obtained at the TeVatron, RHIC, and at the LHC.

\subsection{Feed-down contributions}
In this paper we will focus on the $p_T$ differential yields of $J/\psi$,
$\Upsilon(1S)$, $\Upsilon(2S)$, and $\Upsilon(3S)$ mesons.
Section~\ref{section:Formalism} gives expressions for the direct production
cross sections of these and for the $p-$wave quarkonia. Excited states of the
mesons decay to the states of lower energy on a short time scale and therefore
we include these feed-down contributions to obtain what is called the prompt
yield.

For example, $\chi_{cJ}$ and $\psi(2S)$ contribute to the prompt yields of
$J/\psi$.  Therefore we need the color-singlet and color-octet matrix elements
for each of these species, which we give in Section~\ref{section:Matrix}.
(Details about the fitting procedure to obtain the color-octet matrix elements
and the feed-down contributions to prompt $J/\psi$ are discussed in
Appendix~\ref{appendix:LHCbaseJ}.)

$B$ hadrons can also decay to $J/\psi$ with a net effective branching fraction
$B(H_b\rightarrow J/\psi+X)_{\rm {eff}}
=1.16\times10^{-2}$~\cite{Acosta:2004yw}. In particular, at high $p_T$,
the contribution to the inclusive yield from the decay of $B-$hadrons is
substantial, and can possibly even dominate production.  At the LHC and the
TeVatron the $B-$decay contributions have been separately
measured~\cite{Acosta:2004yw,Aad:2011sp}, while
RHIC~\cite{Adare:2009js,Abelev:2009qaa} reports the inclusive yields. In
Fig.~\ref{fig:Jyields} in Section~\ref{section:ppyields} we will only show
the prompt production yields and discuss the $B$ feed-down contribution in
Appendix~\ref{appendix:Bfeed}.

Similarly, for $\Upsilon(1S)$ production, we consider states up to $\Upsilon(3S)$.
The relevant color-singlet and -octet matrix elements are given in
Section~\ref{section:Matrix}. (Details about the fitting procedure and the
feed-down contributions to prompt $\Upsilon(1S)$ are discussed in
Appendix~\ref{appendix:LHCbaseU}.)

\subsection{Matrix elements for quarkonia production~\label{section:Matrix}}

In this work, following~\cite{Cho:1995ce,Cho:1995vh} we use the values of the
color-singlet operators calculated using the potential model.  The expressions
and the values for the color-singlet operators are given
in~\cite{Cho:1995ce,Cho:1995vh,Eichten:1994gt}. The values are obtained by
solving the non-relativistic wavefunctions:
\begin{equation}
\begin{split}
\begin{array}{ccc}
\langle \calO(c\barc([^3S_1]_{1})\rightarrow J/\psi)\rangle 
=3\langle \calO(c\barc([^1S_0]_{1})\rightarrow J/\psi)\rangle
&=3N_c\frac{|R_{n=1}(0)|^2}{2\pi}
&=1.2\;{\rm GeV^3} \;, \\
\langle \calO(c\barc([^3S_1]_{1})\rightarrow \psi(2S))\rangle 
=3\langle \calO(c\barc([^1S_0]_{1})\rightarrow \psi(2S))\rangle
&=3N_c\frac{|R_{n=2}(0)|^2}{2\pi}
&=0.76\;{\rm GeV^3} \;, \\
\frac{1}{5}\langle \calO(c\barc([^3P_2]_{1})\rightarrow \chi_{c2}(1P))\rangle
=\frac{1}{3}\langle \calO(c\barc([^3P_1]_{1})\rightarrow \chi_{c1}(1P))\rangle 
=&&\\ 
\langle \calO(c\barc([^3P_0]_{1})\rightarrow \chi_{c0}(1P))\rangle 
&=3N_c\frac{|R^\prime_{n=1}(0)|^2}{2\pi}
&=0.054 m_{\charm}^2\;{\rm GeV^3}\; ,
\end{array}
\label{eq:charmsinglet}
\end{split}
\end{equation}
where $R(0)$ is the radial wavefunction at the origin, $R^\prime(0)$ is the
first derivative of the radial wavefunction at the origin, and $n$ refers to
the radial quantum number. We take the mass of the charm quark,
$m_{\charm}=1.4$GeV~\cite{Cooper:LongAndShort}. 

The values of the color-singlet operators for  bottomonia are given
in~\cite{Braaten:2000cm}, which we reproduce here:
\begin{equation}
\begin{split}
\begin{array}{ccc}
\langle \calO(b\barb([^3S_1]_{1})\rightarrow \Upsilon(1S))\rangle &=
3N_c\frac{|R_{n=1}(0)|^2}{2\pi}
&=10.9\;{\rm GeV^3} \;, \\
\langle \calO(b\barb([^3P_0]_{1})\rightarrow \chi_{b0}(1P))\rangle &= 
3N_c\frac{|R^\prime_{n=1}(0)|^2}{2\pi}
&=0.100 m_{\bottom}^2\;{\rm GeV^3}\;, \\
\langle \calO(b\barb([^3S_1]_{1})\rightarrow \Upsilon(2S))\rangle &=
3N_c\frac{|R_{n=2}(0)|^2}{2\pi}
&=4.5\;{\rm GeV^3}\;, \\
\langle \calO(b\barb([^3P_0]_{1})\rightarrow \chi_{b0}(2P))\rangle &= 
3N_c\frac{|R^\prime_{n=2}(0)|^2}{2\pi}
&=0.036 m_{\bottom}^2 \;{\rm GeV^3} \;, \\
\langle \calO(b\barb([^3S_1]_{1})\rightarrow \Upsilon(3S))\rangle &=
3N_c\frac{|R_{n=3}(0)|^2}{2\pi}
&=4.3\;{\rm GeV^3} \; ,
\end{array}
\label{eq:bottomsinglet}
\end{split}
\end{equation}
where we use $m_{\bottom}=4.88$GeV~\cite{Cho:1995ce,Cho:1995vh}.

The color-octet operators can not be related to the non-relativistic
wavefunctions of $Q\barQ$ since it involves a higher Fock state.
Following~\cite{Cho:1995ce,Cho:1995vh,Braaten:2000cm}, we fit them to the data.

For the charm mesons we use data from the
TeVatron~\cite{Abe:1997older,Abe:1997yz,Acosta:2004yw,Aaltonen:2009dm},
RHIC~\cite{Adare:2009js,Abelev:2009qaa}, and the
LHC~\cite{Chatrchyan:2011kc,Aad:2011sp} (see Fig.~\ref{fig:Jyields}, and
Appendix~\ref{appendix:LHCbaseJ} for details).  We obtain the following values:
\begin{equation}
\begin{split}
\langle \calO(c\barc([^3S_1]_{8})\rightarrow J/\psi)\rangle
&=(0.0013\pm0.0013)\;{\rm GeV^3} \;, \\
\langle \calO(c\barc([^1S_0]_{8})\rightarrow J/\psi)\rangle 
&=(0.018\pm0.0087)\;{\rm GeV^3}\;, \\
&=\langle \calO(c\barc([^3P_0]_{8})\rightarrow J/\psi)\rangle/(m_{\charm}^2) 
\;, \\
\langle \calO(c\barc([^3S_1]_{8})\rightarrow \psi(2S))\rangle
&=(0.0033\pm0.00021)\;{\rm GeV^3} \;, \\
\langle \calO(c\barc([^1S_0]_{8})\rightarrow \psi(2S))\rangle 
&=(0.0080\pm0.00067)\;{\rm GeV^3}\;, \\
&=\langle \calO(c\barc([^3P_0]_{8})\rightarrow J/\psi)\rangle/(m_{\charm}^2) 
\;, \\
\langle \calO(c\barc([^3P_1]_{8})\rightarrow J/\psi)\rangle 
&=3\times\langle \calO(c\barc([^3P_0]_{8})\rightarrow J/\psi)\rangle \;, \\
\langle \calO(c\barc([^3P_2]_{8})\rightarrow J/\psi)\rangle 
&=5\times\langle \calO(c\barc([^3P_0]_{8})\rightarrow J/\psi)\rangle\;, \\
\langle \calO(c\barc([^3S_1]_{8})\rightarrow \chi_{c0}(1P))\rangle 
&=(0.00187\pm0.00025)\;{\rm GeV^3}\;, \\
\label{eq:charmoctet} \;.
\end{split}
\end{equation}
Since the shape of the short distance part as a function of $p_T$ is very
similar for $[^1S_0]_8$ and $[^3P_0]_8$
contributions~\cite{Cho:1995ce,Cho:1995vh}, we do not attempt to fit the two
long distance matrix elements separately and only fit a linear combination. The
$\chi^2/{\rm{dof}}=4.56$ for the $\chi_{cJ}$ feed-down contribution,
$\chi^2/{\rm{dof}}=5.6$ for the $\psi(2S)$ contribution, and
$\chi^2/{\rm{dof}}=5.2$ for the $J/\psi$ direct production.

The octet matrix elements for the bottomonia are obtained by fitting
TeVatron~\cite{Acosta:2001gv} and the LHC~\cite{Khachatryan:2010zg} (see
Fig.~\ref{fig:Uyields}, and Appendix~\ref{appendix:LHCbaseU} for details) data
and are as follows:
\begin{equation}
\begin{split}
\langle \calO(b\barb([^3S_1]_{8})\rightarrow \Upsilon(1S))\rangle
&=(0.0477\pm0.0334)\;{\rm GeV^3} \;,\\
\langle \calO(b\barb([^1S_0]_{8})\rightarrow \Upsilon(1S))\rangle
&=(0.0121\pm0.040)\;{\rm GeV^3}\;, \\
&=\langle \calO(b\barb([3P_0]_{8})\rightarrow \Upsilon(1S))\rangle
/(5m_{\bottom}^2) \;, \\
\langle \calO(b\barb([^3S_1]_{8})\rightarrow \chi_{b0}(1P))\rangle
&=(0.1008)\;{\rm GeV^3}\;, \\
\langle \calO(b\barb([^3S_1]_{8})\rightarrow \Upsilon(2S))\rangle
&=(0.0224\pm0.02)\;{\rm GeV^3}\;, \\
\langle \calO(b\barb([^1S_0]_{8})\rightarrow \Upsilon(2S))\rangle
&=(-0.0067\pm0.0084)\;{\rm GeV^3}\;, \\
&=\langle \calO(b\barb([3P_0]_{8})\rightarrow \Upsilon(2S))\rangle
/(5m_{\bottom}^2)\\
\langle \calO(b\barb([^3S_1]_{8})\rightarrow \chi_{b0}(2P))\rangle
&=(0.0324)\;{\rm GeV^3}\; ,\\
\langle \calO(b\barb([^3S_1]_{8})\rightarrow \Upsilon(3S))\rangle
&=(0.0513\pm0.0085)\;{\rm GeV^3}\;, \\
\langle \calO(b\barb([^1S_0]_{8})\rightarrow \Upsilon(3S))\rangle
&=(0.0002\pm0.0062)\;{\rm GeV^3}\;, \\
&=\langle \calO(b\barb([3P_0]_{8})\rightarrow \Upsilon(3S))\rangle
/(5m_{\bottom}^2)
\label{eq:bottomoctet}
\end{split}
\end{equation}
The $\chi^2/{\rm{dof}}=1.3$ for $\Upsilon(3S)$, $3.5$ for $\Upsilon(2S)$ and
$3.8$ for $\Upsilon(1S)$.

For a more sophisticated fitting of the color-octet matrix elements including
NLO effects, see~\cite{Butenschoen:2010rq, Butenschoen:Long,
Butenschoen:polarised}. The matrix elements including feed-down effects
obtained in~\cite{Butenschoen:polarised} are similar, albeit slightly smaller
than our estimates.

\subsection{Quarkonia yields in p$+$p and p+${\bar{\rm p}}$ 
collisions~\label{section:ppyields}}

\begin{figure}[!ht]
\vspace*{.2in} 
\includegraphics[width=3.38in,angle=0]{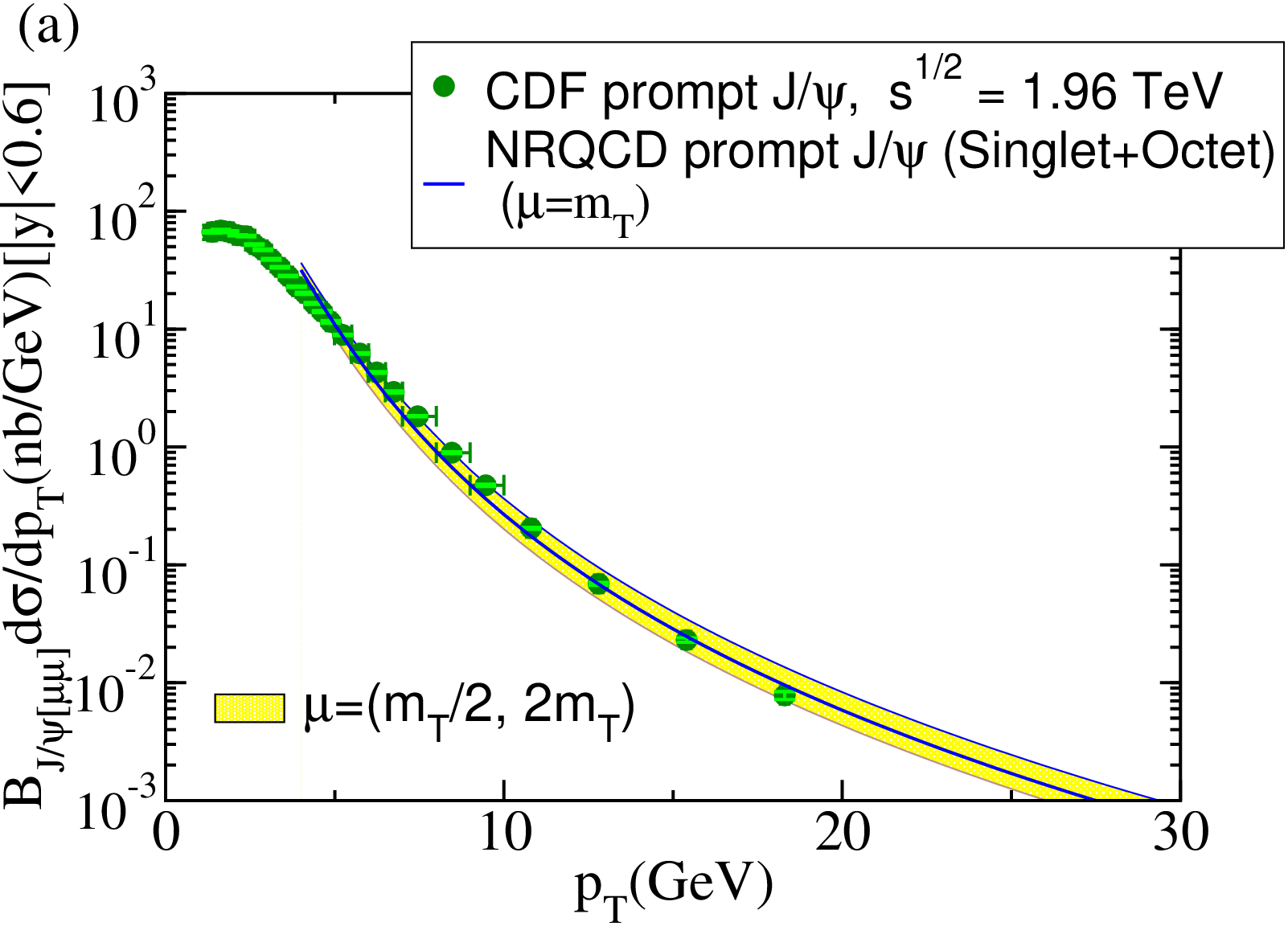} 
\includegraphics[width=3.38in,angle=0]{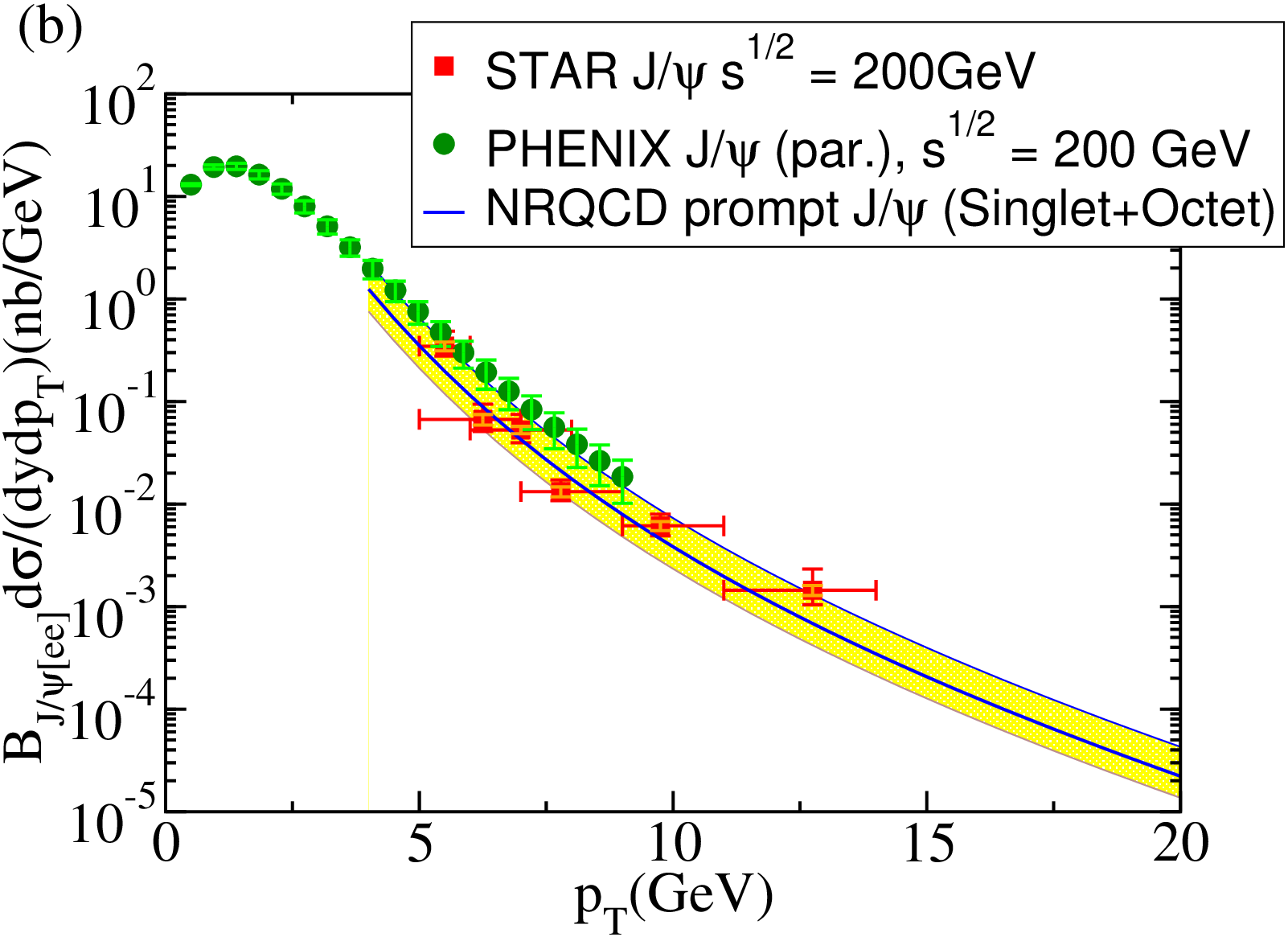} 
\includegraphics[width=3.38in,angle=0]{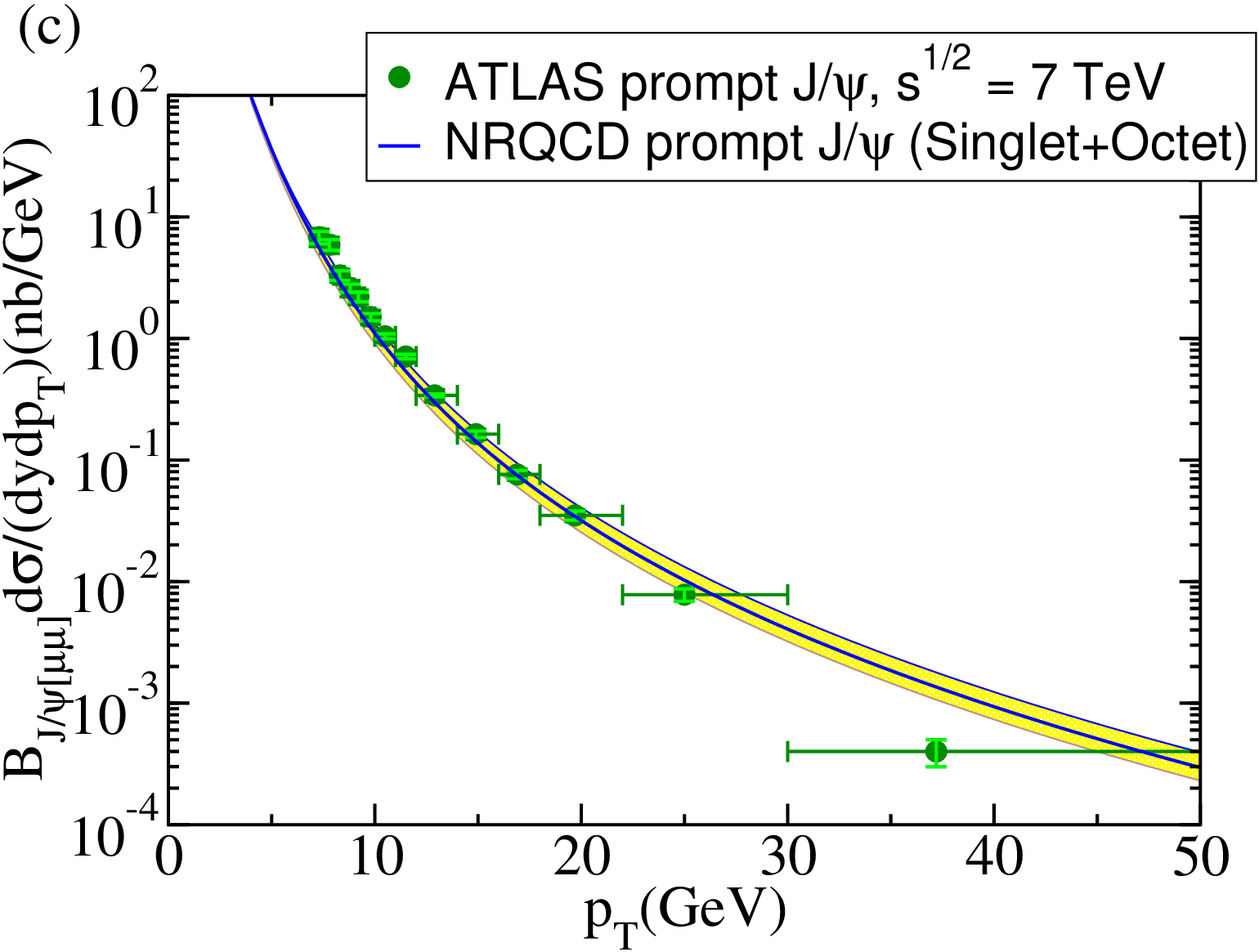} 
\caption{(Color online) $J/\psi$ production yields multiplied by the branching ratios,
$B(J/\psi\rightarrow ee) \simeq B(J/\psi\rightarrow \mu\mu) \simeq 5.93\%$.
The upper left panel corresponds to data for the yields of $J/\psi$ from
CDF at $\sqrt{S}=1.96$~TeV~\cite{Acosta:2004yw}. The upper right panel is for
RHIC at $\sqrt{S}=0.2$~TeV with inclusive data from the PHENIX
experiment~\cite{Adare:2009js} (using the fit given in ~\cite{Adare:2009js}) and the STAR
experiment~\cite{Abelev:2009qaa}. The bottom panel has prompt data from the
LHC at $\sqrt{S}=7$~TeV from the ATLAS collaboration~\cite{Aad:2011sp}
The darker colored error bars (dark green or red) give the systematic
errors and the lighter colored give the statistical errors in the experimental
data. The solid line (blue) are the theoretically calculated prompt
yields. We show the uncertainty associated with changing the renormalization 
scale with a band (yellow). The upper curves are obtained for $\mu_R=\mu_F=m_T/2$,
the central for $\mu_R=\mu_F=m_T$ and the lower for $\mu_R=\mu_F=2m_T$. The CDF
results are for rapidity $|y|<0.6$, and the RHIC and LHC results are quoted per
unit rapidity at mid rapidity. \label{fig:Jyields}} 
\end{figure}

Shown in Fig.~\ref{fig:Jyields} are yields for $J/\psi$ production at the
TeVatron, RHIC and the LHC at $\sqrt{S}=1.96,\; 0.2$, and $7$~TeV,
respectively.  The data from CDF~\cite{Acosta:2004yw} at the TeVatron and from
ATLAS~\cite{Aad:2011sp}  at the LHC shows the prompt production and can,
therefore, directly be compared to the theoretical curves of the prompt yields
obtained from NRQCD and shown in Fig.~\ref{fig:Jyields}. Comparison with prompt
production yields measured by CMS~\cite{Chatrchyan:2012np} at $\sqrt{S} =
2.76$~TeV is given in Fig.~\ref{fig:LHC2760Jcocktail} in
Appendix~\ref{appendix:LHCbaseJ} where the feed-down contributions from
$\chi_{cJ}$ and $\psi(2S)$ are also shown explicitly.

The PHENIX~\cite{Adare:2009js} and STAR~\cite{Abelev:2009qaa} data from RHIC
shows the inclusive yields. The theoretical curve is for the prompt yield.
Including the $B$ feed-down contribution at RHIC energies gives an inclusive
yield roughly $20-50\%$ larger, slightly improving agreement with PHENIX data.
The ratios of the $B$ feed-down contributions to the prompt yields are shown in
Appendix~\ref{appendix:Bfeed}.

We also show the uncertainty associated with the factorization and renormalization 
scales $\mu_R$, $\mu_F$ by taking $\mu_R=\mu_F=m_T/2$ (upper curves),
$\mu_r=\mu_F=m_T$ (central curves) and $\mu_R=\mu_F=2m_T$ (lower curves). This
scale variation affects the cross section in the following two ways (as can be
seen in Eq.~\ref{eq:dsigmabydpT}). First, the parton distribution functions at
a higher scale for the same value of $x$ are smaller in the regime of interest.
Second, the strong coupling constant $\alpha_s$ decreases with increasing
scale. These effects give a variation in the yields as shown in
Fig.~\ref{fig:Jyields}.

We see that the NRQCD results are a bit less steep than the data. The results
for RHIC at $\sqrt{S}=0.2$~TeV provide the baseline for the p$+$p yield, which
we use to calculate $R_{AA}$. For the LHC, we need the baseline at a
center-of-mass energy $\sqrt{S}=2.76$~TeV, which is given in
Fig.~\ref{fig:LHC2760Jcocktail} in Appendix~\ref{appendix:LHCbaseJ}.

\begin{figure}[!ht]
\vspace*{.2in}
\includegraphics[width=3.38in,angle=0]{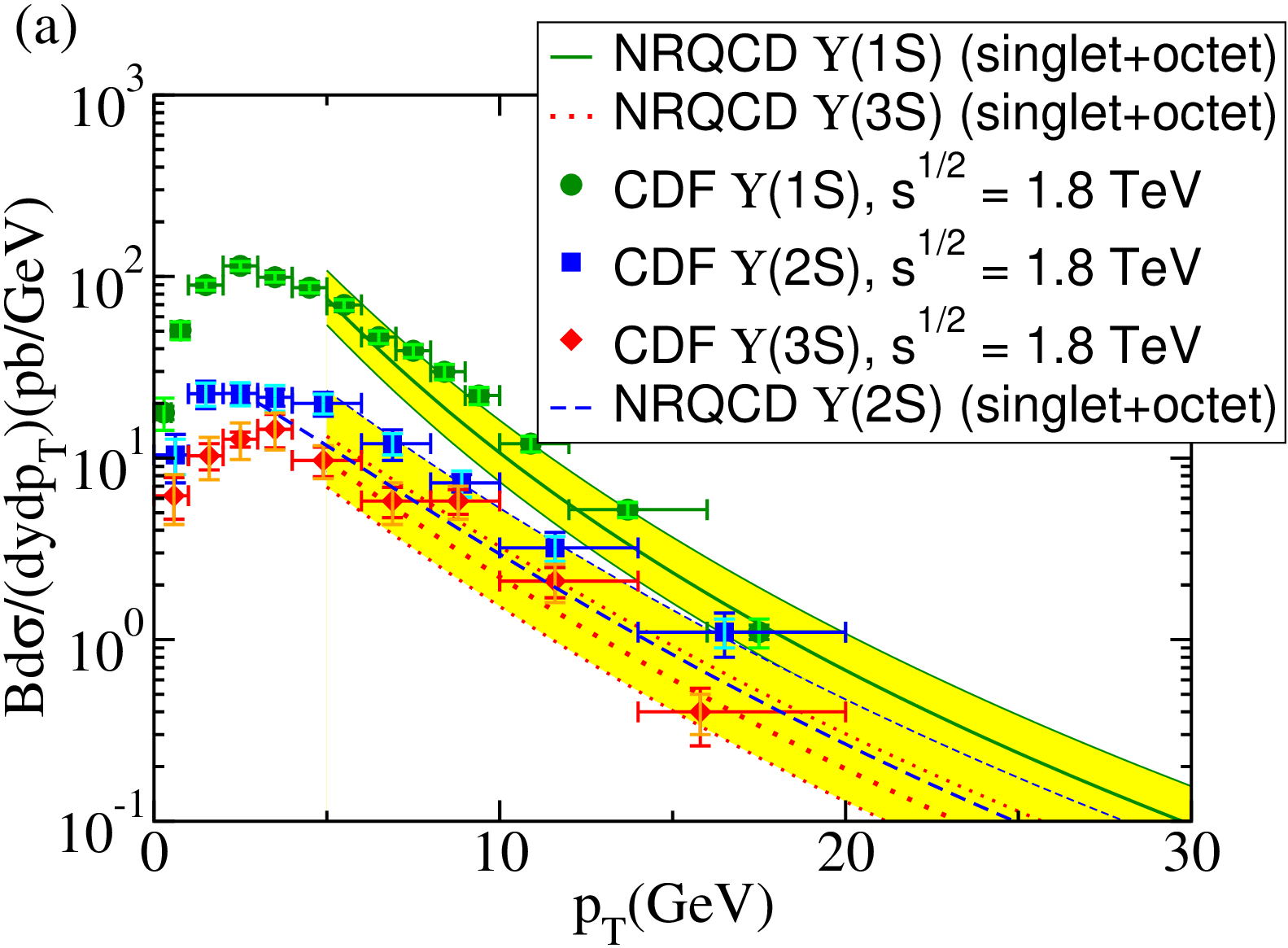} 
\includegraphics[width=3.38in,angle=0]{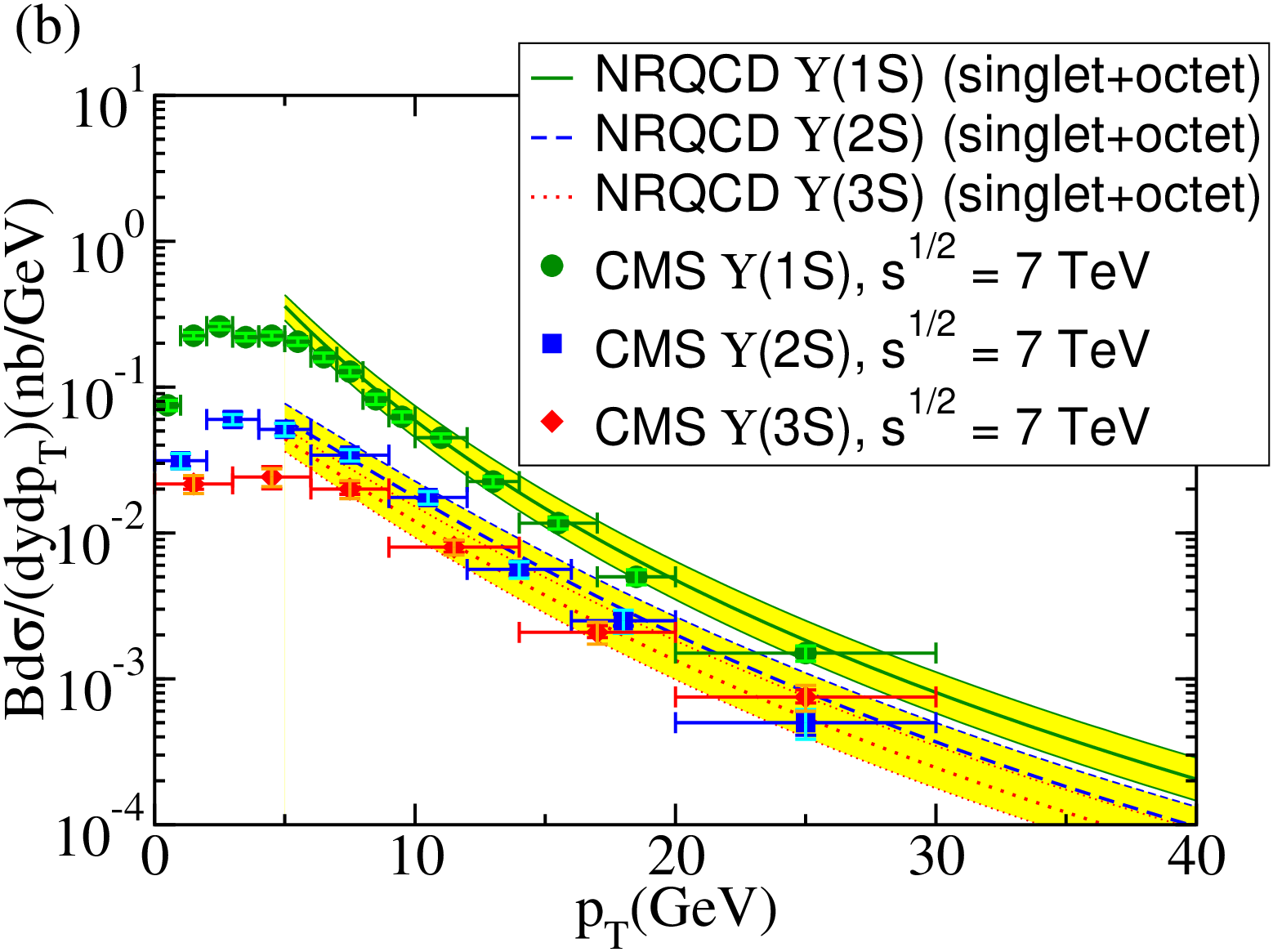}
\caption{(Color online) $\Upsilon$ yields multiplied by the corresponding
$B(\Upsilon\rightarrow\mu\mu)$. The left panel shows data from the TeVatron at
$1.8$~TeV~\cite{Acosta:2001gv}.  The right panel corresponds to data from CMS
at $7$~TeV~\cite{Khachatryan:2010zg}. The solid line (green) is for
$\Upsilon(1S)$, the dashed (blue) for $\Upsilon(2S)$ and the dotted (red) for
$\Upsilon(3S) $. The yellow band shows the uncertainty associated with varying
the factorization and the renormalization scale from $m_T/2$ to
$2m_T$.~\label{fig:Uyields}}
\end{figure}

The yields of $b\barb$ states are shown in Fig.~\ref{fig:Uyields}. For the
$\Upsilon(3S)$ we only consider the direct production and ignore feed-down.
For $\Upsilon(2S)$ we include feed-down from $\Upsilon(3S)$ and $\chi_b(2)$.
For $\Upsilon(1S)$ there is additional feed-down from $\Upsilon(2S)$ and
$\chi_b(1)$. 

We see that the values of matrix elements that give a good agreement with the
LHC data~\cite{Khachatryan:2010zg} give slightly smaller than measured yields
at the TeVatron~\cite{Acosta:2001gv}.  We expect that the deviation from the
measured yields will be smaller at $\sqrt{S}=2.76$~TeV, for which we calculate
$R_{AA}$. The baseline for this energy is shown in
Fig.~\ref{fig:LHC2760Ucocktail} in Appendix~\ref{appendix:LHCbaseU}, where the
feed-down contributions are also given.

\section{Cold Nuclear Matter  effects~\label{section:CNM}}

In heavy ion reactions, the production yields of energetic particles are always
affected by cold nuclear matter (CNM) effects.  These include nuclear
shadowing, initial state energy loss and transverse momentum broadening (also
interpreted as the origin of the Cronin effect).  All these effects are
grow linearly with the system size $L$, which for CNM effects
is proportional to the nuclear size $R\approx 1.2 A^{1/3}$~fm. In our
calculation, the CNM effects are calculated from the elastic, inelastic and
coherent scattering processes of partons in large nuclei~\cite{Vitev:2006bi}.

\begin{enumerate}
\item{Nuclear shadowing: Shadowing generally refers to the suppression  of the
inclusive deep-inelastic scattering (DIS) cross section per nucleon   in
reactions with  a nuclear target  relative to the corresponding cross section
in reactions with a proton target. For Bjorken-$x$ larger than $0.25$
(EMC and the Fermi motion regions) the effect is mainly due to collective effects in the
nucleus~\cite{Norton:2003}. In this region shadowing can be parametrized as a
modification of the parton distribution function $\phi(x)$ by a factor which
depends on $x$. We use the EKS98 modification factor provided
in~\cite{eks1:98}. For $x<0.1$,  power-suppressed
resummed~\cite{Qiu:2003vd,Qiu:2004da} coherent final-state scattering of the
struck partons leads to suppression in the observed cross sections. These are
included in our calculation and modify  the momentum fractions of the incident
partons in Eq.~\ref{eq:dsigmabydpT} as follows~\cite{Vitev:2006bi}:
\begin{equation}
   \tilde{x}_{a} =
   x_a\left[1+\frac{\xi_d^2(A^{1/3}-1)}{-\hat{t}+m_d^2}\right]\;,\qquad
  \tilde{x}_{b} = 
   x_b\left[1+\frac{\xi_c^2(A^{1/3}-1)}{-\hat{u}+m_c^2}\right]\; , 
\label{shadb}
\end{equation}
where $x_{a,\;b}$ are the Bjorken-$x$ of the colliding partons.  In
Eqs.~\ref{shadb}  $\hat{t}, \; \hat{u}$ are the relevant Mandelstam variables
at the partonic level and $m_c$, $m_d$ are the masses of the struck partons
labeled as in Eq.~\ref{eq:dsigmabydpT}. In this case we treat the octet $Q\barQ$
state as a massive gluon  with $m_c=m_H$. Immediately after the hard scattering
the unexpanded color-singlet state does not couple to the medium.
(Equivalently, its color factor is 0.) Following previous
studies~\cite{Sharma:2009hn}, we use $ (\xi_{q,g}^2 A^{1/3}) \approx ( 2 \mu^2
L/ \lambda_{q,g} )$, which yields $(\xi_q^2) \approx 0.12$~GeV$^2$ and
$(\xi_g^2) \approx 0.27$~GeV$^2$. Here $\mu^2$ is related to the gluon density
in the nucleus~\cite{Vitev:2006bi}, and $\lambda_{q(\;g)}$ are the quark
(gluon) mean free paths respectively. The parameters $\mu$ and
$\lambda_{q,\;g}$ also determine the transverse momentum broadening as we shall
see below in item~3.}

\item{Initial state energy loss: Before the large $Q^2$ parton scattering process, 
the incoming partons lose radiatively a fraction of their energy due to 
multiple interactions in the target nucleus. If the colliding partons $a$, $b$ lose
fractional energy $\epsilon_{a,\;b}=\frac{\Delta E_{a,\;b}}{E_{a,\;b}}$, to
satisfy the same final-state kinematics they must initially carry a larger
fraction of the colliding hadron momentum and, correspondingly, a larger value
of $x$. 
This can be implemented in Eq.~\ref{eq:dsigmabydpT} by the following
modification
\begin{equation}
\phi_a({\tilde{x}}_{a}) \rightarrow \tilde{\phi}_a(x_a) 
= \phi_a\left(\frac{ {\tilde{x}}_{a}}{1-\epsilon_{a}}\right) \;,\quad
\phi_b({\tilde{x}}_{b}) \rightarrow \tilde{\phi}_b(x_b)
=\phi_b\left(\frac{ {\tilde{x}}_{b}}{1-\epsilon_{b}}\right)\;,
\quad 
{\tilde{x}}_{a,b} \leq 1 \; , 
\end{equation}
in the parton distribution functions $\phi_{a,b/N}({\tilde{x}}_{a,b}, \mu_F)$. 
The medium-induced radiative corrections factorize as a standard integral
convolution~\cite{Ovanesyan:2011kn}. If $P_{q,g}(\epsilon)$ is the probability
density for quarks and gluons to lose a fraction $\epsilon$ of their energy, it
can be implemented in the cross section calculation as
follows~\cite{Neufeld:2011}:  
\begin{equation}
\int dx \, \phi_{q}(x) \cdots \rightarrow \int d\tilde{x} \int d\epsilon \,
\phi_q\left(\frac{\tilde{x}}{1-\epsilon}\right)P_q(\epsilon) \;,  \qquad
\int dx \, \phi_{g}(x) \cdots \rightarrow \int d\tilde{x} \int d\epsilon \,
\phi_g\left(\frac{\tilde{x}}{1-\epsilon}\right)P_g(\epsilon) \; .
\end{equation}
The calculation of $P(\epsilon)$ is described in~\cite{Vitev:2006}.}

\item{Cronin effect: In p$+$A and A$+$A reactions, Cronin effect can be modeled at the
level of the $p_T$-differential cross sections by including the $k_T$
(transverse momentum) broadening of incoming partons that arises from
initial-state scattering~\cite{Accardi:2002ik,Vitev:2003xu}. This involves
relaxing the assumption that the incident partons $a,\;b$ in
Eq.~\ref{eq:dsigmabydpT} are collinear and allowing them to carry a
transverse momentum $k_{Ta,\;b}$ with model distributions $f(k_{Ta,\;b})$. The
advantage of using a simple normalized Gaussian form for $f({k}_{Ta,\;b})$ is
the additive variance property:   
\begin{equation}
\langle {k}_{Ta,\;b}^2 \rangle_{AB} = \langle { k}_{Ta,\;b}^2 \rangle_{NN}
+  \langle {k}_{Ta,\;b}^2 \rangle_{\rIS}\, , \;  
\langle {k}_{Ta,\;b}^2 \rangle_{\rIS} = \left\langle \frac{2 \mu^2 L}{\lambda_{q,g}}  
\right\rangle \xi\label{Cron}\;,
\end{equation}
where $\langle {k}_{Ta,\;b}^2 \rangle_{\rIS}$ accounts for increased momentum
broadening in the initial state ($\rIS$) in collisions with nuclei.
Specifically, in Eq.~\ref{Cron}  $\mu^2 = 0.12$~GeV$^2$ appears also in
the calculation of the shadowing, and is also related to the dynamical mass of
gluons in the nucleus. It sets the scale for the typical transverse momentum
exchanged with the gluons of the incident nucleus. $\lambda_g = (C_F/C_A)
\lambda_q$ is the mean free path of the parton, where $C_F/C_A=4/9$
is the ratio of the color Casimir factors in the fundamental and the adjoint
representation. We take $\lambda_g=1$~fm~\cite{Vitev:2003xu}.
 $\xi$ is a dimensionless numerical factor that accounts for the
enhancement of the broadening coming from the power-law tails of the Moliere
multiple scattering~\cite{Vitev:2002} and is expected to be greater than $1$.
Typical values used are $\xi\sim2-3$~\cite{hep-ph/0611109}. For light
final-state partons, the Cronin effect is well studied
phenomenologically~\cite{Vitev:2003xu}. For $D$-meson and $B$-meson production
the Cronin effect was discussed in~\cite{Sharma:2009hn}.  We note that a 33\%
reduction of the logarithmic enhancement factor $\xi$ relative to the one used
in~\cite{Sharma:2009hn} gives a better description of the  $p_T \sim2-10$~GeV
light hadron production and this is what we use in our work. For quarkonia, the
effect of initial-state transverse momentum broadening is not well understood.
To our knowledge, this is the first attempt to calculate the Cronin effect for
$Q\barQ$ production. We do not try to constrain its
magnitude phenomenologically and point out that data from p$+$A reactions is needed to constrain
all CNM effects.   We find that including transverse momentum broadening in the
same way as is done for light final states reduces the suppression for $p_T$
between $3$ and $10$~GeV significantly and may actually lead to a small
enhancement of the charmonium cross section.  It is not clear, given the large
error bars, whether the Au+Au and Cu$+$Cu data at  RHIC are better described by
the Cronin calculation. In contrast, it is evident that the Cronin effect is
not consistent  with Pb$+$Pb data at the LHC, which sees a large attenuation at
$p_T\sim8$~GeV.  In Section~\ref{section:NoCronin} we will discuss the yields
without the Cronin effect. We will discuss the phenomenology of the Cronin
effect in Section~\ref{section:Cronin}.}
\end{enumerate}

\section{Quarkonium Wavefunctions~\label{section:Wavefunction}}

The color-singlet matrix elements are proportional to the square of
the value of the $l^{th}$ derivative of the radial wavefunction of the $Q\barQ$
Fock state at the origin, where $l$ refers to the orbital angular momentum of
the bound state. Also, for the calculation of the collisional dissociation
rate, it is convenient to use wavefunctions in the light-cone coordinate
system. We obtain these by taking the Fourier transformation of the spatial
wavefunctions and changing the momentum variables to light-cone coordinates.
In this section we describe the calculation of these bound state wavefunctions
of the heavy mesons. In the calculations, we will assume that high $p_T$
quarkonia do not thermalize and, therefore, we will use only the zero
temperature wavefunctions.

\subsection{The instantaneous wavefunctions~\label{section:Potential}}
Making the usual separation of the radial and angular parts of the
wavefunction, $\psi(\bfr)=Y_l^m(\hatr)  R_{nl}(r)$, the Schr\"{o}dinger
equation for the radial part of the $Q\barQ$ state can be written as
\begin{equation}
\left[-\frac{1}{2\mu_{\rm{red}}}\frac{\partial^2}{\partial r^2}
 +\frac{l(l+1)}{2\mu_{\rm{red}} r^2}
 +V(r)\right]rR_{nl}(r) = (E_{nl}-2m_Q) rR_{nl}(r)  \; , 
\end{equation}
where $\mu_{\rm{red}}=\frac{m_Q}{2}$ is the reduced mass, $l$ is the angular
quantum number, $n$ the radial quantum number and $E_{nl}$ is the binding
energy. ($n-1$ is the number of non-trivial nodes of $R_{nl}$.)
$V(r)$ is the potential between the two heavy quarks, which can be estimated
from the lattice. We use the potential described in~\cite{Mocsy:2007jz}. The
results are compactly written in Table~\ref{table:Charm} and
Table~\ref{table:Bottom}, where we approximate the radial wavefunction as
$R_{nl}(r)=N_r r^l \Pi_{i=1, n-1}(1-\frac{r}{r_0^i}) \exp^{(-r^2/(2a^2))}$,
where ${r_0^i}$ are the nodes of the wavefunction in radial coordinates.

For the calculation of the dissociation rate, we also require the form of the
wavefunction in momentum space,
\begin{equation}
\tilpsi(\bfk) = \int d^3 r \psi(\bfr) = Y_l^m(\hatk) \tilR_{nl}(k)\;.
\end{equation}
$\tilpsi(\bfk)$ satisfies the normalization condition 
$\int \frac{d^3 k}{(2\pi)^3}|\tilpsi(\bfk)|^2
=\frac{1}{(2\pi)^3}\int dk k^2 |\tilR_{nl}(k)|^2=1$.  The momentum space
wavefunctions are readily related to the light cone wavefunction as described
in~\cite{Sharma:2009hn}. In order to simplify the calculations of the dissociation rates, we approximate
the momentum space wavefunctions by the form $\tilR_{nl}(k)=N_kk^l\Pi_{i=1,
n-1}(1-\frac{k}{k_0^i}) \exp^{(-k^2/(2b^2))}$, where ${k_0^i}$
are the location of the nodes in the momentum space wavefunction. Our results are again presented  in 
Table~\ref{table:Charm} and Table~\ref{table:Bottom}.
The color-singlet operators are given by the expressions Eq.~\ref{eq:charmsinglet}.

\begin{table*}[h]
\begin{tabular}{cccccccccc}
$l$ & $n$ & $E_{nl}$~(GeV) & $a_\perp=\sqrt{\frac{2}{3}\langle r^2\rangle}$~(GeV$^{-1}$) & $k^2$~(GeV$^2$) & $N_r$~(GeV$^{3/2}$)  &$N_k$~(GeV$^{-3/2}$)    & ${r_0}$~(GeV$^{-1}$) & ${k_0}$~(GeV) & Meson \\ 
\hline \hline
0  & 1  & 0.7003    & 1.829           & 0.2988         & 0.6071              & 58.55                 & { }                 & { }           & $J/\psi$         \\
0  & 1  & 0.0858    & 2.978           & 0.1127         & 0.4020              & 296.0                 & {2.375}             & {0.5219}      & $\psi(2S)$      \\
1  & 1  & 0.2678    & 2.216           & 0.2036         & 0.1677              & 141.3                 & {0}                 & {0}           & $\chi_{c0}$, $\chi_{c1}$, $\chi_{c2}$\\
\end{tabular}
\caption{Charmonia wavefunctions. $l$ refers to the angular momentum of the
$Q\barQ$ state, while $n$ is the radial quantum number. $\sqrt{\langle r^2\rangle}$ is the root mean square 
radius of the quarkonium state, $k^2$ is the mean square momentum. $N_r$ is the
normalizaton factor and ${r_0}$ is the set of roots for the radial
wavefunction. $N_k$ and ${k_0}$ are corresponding quantities for the momentum
space wavefunctions.~\label{table:Charm}}
\end{table*}

\begin{table*}[h]
\begin{tabular}{cccccccccc}
 $l$ & $n$ & $E_{nl}$~(GeV) & $a_\perp=\sqrt{\frac{2}{3}\langle r^2\rangle}$~(GeV$^{-1}$) & $k^2$~(GeV$^2$) & $N_r$~(GeV$^{3/2}$)  & $N_k$~(GeV$^{-3/2}$) & ${r_0}$~(GeV$^{-1}$) & ${k_0}$~(GeV)     & Meson\\ 
 \hline \hline
 0  & 1  & 1.122     & 1.007           & 0.9854         & 1.486               & 23.92                 & { }                 & { }               & $\Upsilon(1S)$\\
 0  & 2  & 0.5783    & 1.446           & 0.4784         & 1.479               & 103.7                 & {   1.304}          & {   1.012}        & $\Upsilon(2S)$\\
 0  & 3  & 0.2139    & 1.768           & 0.3199         & 1.813               & 236.4                 & {   1.106, 2.751}   & {   0.615, 1.457} & $\Upsilon(3S)$\\
 1  & 1  & 0.7102    & 1.309           & 0.5832         & 0.6251              & 37.9                  & {0}                 & {0}               & $\chi_{b0,1,2}(1P)$\\
 1  & 2  & 0.325     & 1.588           & 0.3967         & 1.004               & 196.9                 & {0, 2.122}          & {0, 1.105}        & $\chi_{b0,1,2}(2P)$\\
 1  & 3  & 0.05109   & 2.14            & 0.2183         & 0.5635              & 758.4                 & {0, 1.814, 3.621}   & {0, 0.694, 1.453} & $\chi_{b0,1,2}(3P)$\\
\end{tabular}
\caption{Bottomonia wavefunctions. The notation is the same as Table~\ref{table:Charm}.~\label{table:Bottom}}
\end{table*}

Note that, as mentioned earlier, the $Q\barQ$ pair produced in the hard
collision in a color-octet state has to emit soft gluons to overlap with the
wavefunction of the color neutral quarkonium. Eventually, all the components of
the $Q\barQ$ pair give rise to a color neutral quarkonium state, with a
hierarchy of contributions given in Eq.~\ref{eq:Jfock}. We will assume that after
formation, the dynamics of the quarkonium are dominated by the color singlet 
component. To our knowledge
exact solutions for  the $n\geq 3$ Fock components of mesons do not exist.

\subsection{The light-cone wavefunction}

Simulations of high-$p_T$ quarkonium dissociation in the QGP require
knowledge of the light-cone wavefunctions of the state, which
can be represented as follows~\cite{hep-ph/0611109,Sharma:2009hn}:  
\begin{eqnarray}
 |\vec{P}^+;J \rangle & = &  {a}_h^\dagger(\vec{P}^+;J)  |0 \rangle
= \sum_{n = 2(3)}^\infty  \int \prod_{i=1}^n  \frac{d^2{\bf k}_i}{\sqrt{(2\pi)^{3}}}
\frac{dx_i}{ \sqrt{2 x_i}} 
 \times \, \psi(x_i,{\bf k}_i ; \{\nu_i\}) \,  \delta \left(\sum_{j=1}^n x_j -1 \right) \, 
\delta^2  \left( \sum_{j=1}^n{{\bf k}_j} \right)  \nonumber \\
&&  \times |\cdots  {a}^\dagger_{q_i}(x_{q_i}\vec{P}^+ + {\bf k}_{q_i}, \nu_{q_i} ) \cdots 
 {b}^\dagger_{\bar{q}_j}(x_{\bar{q}_j}\vec{P}^+ + {\bf k}_{\bar{q}_j}, \nu_{{\bar{q}_j} })
\cdots   
 \cdots   {c}^\dagger_{{g}_k}(x_{{g}_k}\vec{P}^+ + {\bf k}_{{g}_k}, \nu_{{{g}_k} } ) 
\cdots \rangle \; .
\label{Mp}
\end{eqnarray}   
Here,  $\vec{P^+} \equiv (P^+,{\bf P})$  are the large light-cone momentum 
and transverse momentum components of the hadron, and the momenta of the partons 
are given by $(x_iP^+,x_i{\bf P}+{\bf k}_i)$.  In Eq.~\ref{Mp}
$\nu_{q_i}$ is a set of additional relevant quantum numbers, such as helicity 
and color. From the normalization of the meson state of fixed projection $\lambda$
of the total angular momentum $J$: 
\begin{equation}
\langle \vec{P}^+;J  |\vec{P}^{+ \prime};J \rangle
= 2P^+ (2\pi)^3 \delta(P^+-P^{+ \prime}) \delta^2 ({\bf P}-{\bf P}^\prime)
\delta_{\lambda \lambda^\prime}\;,~\label{mesonnorm}
\end{equation}
an integral constraint  on the norm of the light-cone wavefunctions 
$\psi(x_i,\bfk_i;\{\nu_i\})$ can be derived.

In the context of Eq.~\ref{Mp},  the color-singlet contribution to quarkonium 
production can be understood  as one that is the dominant, lowest order ($n=2$), 
Fock component in the hard scattering.  A color-octet contribution requires 
the emission of at least one gluon for a color neutral hadron to be
produced. In either case the hadron state can be 
approximated as:
\begin{eqnarray}       
 |\vec{P}^+ \rangle 
&=& \int \frac{d^2{\bf k}}{(2\pi)^{3}} \frac{dx}{ 2\sqrt{ x(1-x)}}
  \frac{\delta_{c_1c_2}}{\sqrt{3}} \,
\psi(x,{\bf k})
 a_Q^{\dagger\;  c_1 }(x\vec{P}^++{\bf k})  b_{\bar{Q}}^{\dagger \;  c_2  }
((1-x)\vec{P}^+-{\bf k})  |0 \rangle  \; , 
\label{Mp1}
\end{eqnarray}       
where, for the color-octet states, $a^\dagger$ ($b^\dagger$) represent an 
``effective'' heavy quark (anti-quark) in the $3$ ($\bar{3}$) state and~\cite{Sharma:2009hn}  
\begin{equation}
  \psi(x,{\bf k}) = {\rm Norm} \times \exp\left( - \frac{{\bf k}^2 + m_Q^2 }{2 \Lambda^2 x(1-x) }   
  \right) \;   , \qquad 
\frac{1}{2 (2\pi)^{3} } \int dx d^2{\bf k}  \;
| \psi(x,{\bf k}) |^2 = 1 \; .
\label{lonorm}
\end{equation}
The light cone wavefunction $\psi(x, {\bf{k}})$ is obtained from the
instantaneous wavefunction $\psi(r)$ by taking the Fourier transform and
changing the longitudinal component of the ${\bf{k}}$ vector to the light cone
coordinates. Details of this change of variables are given
in~\cite{Sharma:2009hn}, and here we have used the fact that for quarkonia the
masses of the constituent quarks are equal and $(1-x)m_Q +x m_{\bar{Q}} =
m_{Q}$.

If we introduce the notation $\Delta {\bf k} = {\bf k}_1-{\bf k}_2 = 2 {\bf
k}$, the transverse width  $\Lambda$ of the light-cone wavefunction
$\psi(x,{\bf k})$  is determined from the condition: 
\begin{equation}
\frac{1}{2 (2\pi)^{3} } \int dx d^2{\bf k}  \;
{\Delta \bf k}^2  | \psi(x,{\bf k}) |^2 =  4\langle {\bf k^2} \rangle =  \frac{2}{3} \kappa^2  \; .
\label{lonorm2}
\end{equation}
The factor $2/3$ comes from the 2D projection of the mean squared  transverse
momentum $\kappa^2$ from the instantaneous wavefunction form calculated in
Tables~\ref{table:Charm},~\ref{table:Bottom}.

\section{Quarkonium dynamics at high transverse
momentum~\label{section:QGPdynamics}}

In this section we present  details of how we treat  the propagation of quarkonia
through the QGP. There are two important ways in which the medium
affects the yields of quarkonia.

First, on time scales shorter than the formation time of the quarkonia,
the color-octet component of the proto-quarkonium state undergoes energy loss
as it passes through the QGP.  Second, on time scales longer than  the formation
times of the heavy mesons, the meson can undergo dissociation due to collisions
with gluons in the thermal medium.  The essence of the dissociation model for
heavy mesons is that they have short formation times and can therefore form in
the medium on a time scale $t_{\rm form}$. Interactions with the thermal medium
can dissociate the mesons on a time scale $t_{\rm diss}$. The final yields are
determined by rate equations which take into account the formation and
dissociation processes. In the next section we first discuss the rate equations
abstractly, using $t_\rf$ and $t_\rd$ as parameters. In the later sections we
will estimate $t_\rf$ and calculate $t_\rd$. For more details on the
dissociation model and its application to the phenomenology of open heavy
flavor, see~\cite{Sharma:2009hn}.  
  
\subsection{The rate equations}
Let us denote by $N^{\rm hard}_{Q\bar{Q}}(p_T, \nu)$ the number of
perturbatively produced  point-like $Q\bar{Q}$ states at transverse momentum
$p_T$. Up to an overall multiplicative Glauber scaling factor $T_{AB}$, $N^{\rm
hard}_{Q\bar{Q}}(p_T,\nu)$ ($\nu$ represent the quantum numbers of the $Q\barQ$
state) are linearly related to the cross sections discussed in
Section~\ref{section:ppProduction}. The time scale for the hard QCD process is
given by  $t \simeq 1/m_T$, where $m_T = \sqrt{p_T^2+m_H^2}$.  For transverse
momenta  above a few GeV and $m_{c\bar{c}}> 3$~GeV, $m_{b\bar{b}}> 9$~GeV,
respectively, this production time is very short. Thus, we take this time to be
the starting point for the evolution of the $Q\bar{Q}$ state. The rate of
formation of the corresponding hadronic state is given by the inverse formation
time $1/t_{\rm form}(p_T)$. In the presence of a medium, the meson
multiplicity, which we denote by $N^{\rm meson}_{Q\bar{Q}}(p_T,\nu)$, is
reduced by collisional dissociation processes at a rate $1/t_{\rm
diss}(p_T)$. Finally, the number of dissociated $Q\bar{Q}$ pairs with a net
transverse momentum $p_T$ is $N^{\rm diss.}_{Q\bar{Q}}(p_T,\nu)$. At the
transverse momenta that we consider, the probability for the  heavy quark $Q$
or antiquark $\bar{Q}$ to pick up a thermal partner and reform a quarkonium
state is negligible. The heavy (anti)quark fragmentation contribution to
quarkonia is also negligible. Heavy quarks fragment primarily into open heavy
flavor mesons.

The dynamics of such a system is governed by the following set of ordinary 
differential equations:
\begin{eqnarray}
\label{rateq11}
\frac{d\,N^{\rm hard}_{Q\bar{Q}}(t;p_T,\nu)}{dt} 
&=& - \frac{1}{t_{\rm form}(t;p_T)} N^{\rm hard}_{Q\bar{Q}}(t;p_T,\nu) \, , \\
\frac{d\, N^{\rm meson}_{Q\bar{Q}}(t;p_T)}{dt} 
\label{rateq12}
&=&  \frac{1}{t_{\rm form}(t;p_T)} N^{\rm hard}_{Q\bar{Q}}(t;p_T,\nu) 
- \frac{1}{t_{\rm diss.}(t;p_T)} N^{\rm meson}_{Q\bar{Q}}(t;p_T,\nu) \, , \\
\frac{d\, N^{\rm diss.}_{Q\bar{Q}}(t;p_T,\nu)}{dt} 
&=&  \frac{1}{t_{\rm diss}(t;p_T)} N^{\rm meson}_{Q\bar{Q}}(t;p_T,\nu) \, , 
\label{rateq13}
\end{eqnarray}
subject to the constraint $N^{\rm hard}_{Q\bar{Q}}(t;p_T,\nu) + N^{\rm meson}_{Q\bar{Q}}(t;p_T,\nu) + 
N^{\rm diss.}_{Q\bar{Q}}(t;p_T,\nu) = N^{\rm hard}_{Q\bar{Q}}(p_T,\nu)$, and is uniquely 
determined by the initial conditions 
\begin{eqnarray}
\label{rateq1}
N^{\rm hard}_{Q\bar{Q}}(t=0;p_T,\nu) &=& N^{\rm hard}_{Q\bar{Q}}({\rm{quenched}};p_T,\nu)\,,\\ 
N^{\rm meson}_{Q\bar{Q}}(t=0;p_T,\nu) &=& 0 \,,\\ 
N^{\rm diss.}_{Q\bar{Q}}(t=0;p_T,\nu) &=& 0 \,. 
\end{eqnarray}
An important point to note here is that we have incorporated the quenching of
the color-octet proto-quarkonium state by using the quenched value of the
distributions $N^{\rm hard}_{Q\bar{Q}}({\rm{quenched}};p_T,\nu)$ as the initial
state to the rate equations. In Section~\ref{section:quenched} we will describe
how the quenched distributions are obtained.  Finally, we note that in
Eqs.~\ref{rateq13} the evolution of the dissociated  $\bar{Q}Q$ pair into
$D$- or $B$- mesons is not shown since it does not couple back to
Eqs.~\ref{rateq11}, \ref{rateq12}.

Realistic simulations include the velocity dependence of the formation rate of
all quarkonium states and the velocity, time, and position dependence of their
dissociation rate. (We do not write the meson dependence of $t_{\rf}$ and
$t_{\rd}$ to avoid cluttered notation.) It is, however, useful to integrate the
system of equations analytically for a simple test case.  Our simplified test
case assumes that the dissociation time is constant in the interval $ 0 \leq t
\leq L_{QGP}$ and 0 if $t > L_{QGP}$, where $L_{QGP}$ is the linear size of the
fireball.  The solution for the $Q\bar{Q}$ mesons as a function of time is:
\begin{eqnarray}
\nonumber
N^{\rm meson}_{Q\bar{Q}}( 0 \leq t \leq L_{QGP};p_T,\nu) 
  & = &  N^{\rm hard}_{Q\bar{Q}}({\rm{quenched}};p_T,\nu)
\frac{ t_{\rm diss.}(p_T)} {t_{\rm diss.}(p_T) -t_{\rm form}(p_T)
}\\
\label{an1} 
& &\left( e^{-t/  t_{\rm diss.}(p_T)} -  e^{-t/  t_{\rm form}(p_T)}  \right) 
\;, \\
\nonumber
 N^{\rm meson}_{Q\bar{Q}}(t>L_{QGP};p_T,\nu) 
  & = &  N^{\rm hard}_{Q\bar{Q}}({\rm{quenched}};p_T,\nu) \Bigg[
\frac{ t_{\rm diss.}(p_T,\nu)} {t_{\rm diss.}(p_T) -t_{\rm form}(p_T)
}\\
\label{an2}
\left( e^{-L_{QGP}/  t_{\rm diss.}(p_T)} -  e^{-L_{QGP}/  t_{\rm form}(p_T)}  \right)
&+&\left(   e^{-L_{QGP}/  t_{\rm form}(p_T)}  -   e^{-t/  t_{\rm form}(p_T)}    \right)   \Bigg]  \, . 
\end{eqnarray}
The interested reader can easily deduce the solutions for $N^{\rm
hard}_{Q\bar{Q}}(t;p_T,\nu)$ and $N^{\rm diss.}_{Q\bar{Q}}(t;p_T,\nu)$
and verify that the solutions in Eqs.~\ref{an1}, \ref{an2} are finite for $
t_{\rm form}(p_T) = t_{\rm diss.}(p_T) $. Eqs.~\ref{an1}, \ref{an2} 
can be used to understand the qualitative features of the time dependence of quarkonium formation.

\subsection{Formation time of quarkonium states }

The approach to estimating the formation time of quarkonium states differs
considerably from the  approach used for open heavy
flavor\cite{hep-ph/0611109,Sharma:2009hn} or light
particles~\cite{arXiv:0807.1509} that come from the fragmentation of a hard
parton. In the latter case the formation time is inversely proportional to the
virtuality of the parton decay and is governed by longitudinal dynamics. For
quarkonia, the $Q\bar{Q}$ state is prepared instantly ($\sim 1/\sqrt{p_T^2
+m_H^2}$)  in the hard collision and subsequently expands to the spatial extent
determined by the size of the asymptotic wavefunction.  In this case all
spatial directions are important.  The velocity of the heavy quarks in the
meson and a typical upper limit of the meson formation time can be evaluated
as follows:
\begin{equation}
\beta_Q = \sqrt{ \frac{\kappa^2}{\kappa^2 + m_Q^2}    }\, ,  
\qquad    t_{\rm rest\  frame}^{\max} = \frac{a_\perp}{\beta_Q} \;,  
\label{formrest}
\end{equation} 
where the typical momenta, $\kappa$ is related via Eq.~\ref{lonorm2} to $k^2$
given the Tables~\ref{table:Charm},~\ref{table:Bottom}. The transverse sizes,
$a_\perp$, are
also given in Tables~\ref{table:Charm},~\ref{table:Bottom}. In this paper we
are  interested in high transverse momentum mesons, in which case there is a
boost in the direction of propagation and, consequently, time dilation      
\begin{equation}
t_{\rm form}^{\max}(p_T,\nu)=  \gamma \, t_{\rm rest\  frame}^{\max} (\nu) =
 \gamma \frac{a_\perp}{\beta_Q}, \qquad \gamma =  \frac{\sqrt{p_T^2 + m_H^2}}{m_H} \;. 
\label{form}
\end{equation}
For example, for the relevant formation time determined by the expansion of the
$Q\bar{Q}$ state in a direction transverse to the direction of propagation the
transverse size remains the same when boosted back to the laboratory frame but
the velocity transforms by  picking up a factor of $1/\gamma$.  Note that in
Eq.~\ref{form} $|\vec{p}| = p_T$ since in this paper we work at mid-rapidity.
The masses of the quarkonium states $m_H$ are taken from~\cite{pdg}. $\gamma$ is
the meson boost factor.  Since the formation process is non-perturbative and
can not be modeled accurately, the values of $t_{\rm form}$ obtained from
Eq.~\ref{form} should be considered as an estimate. We treat this as the upper
limit of the formation time. In addition to calculating the final yields for
$t_{\rm form}=t_{\rm f max}=\frac{\gamma a_\perp}{\beta_Q}$, we also calculate the
yields for $t_{\rm form}=t_{\rm f min}=\frac{\gamma a_\perp}{2\beta_Q}$ and the
variation gives us an estimate of the uncertainty due to the uncertainty in the
formation time.

\subsection{ Dissociation time of quarkonium states }

The propagation of a  $Q\bar{Q}$ state in matter is accompanied by collisional
interactions mediated at the partonic level, as long as the momentum exchanges
can resolve the partonic structure of the meson. Two effects are related to
these interactions: a) a broadening of the distribution of  quarkonium states
relative to the original direction; b) a modification of the quarkonium
wavefunction. The former effect integrates out as long as we consider inclusive
production. The latter effect leads to the dissociation of the  meson state. 

Let us define:
\begin{eqnarray}
&&\chi \mu_D^2 \xi  = \int_{t_0}^t  d\tau \frac{\mu_D^2(\bf{x}(\tau),\tau) }{\lambda_q({\bf x}(\tau),\tau) } \xi\, , \qquad  
{\bf x}(\tau) = {\bf x}_0 + {\bm \beta} (\tau-\tau_0)\, .
\label{broad}
\end{eqnarray}
In Eq.~\ref{broad} $\mu_D^2$ is the typical squared transverse momentum transfer given by the Debye screening 
scale, $\mu_D=g T$ for a gluon-dominated plasma. $\chi$ is the opacity -- the
average number of collisions that the parton undergoes. $\lambda_q$ is the mean
scattering length of the quark and $\xi\sim$ few is an enhancement factor from
the power law tail of the differential scattering cross section.  (Note that
$\xi$ is unrelated to $\xi_q$, $\xi_g$ in Eq.~\ref{shadb} but appears also in 
transverse broadening due to cold nuclear effects in Eq.~\ref{Cron} since the
basic formalism for momentum broadening is the same there.) Finally,
${\bf x}_0$ is the position of the propagating $Q\bar{Q}$ and  ${\bm \beta}$ is
the velocity of the heavy meson. Note that $|{\bm \beta}| < 1$. For a medium of
uniform parton density and length $L$    $\chi \mu_D^2 \xi = \mu_D^2 (L/\lambda_q)
\xi$. For the realistic expanding medium the parton density and temperature can
be determined as described in~\cite{Vitev:2008rz,Neufeld:2010fj} and the
integral in Eq.~\ref{broad} can be performed numerically.

With the results for the cumulative momentum transfer at hand, the medium-modified  quarkonium wavefunction 
can be evaluated analytically for the functional form specified in
Eq.~\ref{lonorm2}.  The survival probability for the
closed heavy flavor mesons is given by
\begin{eqnarray}
  P_{\rm surv.} (\chi\mu_D^2 \xi) & = & \left|  \frac{1}{2 (2\pi)^{3} }  \int d^{2}{\bf k} dx \,
\psi_{f}^* (\Delta {\bf k},x)\psi_{i}(\Delta {\bf k}, x) \right|^{2} 
\nonumber \\
&=& \left| \frac{1}{2 (2\pi)^{3} }  \int dx \; {\rm Norm}^2 \, \pi x(1-x) \Lambda^2
\,  e^{-\frac{ m_{Q}^{2} }{ x(1-x)\Lambda^{2} } } 
\, \left[ \frac{ 2 \sqrt{ x(1-x)\Lambda^{2}}
\sqrt{\chi\mu_D^{2}\xi+x(1-x)\Lambda^{2}} }
{ \sqrt{x(1-x)\Lambda^{2}}^{2}  + \sqrt{\chi\mu_D^{2}\xi+x(1-x)
\Lambda^{2}}^2 } \right] \; \right|^2 \, . \;\; \quad 
\label{sprob}
\end{eqnarray}
The dissociation rate for a given meson is then given by
\begin{equation}
t_{\rm diss.}(p_T) = \frac{d P_{\rm diss.}(p_T)}{dt} = - \frac{d P_{\rm surv.}(p_T)}{dt} 
\end{equation}
The uncertainty in $t_\rd$ arises from the uncertainty in the coupling between the heavy
quarks and the medium (described by the strong coupling constant $g$)  and
the enhancement that arises from the power law tails of the Moliere
multiple scattering in the Gaussian approximation to transverse momentum 
diffusion~\cite{hep-ph/0611109}
(described by $\xi$).  In our calculations we use two sets 
$g=1.85$, $\xi=2$ and $g=2$, $\xi=3$ to understand the sensitivity of the
results to these parameters.

\begin{figure}[!ht]
\vspace*{.2in} 
\includegraphics[width=3.80in,angle=0]{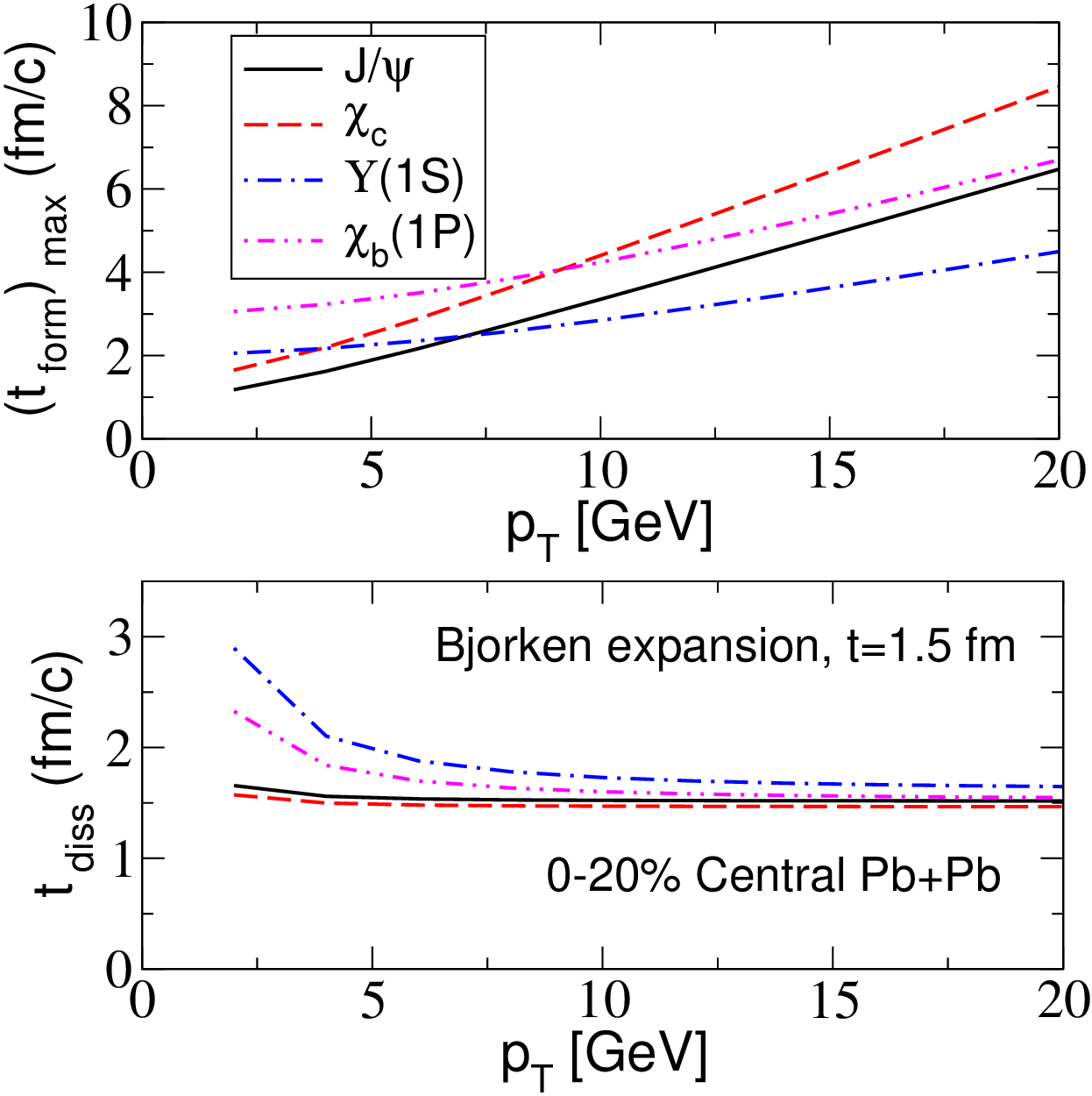} 
\caption{(Color online)Top panel: formation time for selected charmonium
($J/\psi$, $\chi_c$) and bottomonium ($\Upsilon(1S)$, $\chi_b(1P)$) versus
transverse momentum.  Bottom panel: the corresponding dissociation times quoted
at $t=1.5$~fm$/c$ for a hard process in the center of the collision geometry,
0-20\% Pb+Pb reactions at the LHC. \label{fig:Times}} 
\end{figure}

To get a sense of the formation and dissociation times involved in the
quarkonium dynamics in the QGP, we give the values for a fixed transverse
momentum $p_T=10$~GeV for the quarkonium states and consider their production
in the center of the collision geometry. The specific nuclear collisions that
we chose are 0-20\% central Au$+$Au collisions at RHIC at
$\sqrt{S}_{NN}=0.2$~TeV. In the examples in Table~\ref{table:CharmFD} and
Table~\ref{table:BottomFD} $g=2$ and $\xi = 2$. The results include the boost
factor. Since the medium expands after the collision, we quote the formation
and dissociation times at $t = 1.5$~fm$/c$. Results for charmonia and
bottomonia are presented in Table~III and Table~IV, respectively. 

To clarify the $p_T$ dependence of the formation and dissociation times we show
in Fig.~\ref{fig:Times}, $t_\rf$ and $t_\rd$ at $t = 1.5$~fm$/c$ for $0-20$\%
Pb$+$Pb collisions for $2.76$~TeV collisions at the LHC.

\begin{table*}[h]
\begin{tabular}{l||cc}
{\rm Charmonium state \ \  \ }   &   \ \ \ \  $J/\psi$ \ \ \ \  &   $\chi_{c0,1,2}$ \\ \hline
{\rm $(t_{\rf})_{\rm max}$  [fm$/c$]\ \  \ }   & 3.35      &    4.40       \\
{\rm $t_{\rd}$  [fm$/c$]\ \  \ }   &   1.74    &   1.61     
\end{tabular}
\caption{Upper limit on the formation time and dissociation time of quarkonium
states of  $p_T=10$~GeV and produced in the center of the nuclear overlap
region of 0-20\% central Au$+$Au collisions at RHIC.}
\label{table:CharmFD}
\end{table*}

\begin{table*}[h]
\begin{tabular}{l||ccccccc}
 {\rm  Bottomonium state}  \ \   \ \ & \  \  $\Upsilon(1S)$ \ \       &  
 \  \  $\Upsilon(2S)$  \  \    &  \ \  $\Upsilon(3S)$   \  \ &
\ \  $\chi_{b0,1,2}(1)$ \  \ &   \  \ $\chi_{b0,1,2}(2)$ \ \  &
  \ \   $\chi_{b0,1,2}(3)$ \  \   \\ \hline
 {\rm $(t_{\rf})_{\rm max}$  [fm$/c$]\ \  \ }   &     1.44  &  2.85   &  4.17   &    2.36 &  3.45  & 6.23       \\
{\rm  $t_{\rd}$ [fm$/c$]\ \  \ }              &     3.30  &  2.23   &  1.93   &    1.93 &   2.06  &  1.73        
\end{tabular}
\caption{Upper limit on the formation time and dissociation time of bottomonium
states of  $p_T=10$~GeV and produced in the center of the nuclear overlap
region of 0-20\% central Au$+$Au collisions at RHIC.}
\label{table:BottomFD}
\end{table*}

\subsection{Quenching of the color-octet state~\label{section:quenched}}
Before the formation of the overall color neutral wavefunction on a time scale 
of $t_{\rm{form}}$, the $Q\barQ$ state has both color-singlet and color-octet
components. In fact, from Section~\ref{section:ppProduction} we know that the
dominant contribution to quarkonium production comes from the color octet
state. 

Therefore, from the hard proto-quarkonium creation to the formation of the
quarkonium wavefunction, the color octet state undergoes energy loss, and
consequently quenching, as it passes through the medium. This can be thought of
as a  very massive gluon moving through the QGP for $t_{\rm{form}}$.

The quenching for massive partons traversing the QGP has been extensively investigated.
Starting with the discussion of the dead-cone effect~\cite{Dokshitzer:2001}, the effect of mass
has been incorporated in all energy loss approaches. We include the quenching effect in the
calculation of $R_{AA}$ by calculating the quenching factor for a ``gluon'' of
mass $2m_Q$ traversing the medium for a time $t_{\rm{form}}$. The details
of the GLV formalism used for the energy loss calculation can be found
in~\cite{Djordjevic:2003zk}. In our calculation, we incorporate the fluctuations of the energy loss due to the
multiple gluon emission and the diffuseness of the nuclear geometry
(fluctuations in path length). 

Starting with the yields obtained in Section~\ref{section:ppProduction},
(including CNM effects described in Section~\ref{section:CNM}), this energy
loss leads to a quenching of the yield. This quenched yield for the color-octet
states give the initial conditions for the rate equations that take into
account the dissociation dynamics. The color-singlet components are, of course,
unquenched. 

\section{Numerical results for the nuclear modification factors~\label{section:Results}}

\subsection{$R_{AA}$ for $J/\psi$ and $\Upsilon(1S)$~\label{section:NoCronin}}
\begin{figure}[!t]
\vspace*{.2in}
\includegraphics[width=3.38in,angle=0]{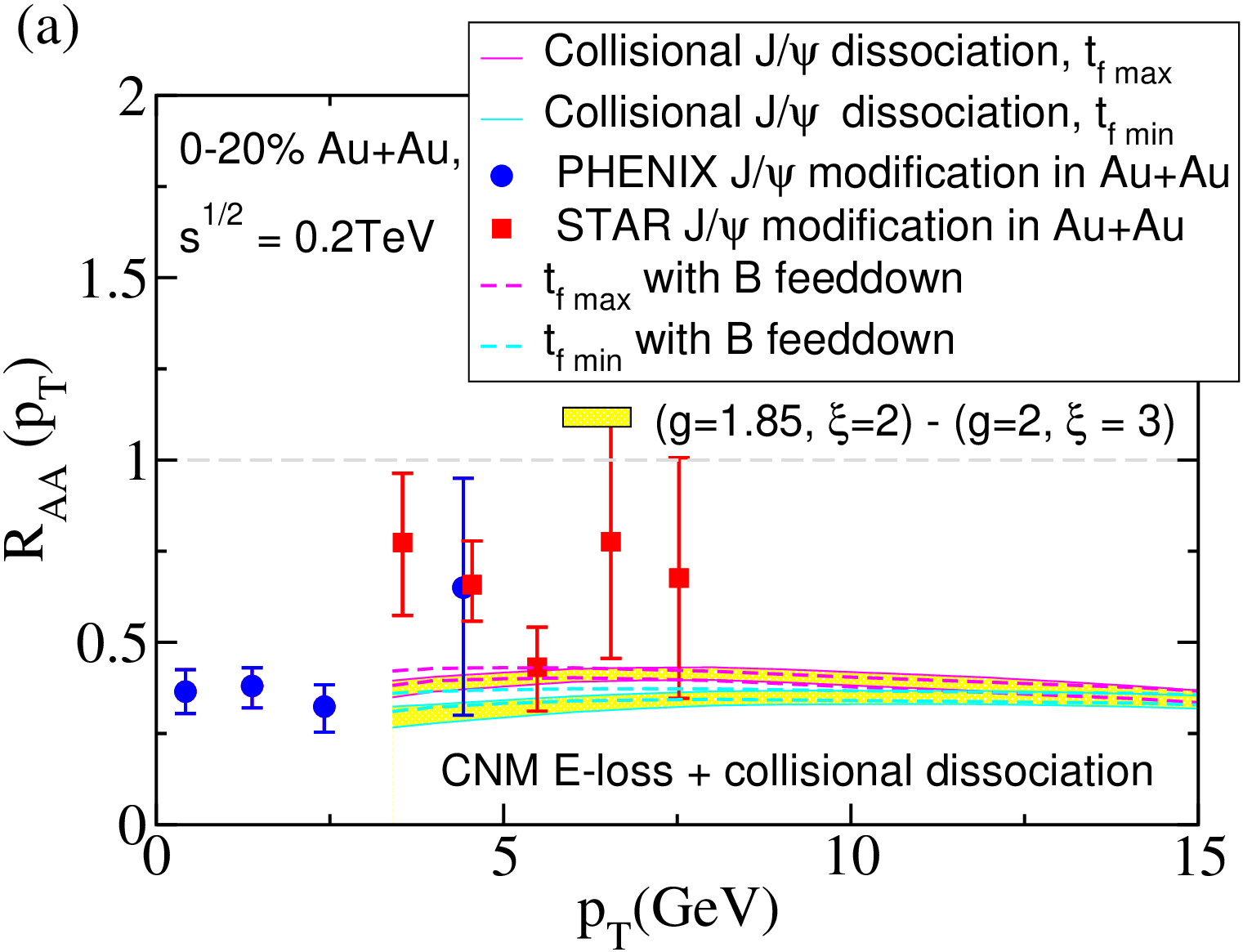} 
\includegraphics[width=3.38in,angle=0]{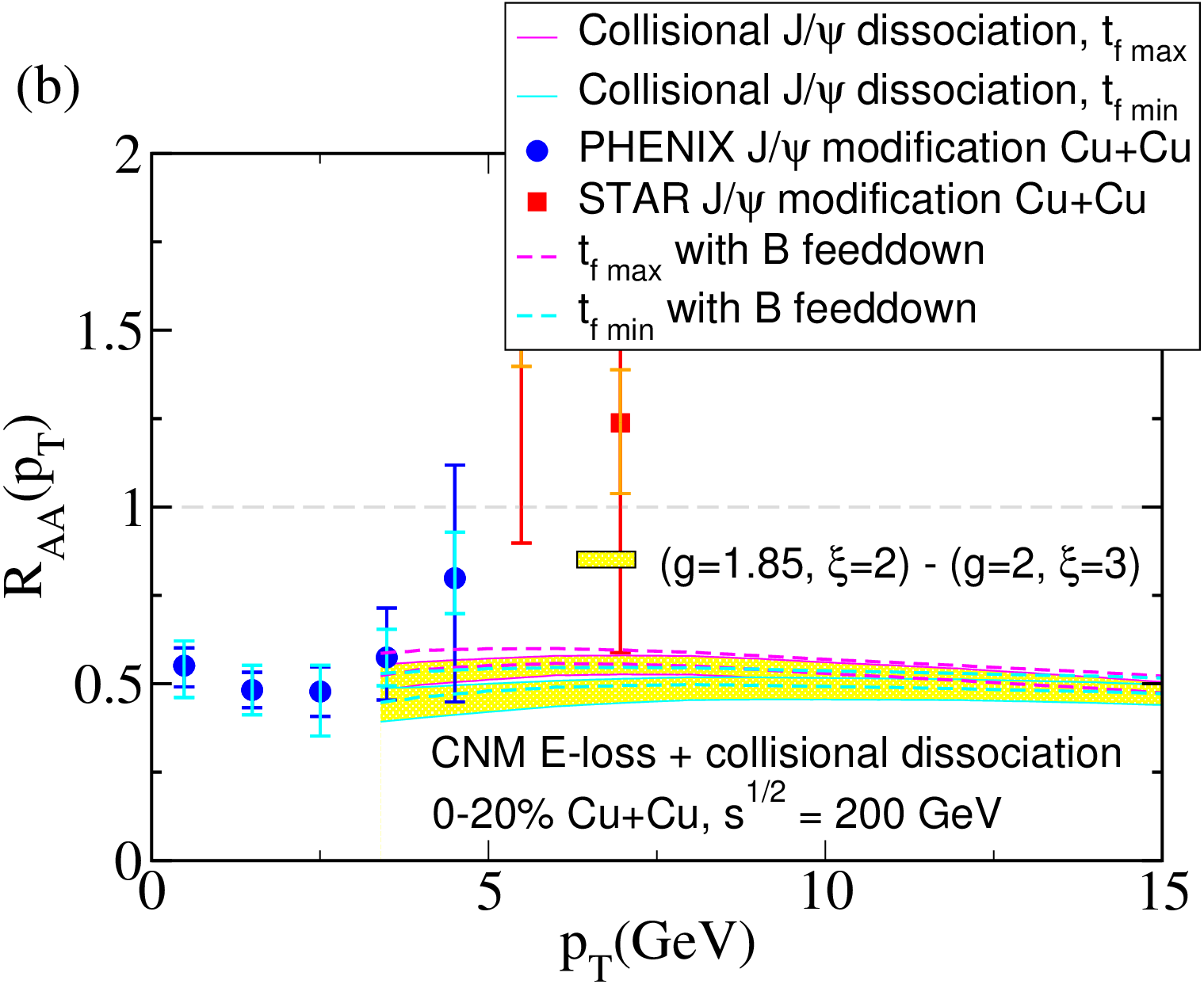} 
\includegraphics[width=3.38in,angle=0]{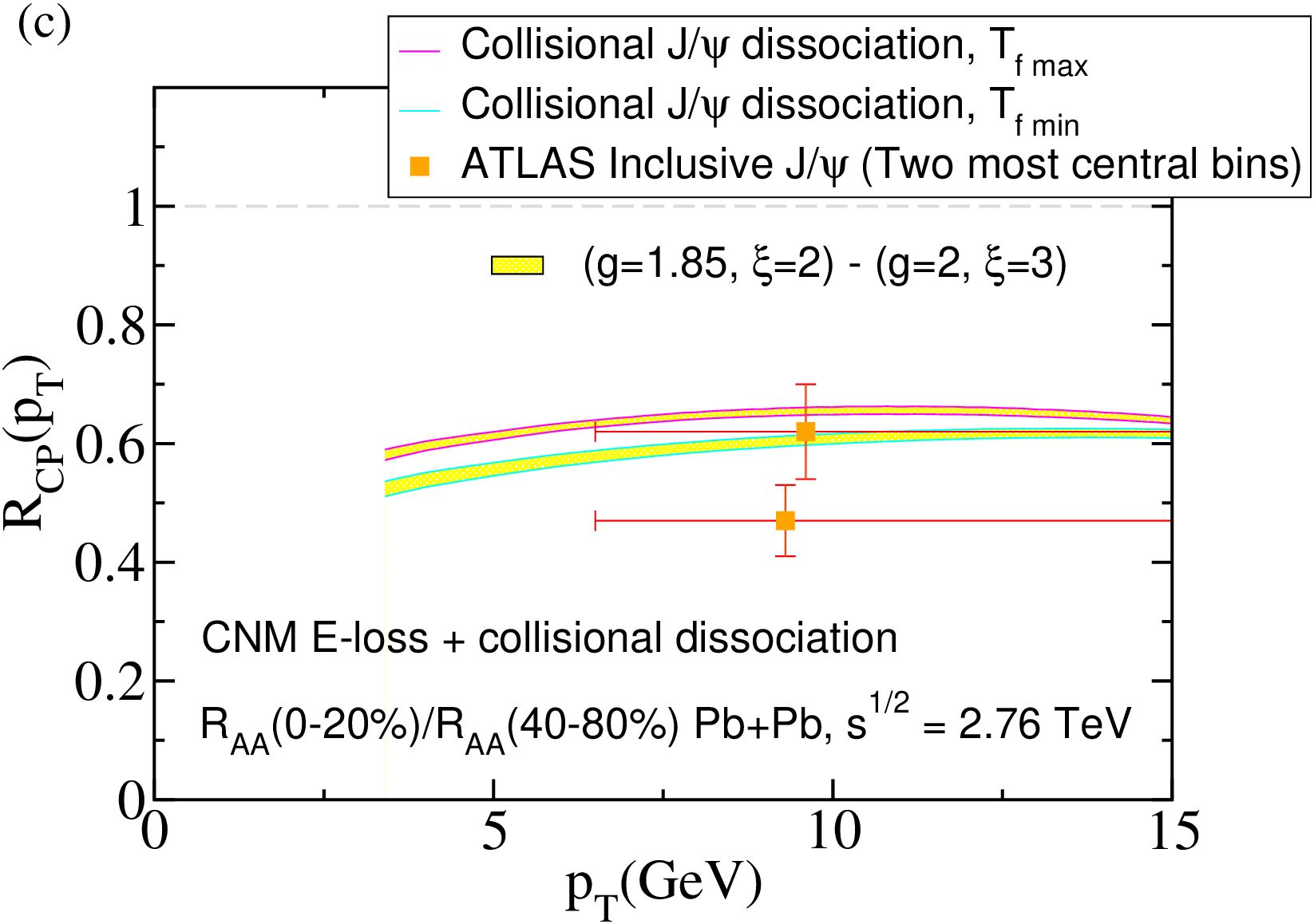}
\includegraphics[width=3.38in,angle=0]{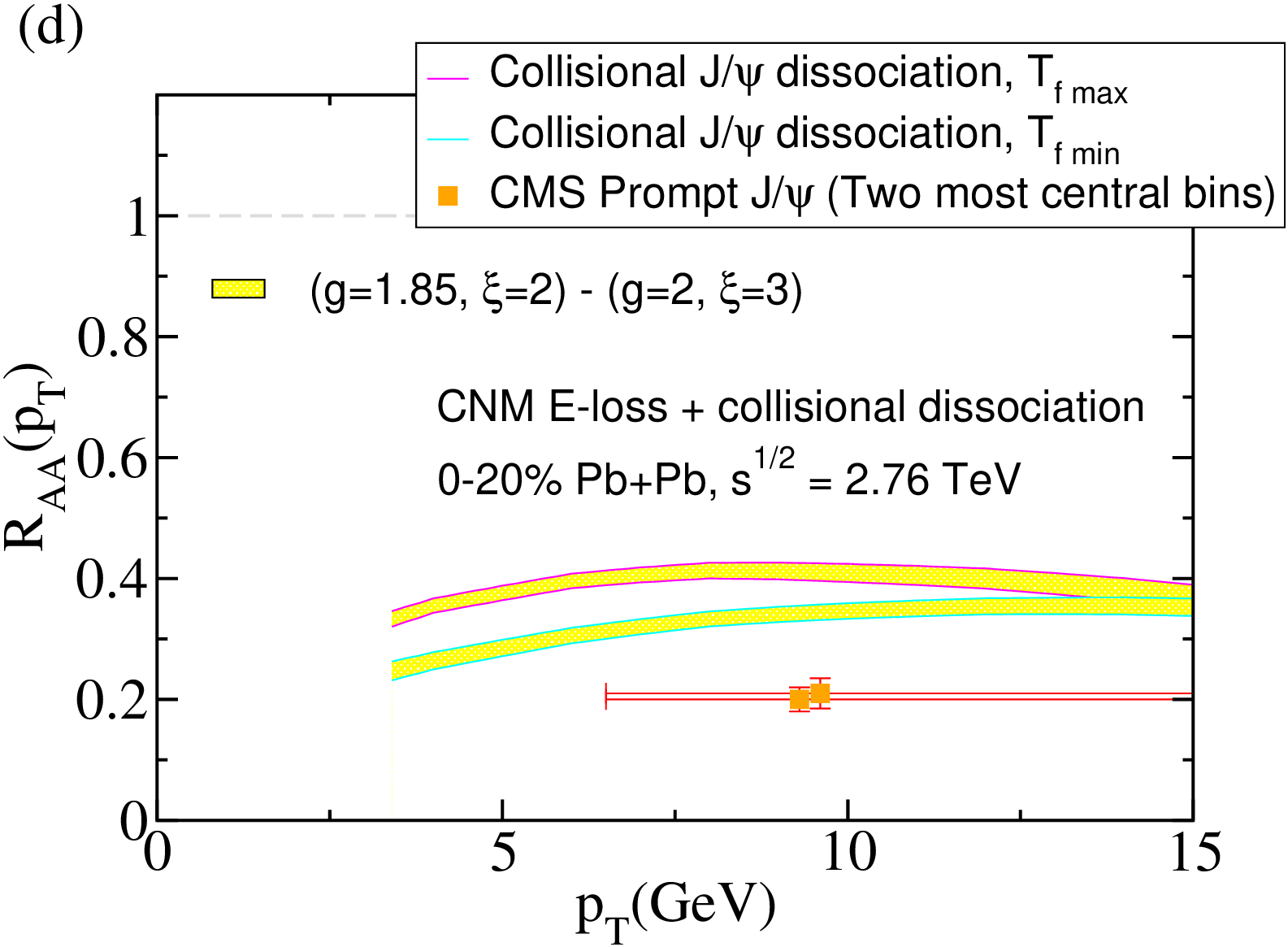}
\caption{(Color online) Theoretical model predictions for $J/\psi$ $R_{AA}$ in
0-20\% central  nucleus-nucleus collisions.  The top left panel is for RHIC Au$+$Au collisions at
$\sqrt{S}=0.2$~TeV~\cite{Tang:2011kr}. The top right panel is for RHIC Cu$+$Cu
collisions at $\sqrt{S}=0.2$~TeV~\cite{Abelev:2009qaa}. The lower  panels
are for LHC Pb$+$Pb collisions at $\sqrt{S}=2.76$~TeV~\cite{Chatrchyan:2012np,:2010px}. The
various curves represent the uncertainty due to variations in $g$ and $\xi$
(taking the sets $g=1.85$, $\xi=2$ and $g=2$, $\xi=3$), and a factor of 2
variation in $t_\rf$. In the lower left panel [panel (c)] the lower data 
point corresponds to $0-10\%$ most central events and the other to $10-20\%$. These results neglect the Cronin effect.
Dashed curves include the $B\rightarrow J/\psi$ feed-down. ~\label{fig:RAAJ1}}
\end{figure}

In this section we neglect the Cronin effect but include initial-state
cold nuclear matter energy loss and  shadowing. Let us first 
consider the nuclear modification factor for $J/\psi$ mesons. From
Fig.~\ref{fig:RAAJ1} we see that the data on $R_{AA}$ in RHIC Au$+$Au
collisions~\cite{Tang:2011kr} shows a $R_{AA}$ roughly between $0.5$ and $0.75$
at $p_T\sim6$~GeV. On the other hand,  Cu$+$Cu collisions~\cite{Abelev:2009qaa}
show a $R_{AA}$ greater than $1$. Both measurements have large error bars.
Within the uncertainty in our model parameters, we obtain results consistent
with RHIC data, albeit systematically slightly smaller than the measured
$R_{AA}$(for $p_T\sim 6$~GeV, $R_{AA}\sim 0.35-0.45$ for Au$+$Au and $R_{AA}
\sim 0.45-0.65$ for Cu$+$Cu). In Fig.~\ref{fig:RAAJ1}, our results for the
prompt yields of $J/\psi$ mesons are marked by upper and lower yellow bands
corresponding to the  upper ($t_\rf^{\rm max}$) and lower ($t_\rf^{\rm
min}=t_\rf^{\rm max}/2$) limits of our formation time estimate respectively.
The bands themselves correspond to our estimate of the uncertainly in the sets
of parameters that determine the coupling of the heavy quarks with the
in-medium partons [$g=1.85$, $\xi=2$ (minimum considered coupling gives the
upper limit of the yellow band) and $g=2$, $\xi=3$ (maximum considered coupling
for the lower limit of the yellow band)]. The pronounced effect of the
variation of the formation time can be intuitively seen as follows. From
Eq.~\ref{rateq13}, we see that the dissociation mechanism is operative only
when $N^{meson}_{Q\bar{Q}}$ is substantial, i.e. after $t_\rf$.  Since the
upper limit for formation time  of quarkonia can be on the order of several
fm$/c$, (see Tables~\ref{table:CharmFD},~\ref{table:BottomFD}), the density of
the medium at $t_\rf^{\rm max}$ is reduced considerably due to Bjorken
expansion, giving weaker dissociation and weaker suppression. This effect is
more pronounced than the details of the coupling of heavy quarks to the
in-medium partons. 

The RHIC experiments report suppression for the inclusive $J/\psi$ yield. For direct
comparisons, we also show the $R_{AA}$ for the inclusive yields in
Fig.~\ref{fig:RAAJ1} with dashed lines. The color scheme for the various
parameters is analogous to the direct production, though we do not color in
yellow the band associated with the uncertainty in coupling to avoid 
cluttering. The $B-$meson yields for p$+$p and A$+$A collisions were taken
from~\cite{Sharma:2009hn}.

The high-$p_T$ suppression of $J/\psi$ mesons in Pb$+$Pb collisions at the LHC,
reported by the ATLAS and CMS experiments, is substantially higher than the RHIC
results.  For example, for $p_T\in(6.5,30)$~GeV, CMS~\cite{Chatrchyan:2012np}
shows $R_{AA}\sim 0.2$ for the most central collisions. For a sharply falling
spectrum we expect the suppression to be dominated by the low momenta in the
$p_T$ bin. (The mean $p_T$ of the observed particles in the bin is roughly
$9.3$~GeV, for p$+$p collisions.) On
general grounds, one expects the suppression at LHC to be larger than at RHIC.
The initial temperature at LHC is higher than at RHIC, and consequently the
gluon density at any given time in the Bjorken expansion is also higher. This
will give rise to more rapid dissociation at LHC both in equilibrium and
non-equilibrium approaches. (The system sizes for Au$+$Au and Pb$+$Pb collisions
are roughly the same.)

\begin{figure}[!t]
\vspace*{.2in}
\includegraphics[width=3.38in,angle=0]{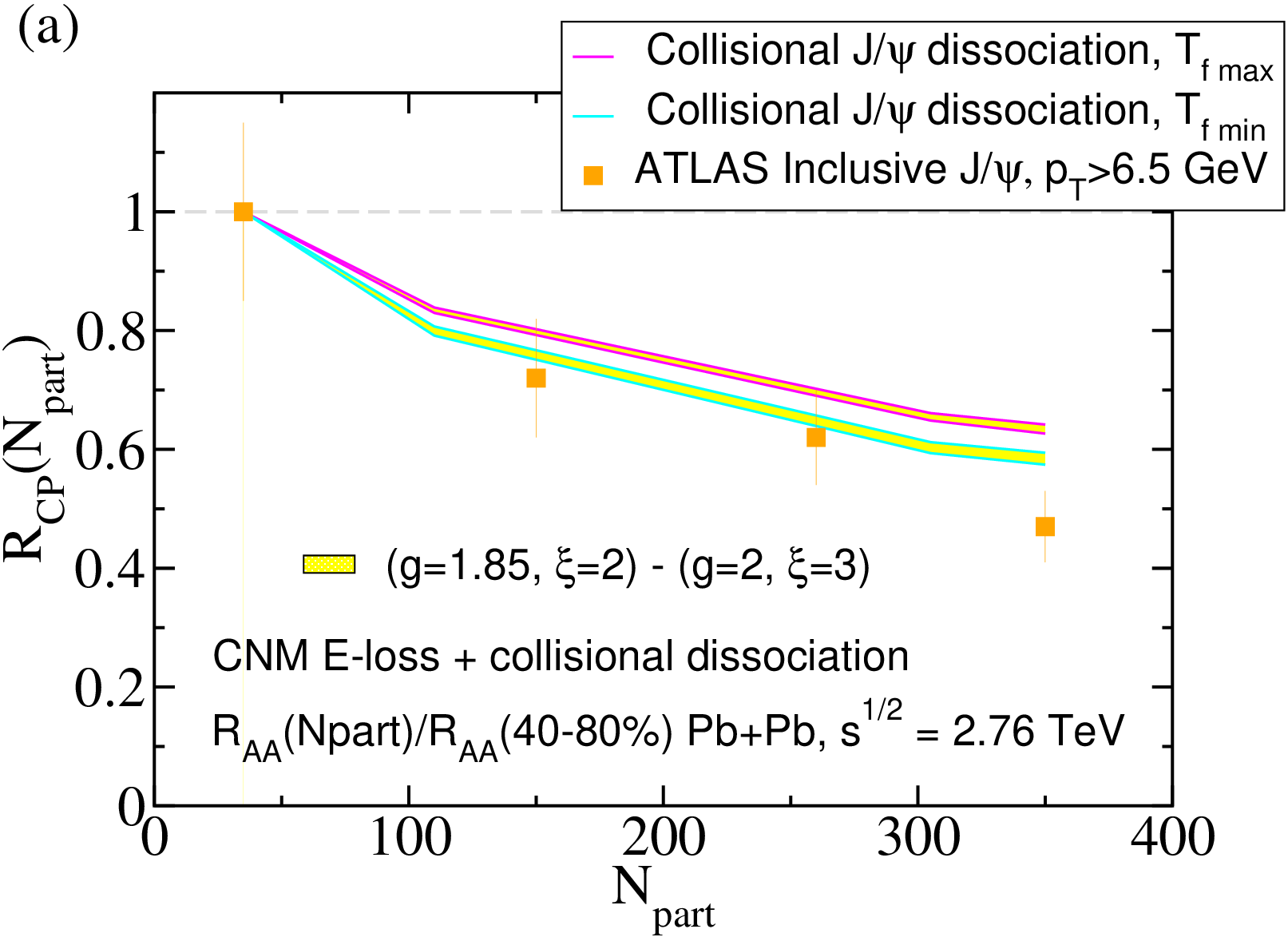} 
\includegraphics[width=3.38in,angle=0]{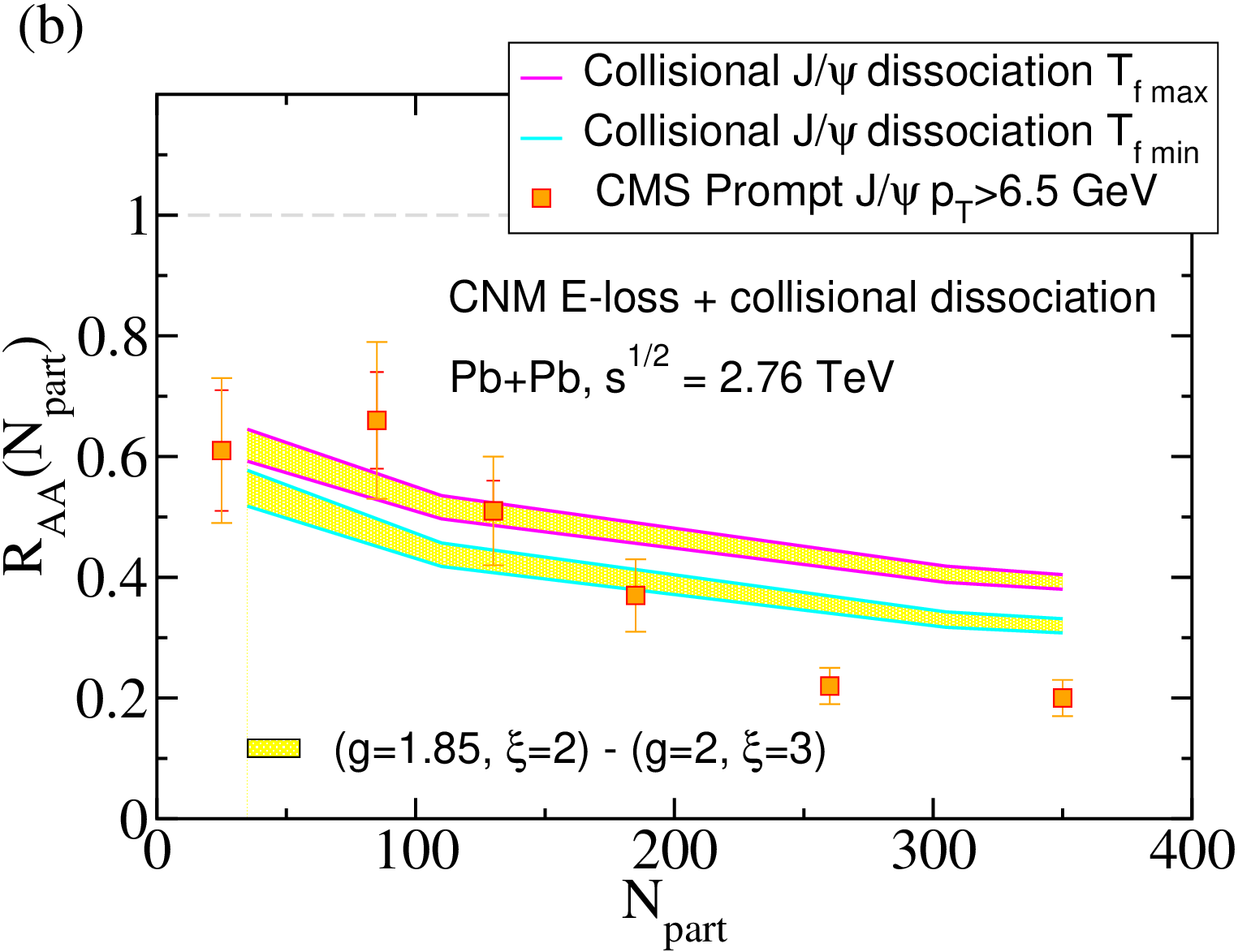} 
\caption{(Color online) Expected $J/\psi$ suppression  versus centrality
($N_{\rm part}$) at $\sqrt{S} = 2.76$~TeV. The left panel compares the
theoretical results to the ATLAS $R_{\rm CP}$ data~\cite{:2010px}.  The right
panel compares the theoretical results to the CMS $R_{\rm AA}$
data~\cite{Chatrchyan:2012np}.
~\label{fig:RNpart}}
\end{figure}

\begin{figure}[!t]
\vspace*{.2in}
\includegraphics[width=3.38in,angle=0]{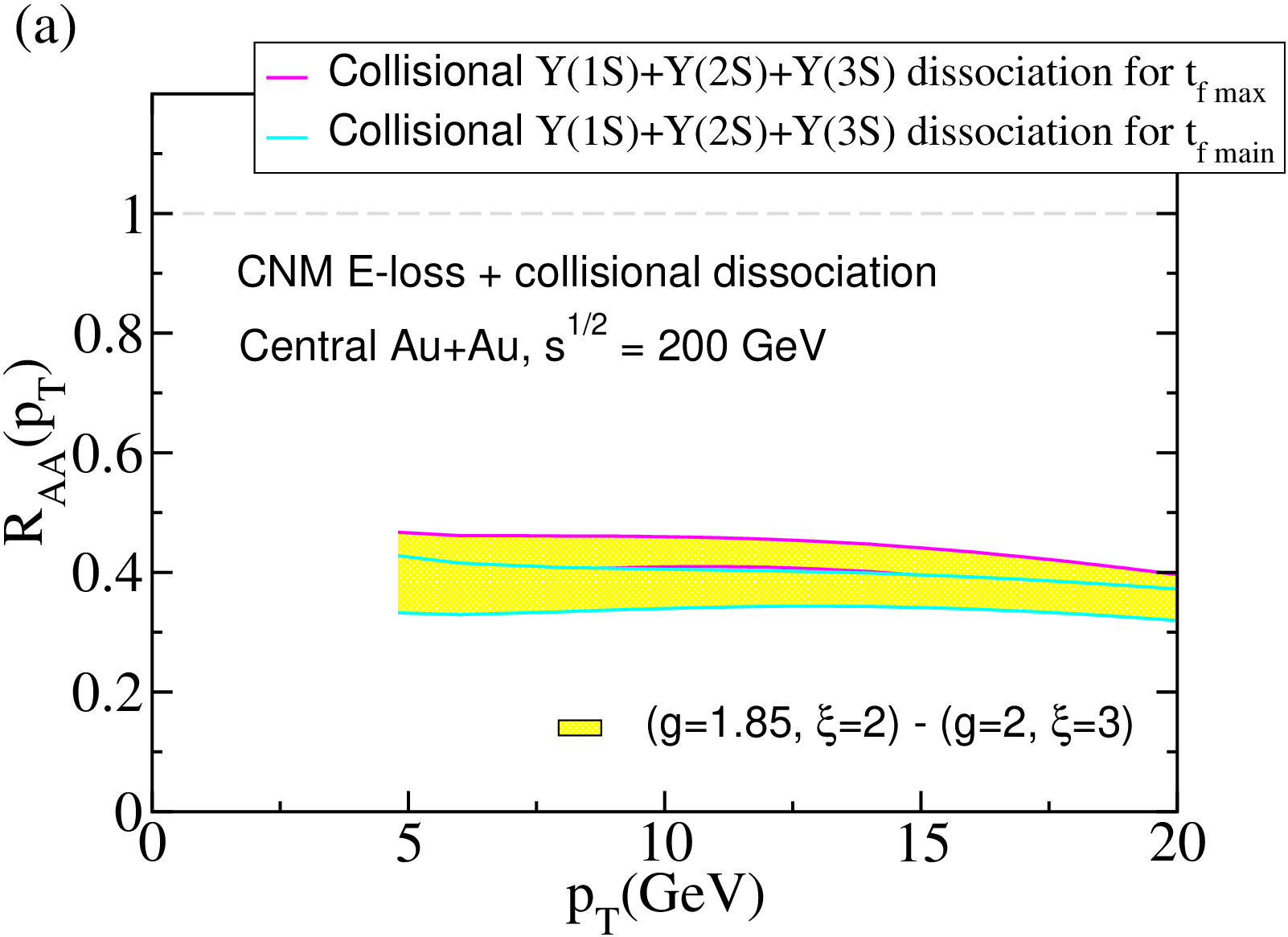} 
\includegraphics[width=3.38in,angle=0]{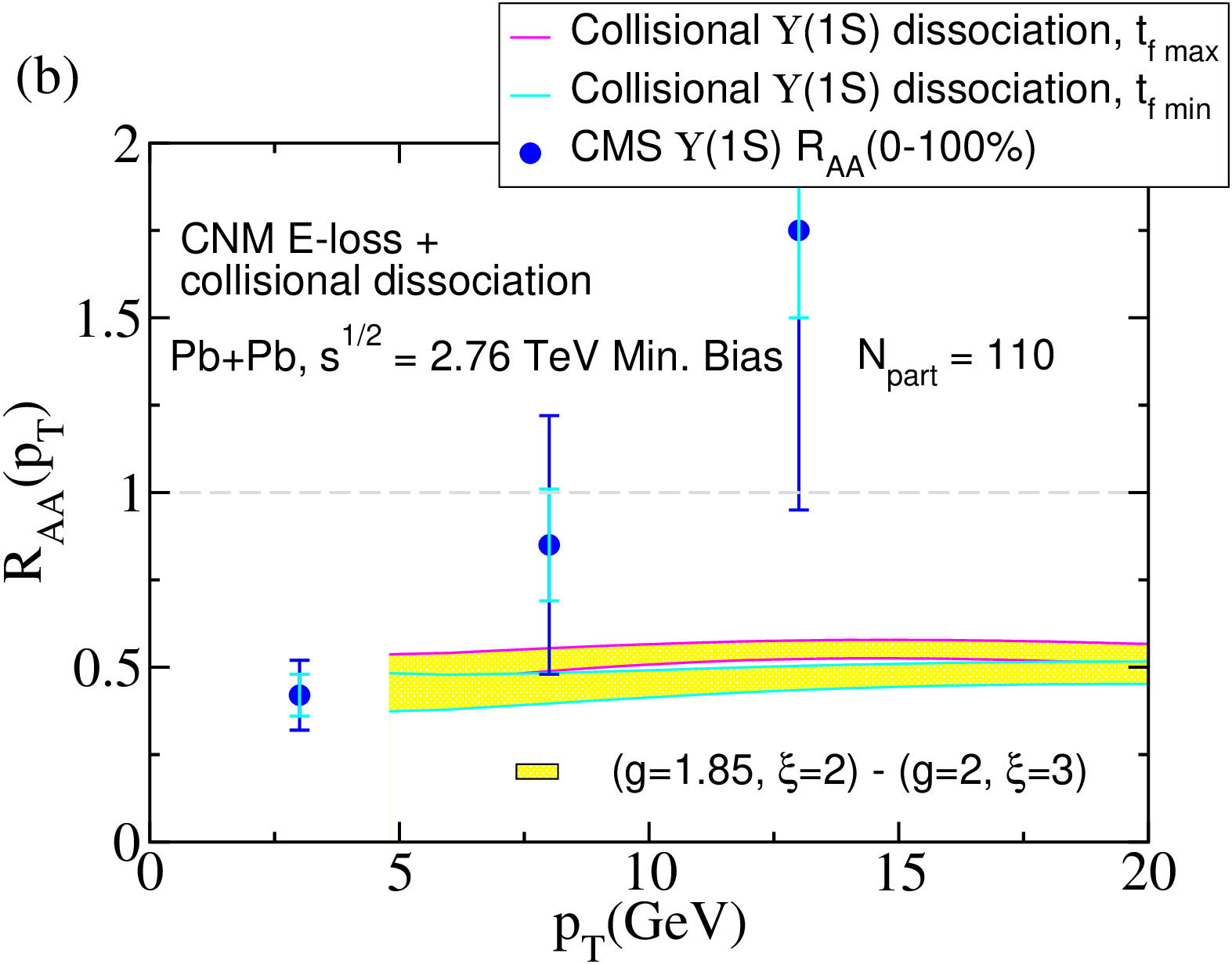} 
\includegraphics[width=3.38in,angle=0]{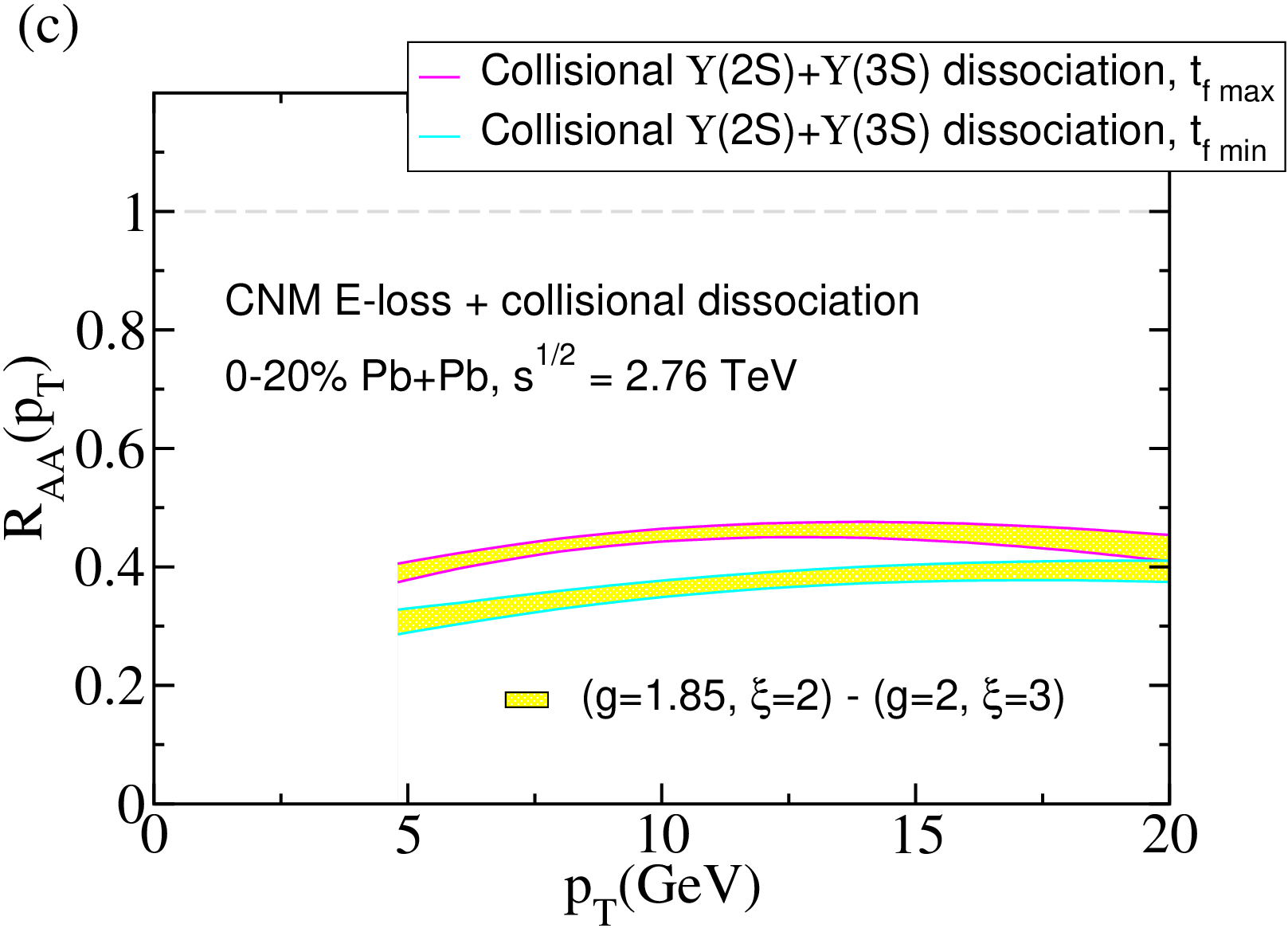}  
\includegraphics[width=3.38in,angle=0]{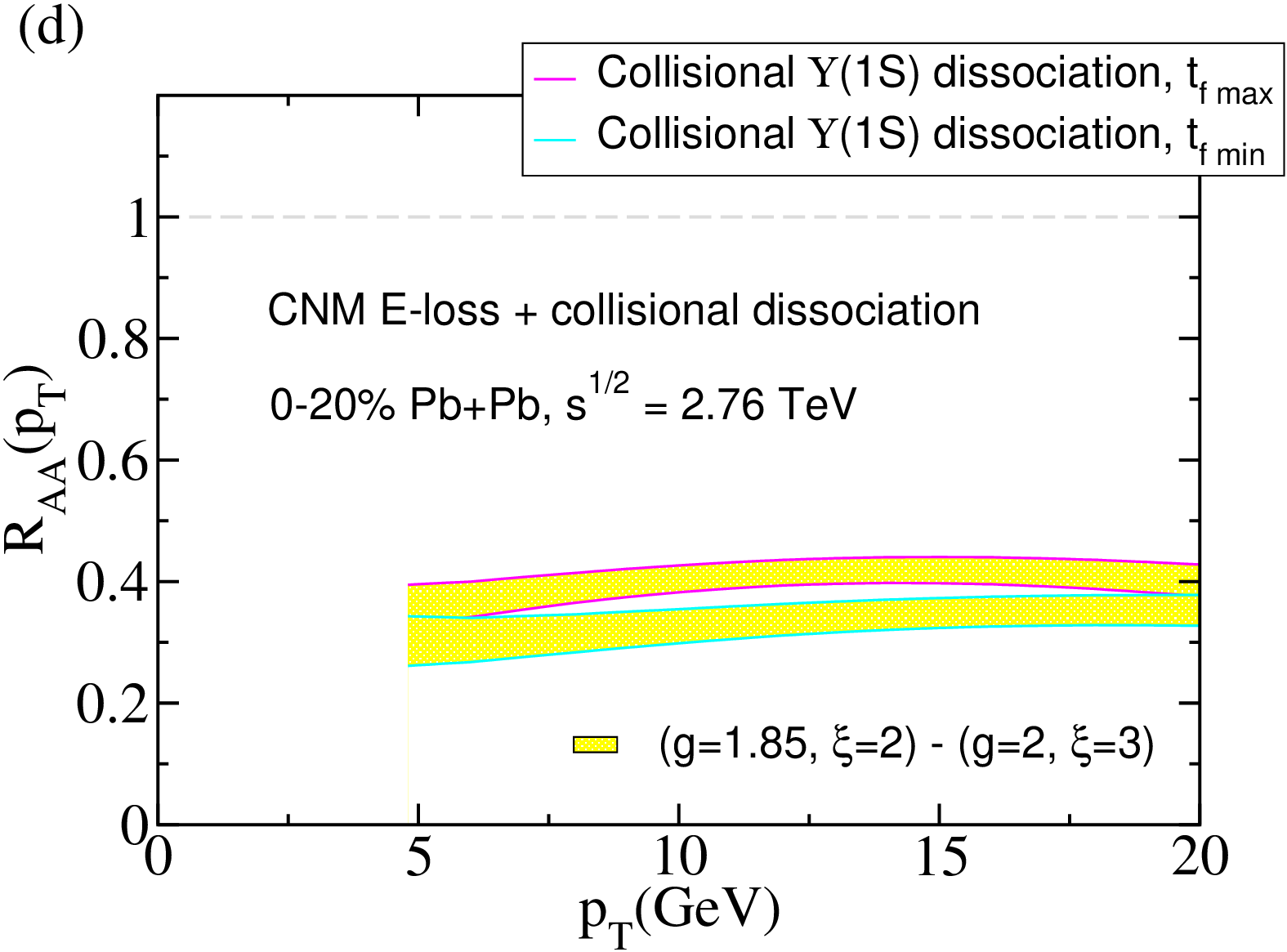}
\caption{(Color online) Theoretical model predictions for $\Upsilon$ $R_{AA}$
in nucleus-nucleus collisions.  The top right panel is for minimum bias collisions
and the rest for 0-20\% central collisions. The top left panel shows
$R_{AA}$ results for   $\Upsilon(1S)+\Upsilon(2S)+\Upsilon(3S)$ in Au+Au at
$\sqrt{S}=0.2$~TeV.  The bottom left and right panels shows $R_{AA}$ results
for   $\Upsilon(2S)+\Upsilon(3S)$ and  $\Upsilon(1S)$, respectively, in
Pb$+$Pb  at $\sqrt{S}=2.76$~TeV. The top right panel data from CMS is for
$0-100\%$ centrality~\cite{Chatrchyan:2012np} compared to the theoretical
prediction at the minimum bias $N_{\rm part} \approx 110$.
~\label{fig:RAAU}}
\end{figure}

For $p_T=9$~GeV at the LHC, our calculations give $R_{AA}\sim 0.3-0.4$, 
which is slightly smaller than at RHIC (at the same $ p_T$) but underestimates the 
observed suppression at  LHC in the most central Pb+Pb collisions.  
The ATLAS experiment has presented the ratio of binary collision scaled 
central-to-peripheral $J/\psi$  yields $R_{\rm CP}$~\cite{:2010px}. More 
specifically, their baseline is given by the 40-80\% peripheral Pb+Pb reactions.
The $p_T$ cuts used to obtain ATLAS results are such that 
$80$\% of the yields come from $p_T>6.5$~GeV. Comparison to the theoretical
model calculation is shown in the left bottom panel of  Fig.~\ref{fig:RAAJ1}.
The two data points are for 0-10\% and 10-20\% centrality. 
The CMS prompt $J/\psi$ $R_{AA}$ result~\cite{Chatrchyan:2012np} is  
shown in right bottom  panel of Fig.~\ref{fig:RAAJ1}. The two data points are
again for 0-10\% and 10-20\% centrality and   
exhibit stronger suppression than our theoretical model predictions. 
The ALICE experiment at LHC has  measured the nuclear modification of 
$J/\psi$ production at forward rapidity~\cite{Abelev:2012rv} and low transverse
momentum. To constrain theoretical models of quarkonium production in heavy
ion collisions, it will be very helpful to extend these forward rapidity 
data to high transverse momentum.   

In our formalism, we have approximated the quarkonium wavefunction by the
vacuum wavefunction, which is valid if the thermal effects on the  quarkonium 
wavefunctions are small.  The thermal wavefunctions  (for example, those obtained by
solving the Schr\"{o}dinger equation with thermal
potentials~\cite{Mocsy:2007jz,Rapp:2009my,Strickland:LongAndShort,Margotta:2011ta})
will be wider in position space at higher temperature and therefore will
dissociate more easily. Therefore a stronger suppression at LHC could be
the evidence for thermalization effects at the level of the quarkonium
wavefunction. We leave a more detailed analysis of thermal effects on
the wavefunctions for future work. Nevertheless, it is important to 
identify at what centrality the discrepancy between the present theoretical 
model predictions and the data appear. 
In Fig.~\ref{fig:RNpart} we show the $p_T$-averaged suppression,
\begin{equation} 
R_{AA}(N_{\rm part}) \left( {\rm or} \;  R_{CP}(N_{\rm part}) \right) = 
\frac{  \int_{p_{\rm min.}} d p_T \, R_{AA}(p_T;N_{\rm part}) \left(  {\rm or} \;  
R_{CP}(p_T;N_{\rm part}) \right) \frac{d\sigma}{dydp_T} } {   \int_{p_{\rm min.}} d p_T \, 
\frac{d\sigma}{dydp_T}   }    \; ,
\label{ptav}
\end{equation} 
of $J/\psi$ mesons versus centrality. We present a comparison to the 
ATLAS central-to-peripheral data~\cite{:2010px} in the left panel. The
deviation between data and theory is only seen for $N_{\rm part} > 300$.  A
comparison to the CMS data~\cite{Chatrchyan:2012np} is shown in the right
panel. In this case the deviation between data and theory is seen for $N_{\rm
part} > 200$.

The CMS  experiment at the LHC has also measured $R_{AA}$ for $\Upsilon(nS)$
states in Pb$+$Pb collisions at $\sqrt{S}=2.76$~TeV per nucleon pair. The
result~\cite{Chatrchyan:2011pe} is presented for decay muons satisfying the
transverse momentum cut  $p_T(\mu^{\pm})>4$~GeV and rapidity $|\eta| < 2.4$.
\begin{equation}
\begin{split}
\frac{\Upsilon(2S+3S)}{\Upsilon(1S)}|_{pp} &=0.76^{+0.16}_{-0.14}\pm0.12 \; ,\\
\frac{\Upsilon(2S+3S)}{\Upsilon(1S)}|_{PbPb} &=0.24^{+0.13}_{-0.12}\pm0.02  \; ,\\
\end{split}
\label{ups1}
\end{equation}
giving
\begin{equation}
\frac{R_{AA}(\Upsilon(2S+3S))}{R_{AA}(\Upsilon(1S))} =
0.32^{+0.19}_{-0.15}\pm0.03\;.
\label{ups2}
\end{equation}
We cannot calculate the equivalent ratio of $R_{AA}$s because our formalism
for the production and propagation of $\Upsilon$s is not applicable to
$p_T(\Upsilon)\lesssim 6$~GeV. In our approach the meson should be boosted
relative to the medium. Furthermore, for static or slowly moving mesons one
may expect thermal effects on the quarkonium wavefunction. These are precisely 
the $\Upsilon$s that determine the total yield ratios in Eqs.~\ref{ups1}, 
(\ref{ups2}).

\begin{figure}[!b]
\vspace*{.2in}
\includegraphics[width=3.38in,angle=0]{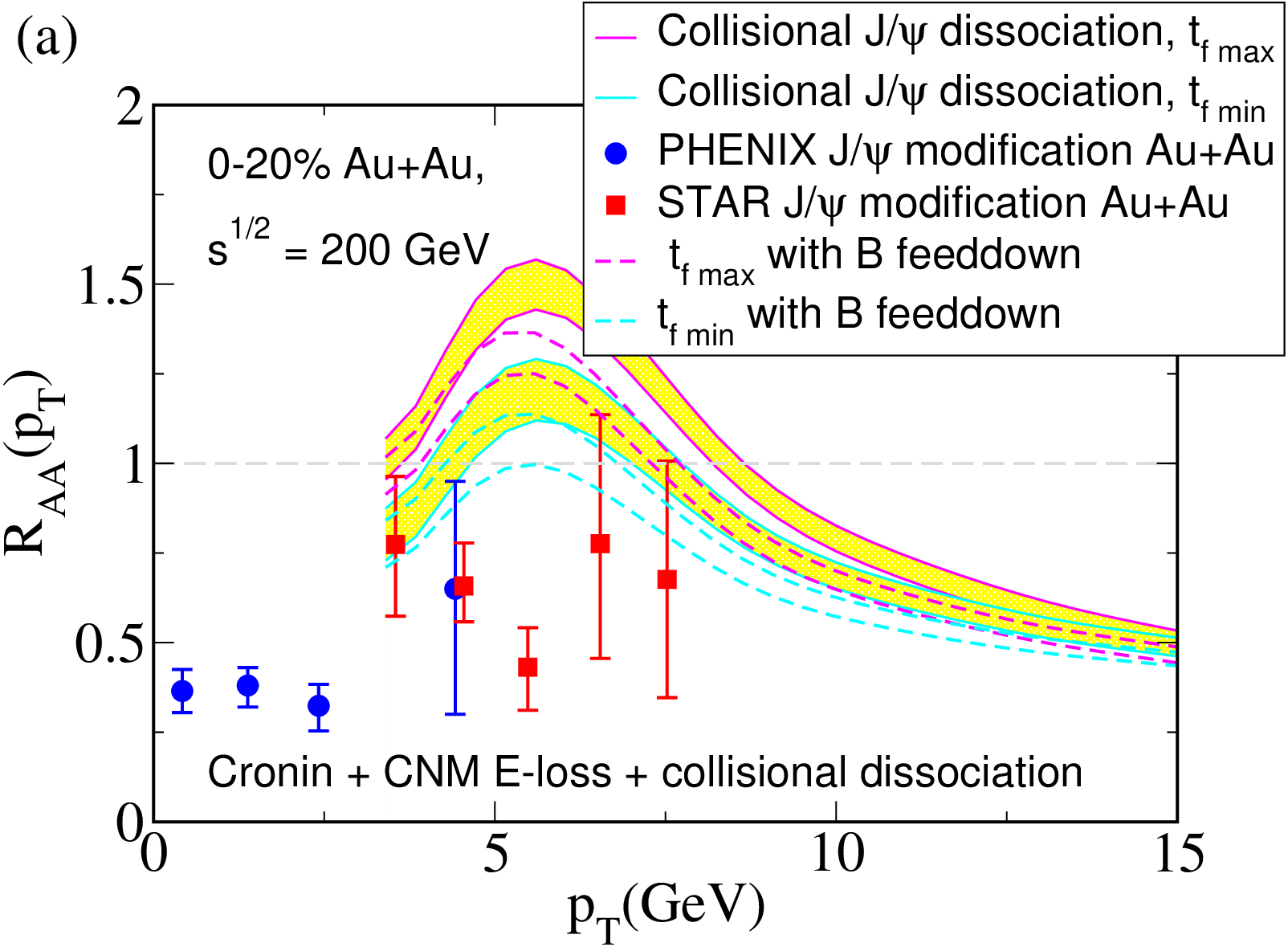} 
\includegraphics[width=3.38in,angle=0]{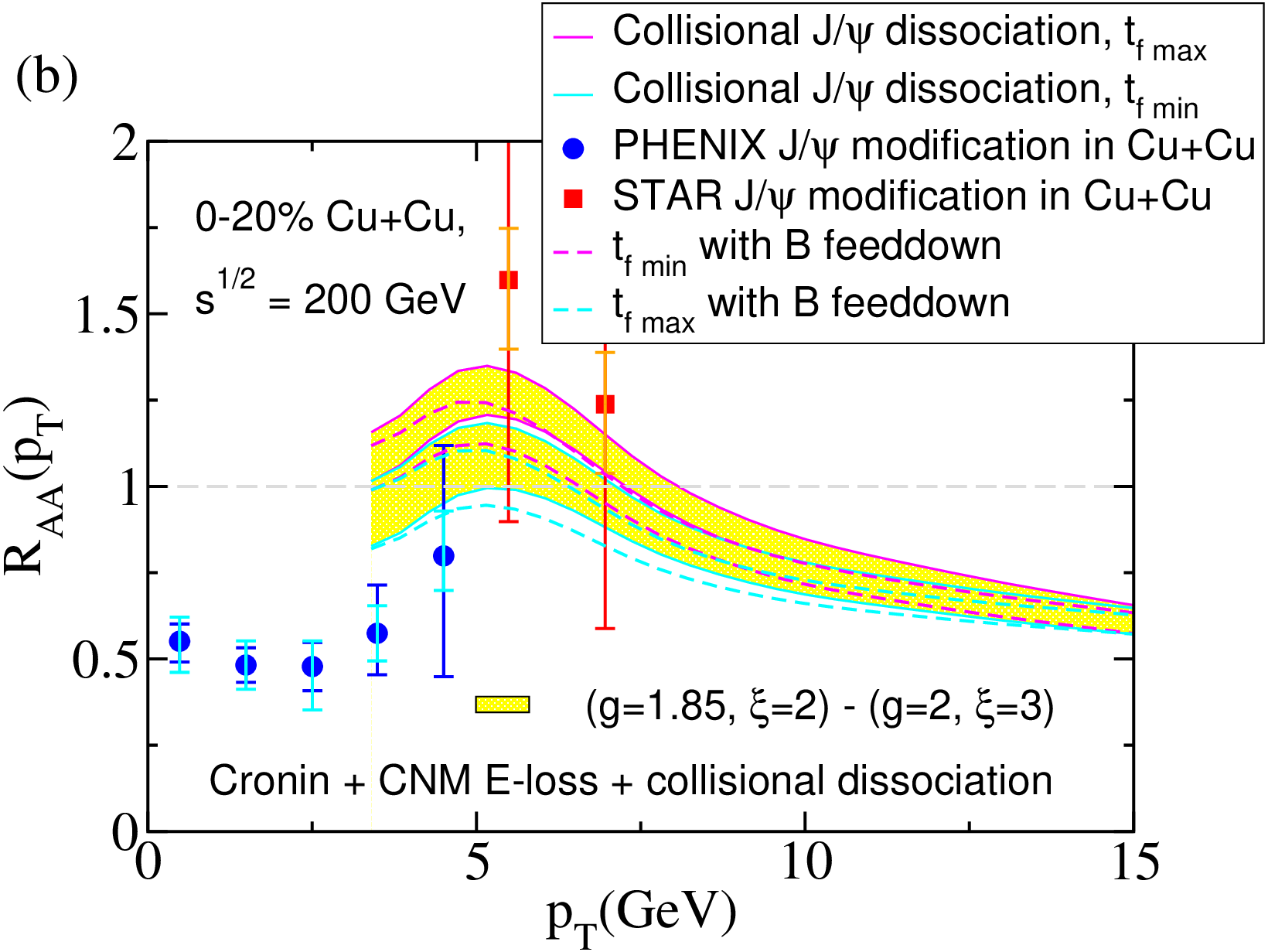} 
\includegraphics[width=3.38in,angle=0]{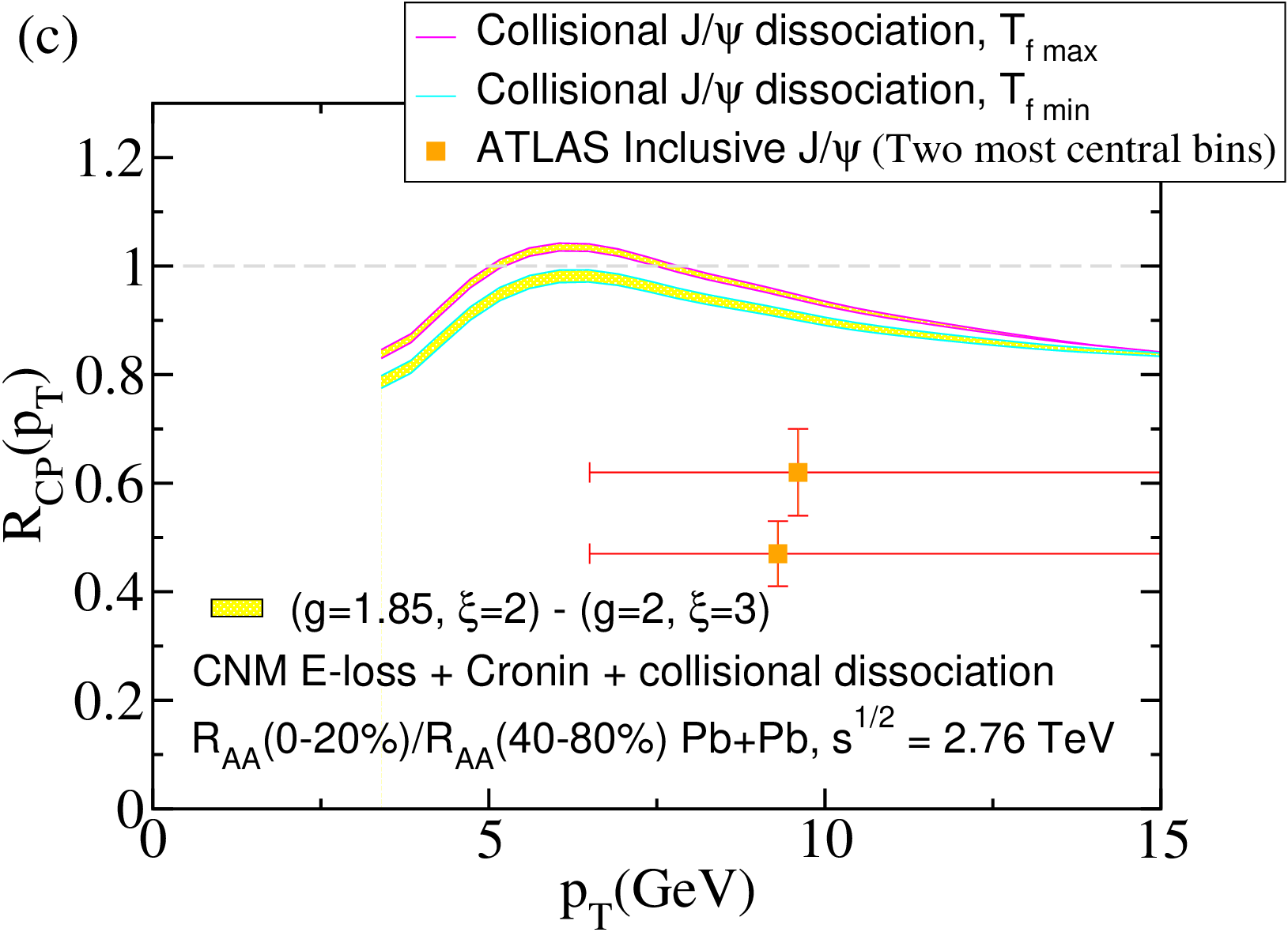} 
\includegraphics[width=3.38in,angle=0]{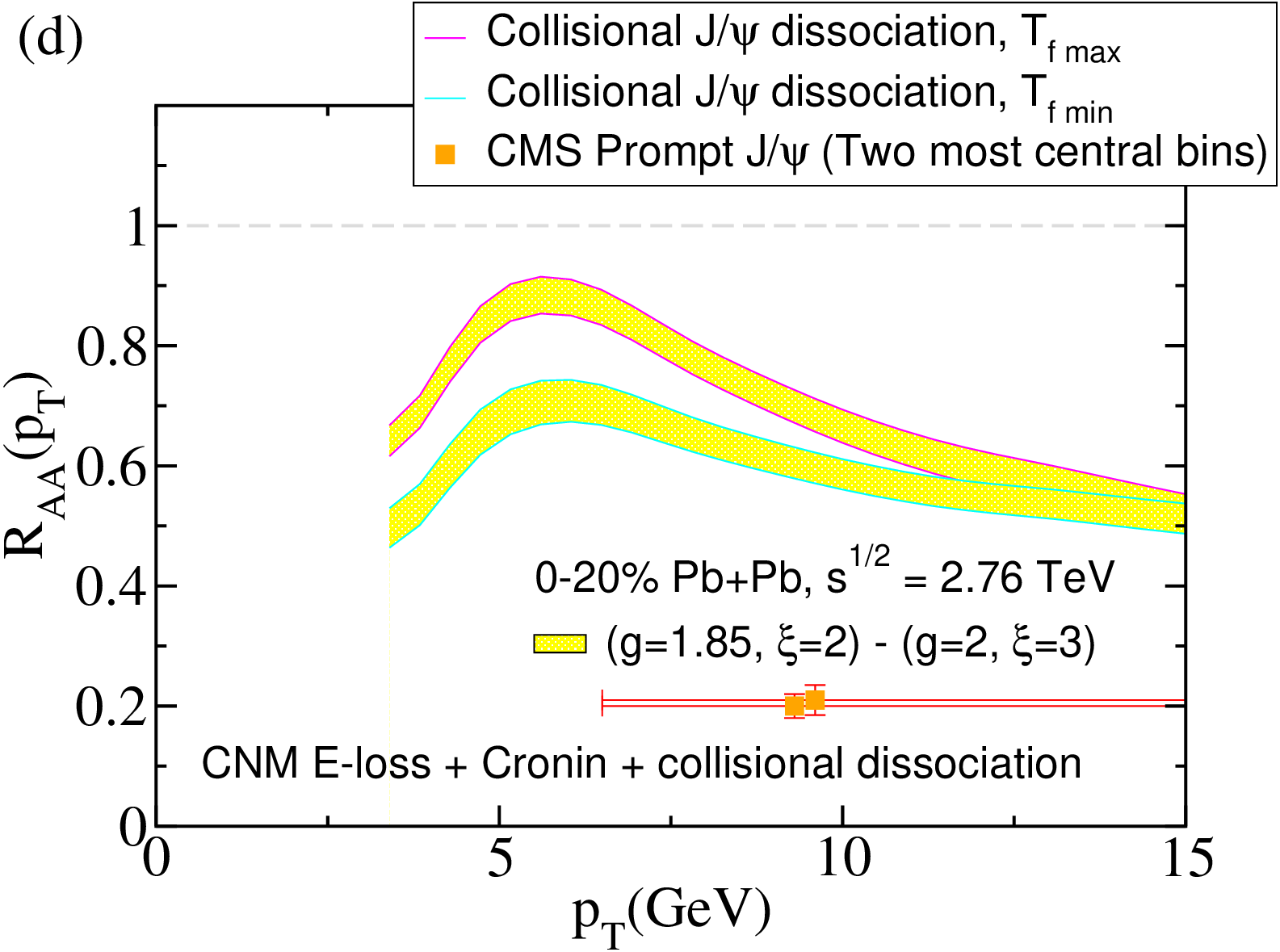} 
\caption{(Color online) Theoretically calculated $R_{AA}$ for $J/\psi$, including the 
Cronin effect. The top left panel
is for RHIC Au$+$Au collisions at $\sqrt{S}=0.2$~TeV. The top right panel is for RHIC
Cu$+$Cu collisions at $\sqrt{S}=0.2$~TeV. The lower two panels are for LHC Pb$+$Pb
collisions at $\sqrt{S}=2.76$~TeV as in Fig.~\ref{fig:RAAJ1}.~\label{fig:RAAJ1Cronin}}
\end{figure}

For $p_T > 7$~GeV at the LHC, our calculations show that
$\frac{R_{AA}(\Upsilon(2S+3S))}{R_{AA}(\Upsilon(1S))}\sim 1$ (the lower panels
of Fig.~\ref{fig:RAAU}). Physically, this is because the higher dissociation
rate for the excited mesons is compensated by their larger formation time,
which means that dissociation becomes dominant after the medium has diluted.
If in future experiments $p_T$-differential yields of $\Upsilon(2S+3S)$ shows a
much stronger suppression compared to $\Upsilon(1S)$ for $p_T\gtrsim 7$~GeV,
that may also support the possibility for thermal effects on the quarkonium
wavefunction at high $p_T$. This is because the thermal wavefunctions of higher
excited states, being broader in position space, are more easily affected by
the QGP, thereby naturally giving even higher dissociation rates.
Fig.~\ref{fig:RAAU} also shows theoretical predictions for the suppression of
the various $\Upsilon$ states in central Au+Au and Pb+Pb collisions at RHIC and
the LHC, respectively. In the top right panel we present a comparison to the
minimum bias $p_T$-differential $\Upsilon(1S)$ CMS nuclear modification data.
In this case, the theoretical calculation is performed for the average number
of participants for minimum bias collisions.

The overall suppression of the $J/\psi$ and the $\Upsilon$ yields in A$+$A
collisions is a combination of CNM and QGP effects. In this section we ignored
transverse momentum broadening effects. Therefore, the suppression is largely due to
cold nuclear matter energy loss and QGP dissociation (the effect of shadowing
is small). In Section~\ref{section:Cronin}, we will show results for A$+$A
collisions including Cronin, and results for p$+$A collisions where QGP effects
are absent. (See Fig.~\ref{fig:RpAJ1} for $J/\psi$ and Fig.~\ref{fig:RpAU1} for
$\Upsilon$.) For $p_T\sim 6$~GeV, $R_{pA}\sim0.8$ for both $J/\psi$ and
$\Upsilon$. Noting that in A$+$A collisions the CNM effects are amplified
relative to p+A, we conclude that a significant part of the suppression in
quarkonium yield in our calculation comes from cold nuclear matter energy loss.
The situation is more complicated when we include transverse momentum
broadening as we discuss next.

\subsection{Transverse momentum broadening effects~\label{section:Cronin}}

\begin{figure}[!b]
\vspace*{.2in}
\includegraphics[width=3.38in,angle=0]{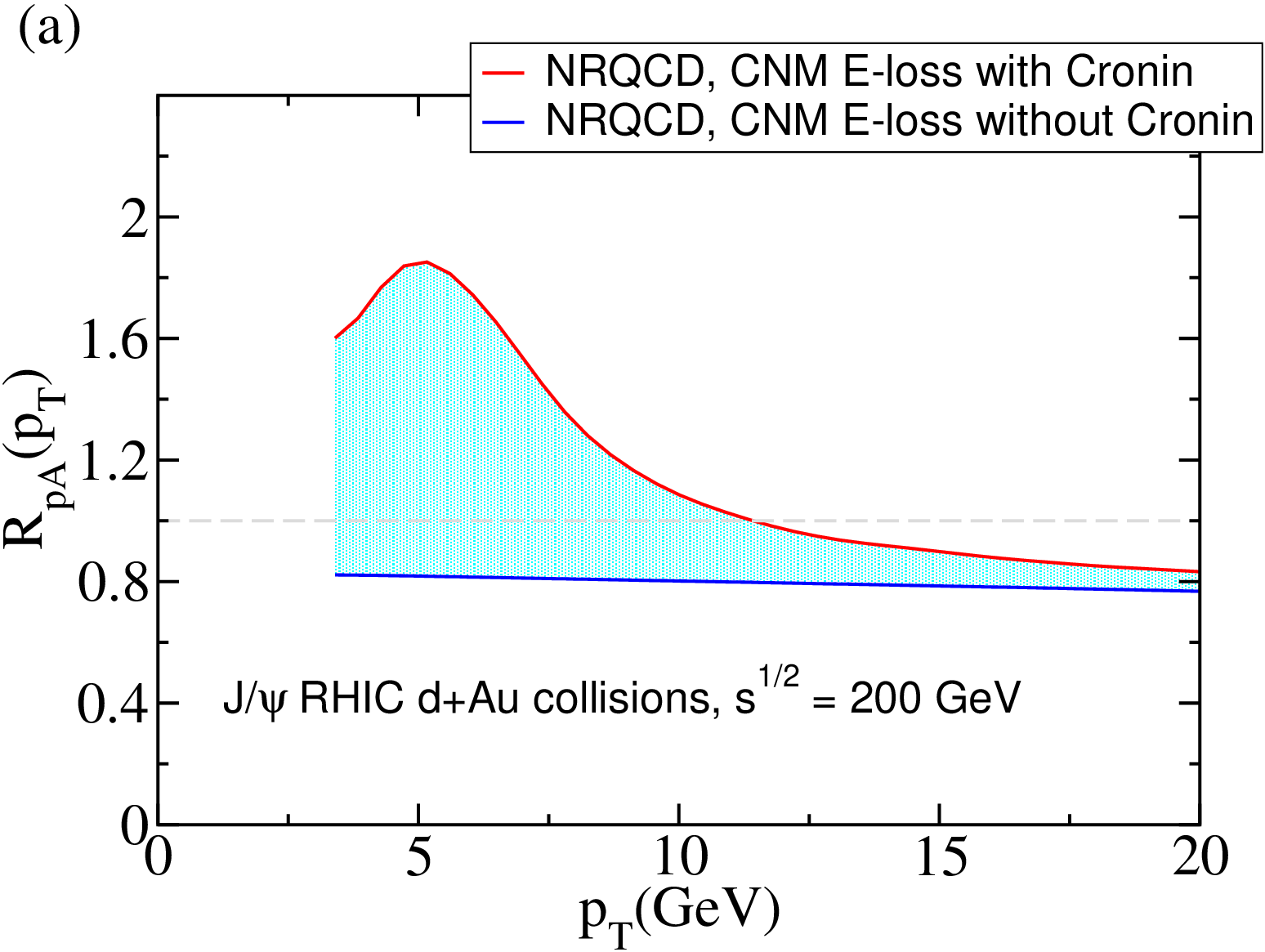} 
\includegraphics[width=3.38in,angle=0]{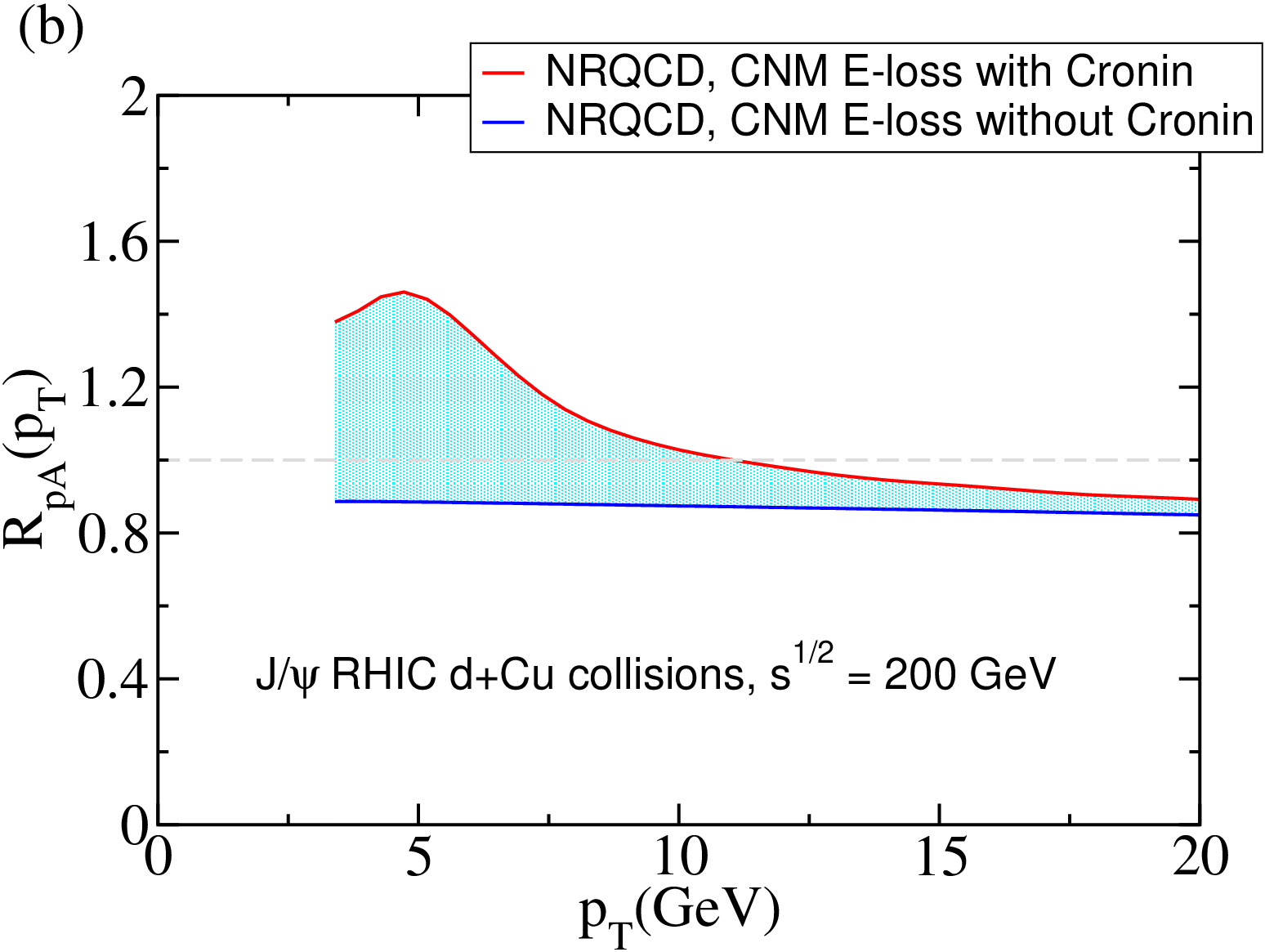} 
\includegraphics[width=3.38in,angle=0]{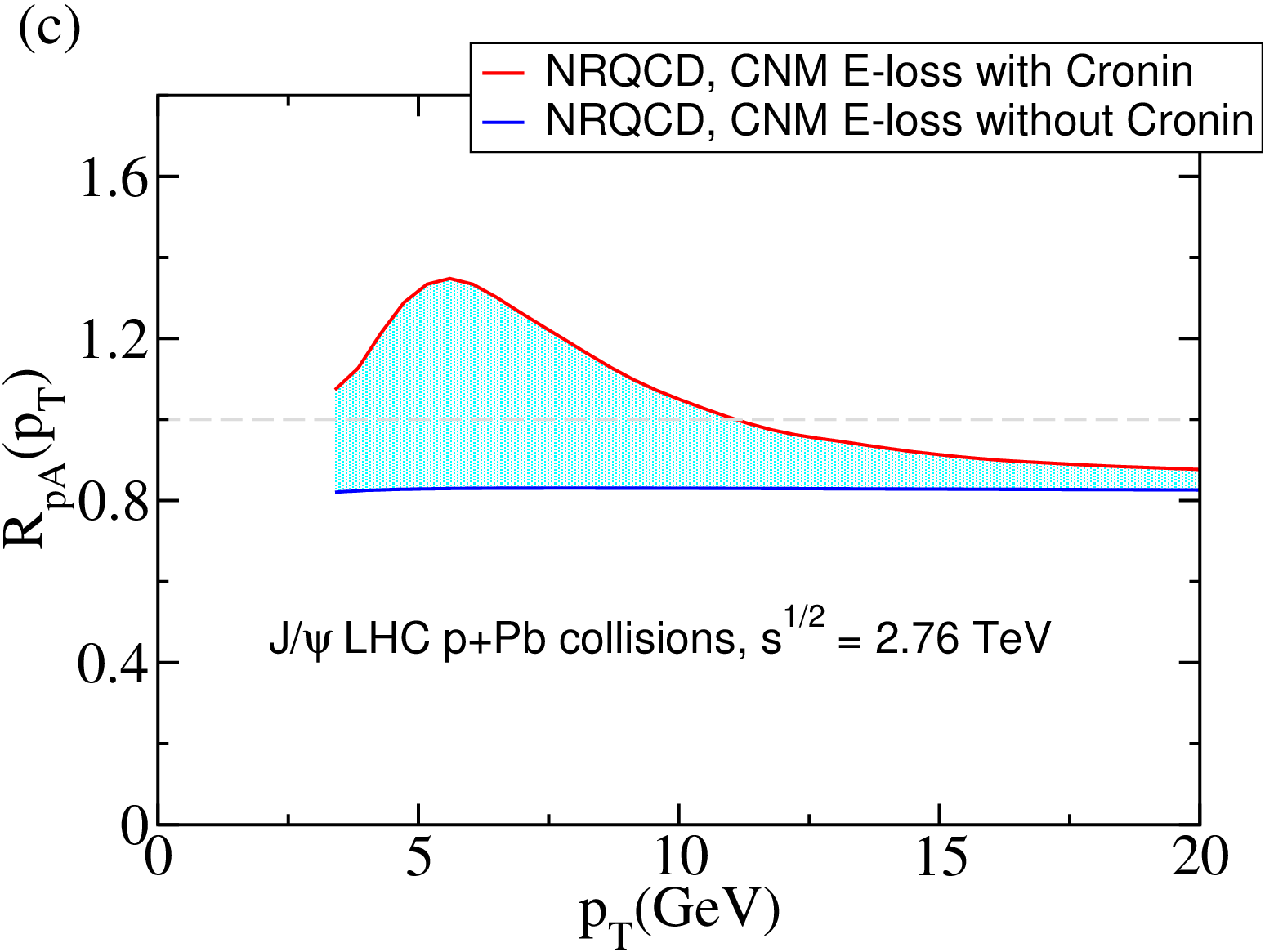} 
\caption{(Color online) Theoretical predictions for $J/\psi$  $R_{pA}$ in
minimum bias collisions with (upper curve, red online) and without (lower
curve, blue online) the Cronin effect. In this plot we only show the
prompt yields. The top left panel is for RHIC d$+$Au collisions at
$\sqrt{S}=0.2$~TeV. The top right panel is for RHIC d$+$Cu collisions at
$\sqrt{S}=0.2$~TeV. The lower panel is for LHC p$+$Pb collisions at
$\sqrt{S}=2.76$~TeV.~\label{fig:RpAJ1}}
\end{figure}

As expected, transverse momentum broadening effects enhance the production of
$J/\psi$ mesons for $p_T \sim 5-10$~GeV in heavy ion collisions
(Fig.~\ref{fig:RAAJ1Cronin}).  The Cronin effect is not very important for
$p_T\gtrsim10$~GeV at these center-of-mass energies. The  main features are as
follows: the Cronin effect can substantially alter the $R_{AA}$  of $J/\psi$
mesons with $p_T\sim 7$~GeV  at RHIC.   For example, for Au$+$Au collisions
$R_{AA}$ increases from $\sim 0.35-0.45$ to $\sim0.8-1.2$. For Cu$+$Cu
collisions $R_{AA}$ increases from $\sim0.45-0.65$ to $\sim0.7-1.0$. The reason
behind this enhancement is that the mean scattering lengths of the
initial-state gluons (which dominate quarkonium production) are considerably
smaller than the mean scattering lengths for quarks.  The Cronin effect at the
LHC is smaller than the  Cronin effect at RHIC because of the harder quarkonium
spectrum. Including broadening only increases $R_{AA}$ at $p_T\sim 7$~GeV from
$\sim0.3-0.4$ to $\sim0.6-0.8$.  The theoretically calculated $R_{AA}$  for
$J/\psi$ may be consistent with the RHIC Cu$+$Cu results but is only marginally
compatible with the RHIC Au$+$Au results for $p_T$ above $4$~GeV.  Also, it is clearly incompatible
with the ATLAS central-to-peripheral  Pb$+$Pb data at the LHC and the
CMS $R_{AA}$ suppression. The differences in the degree of $p_T = 5 -10$~GeV
suppression of quarkonium production between RHIC and the LHC in
Fig.~\ref{fig:RAAJ1Cronin}  suggest  that better understanding of the Cronin
effect (if any) is necessary for  consistent $J/\psi$ phenomenology in heavy
ion collisions.

More specifically,  experimental quarkonium yields in  p$+$A or d$+$A
collisions, where  effects from the QGP are absent, are important to constrain
the cold nuclear matter effects.  Our theoretical predictions for the
high-$p_T$  $J/\psi$ nuclear modification factor  $R_{pA}$ in such collisions
are given in Fig.~\ref{fig:RpAJ1}. The upper (solid red) curve includes the
Cronin effect. The lower (solid blue) curve includes only power corrections and
CNM energy loss.  The upper curve is the maximum Cronin enhancement we obtain
for reasonable parameters and therefore the band should be interpreted as the
plausible region for $R_{pA}$. Comparison with recent PHENIX
data~\cite{Adare:2012qf} suggests that the $R_{pA}$ is closer to the
lower curves (no Cronin). LHC $R_{AA}$ measurements also point to an absence of
the Cronin effect for quarkonia. One interesting point to note is that the
Cronin peak in $AA$ is smaller than $pA$ collisions due to the QGP effects. We
finally point out that including the contribution from $B$ decays reduces the
Cronin enhancement for $p_T \sim 5-10$~GeV slightly. 

\begin{figure}[!t]
\vspace*{.2in}
\includegraphics[width=3.38in,angle=0]{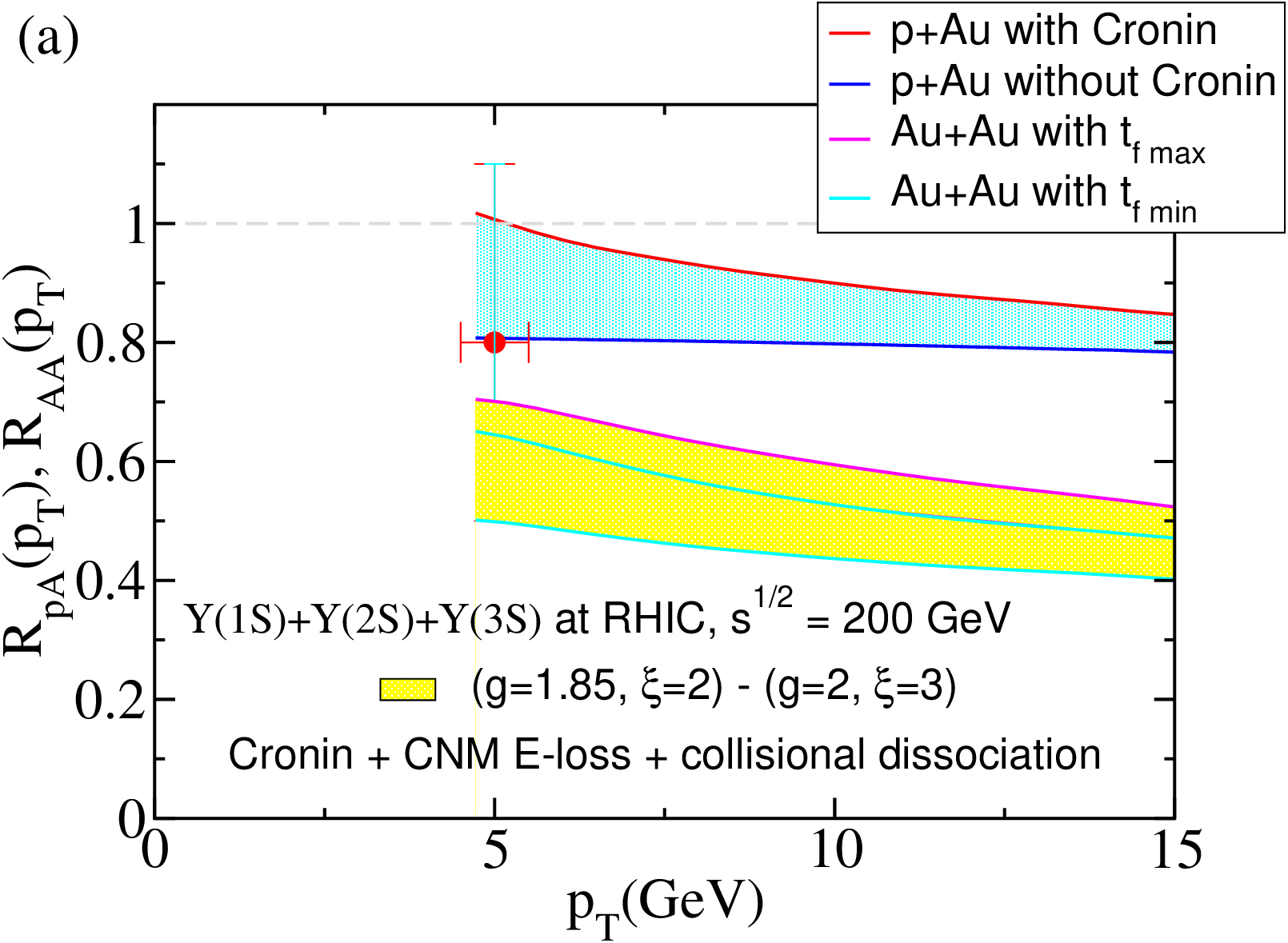}
\includegraphics[width=3.38in,angle=0]{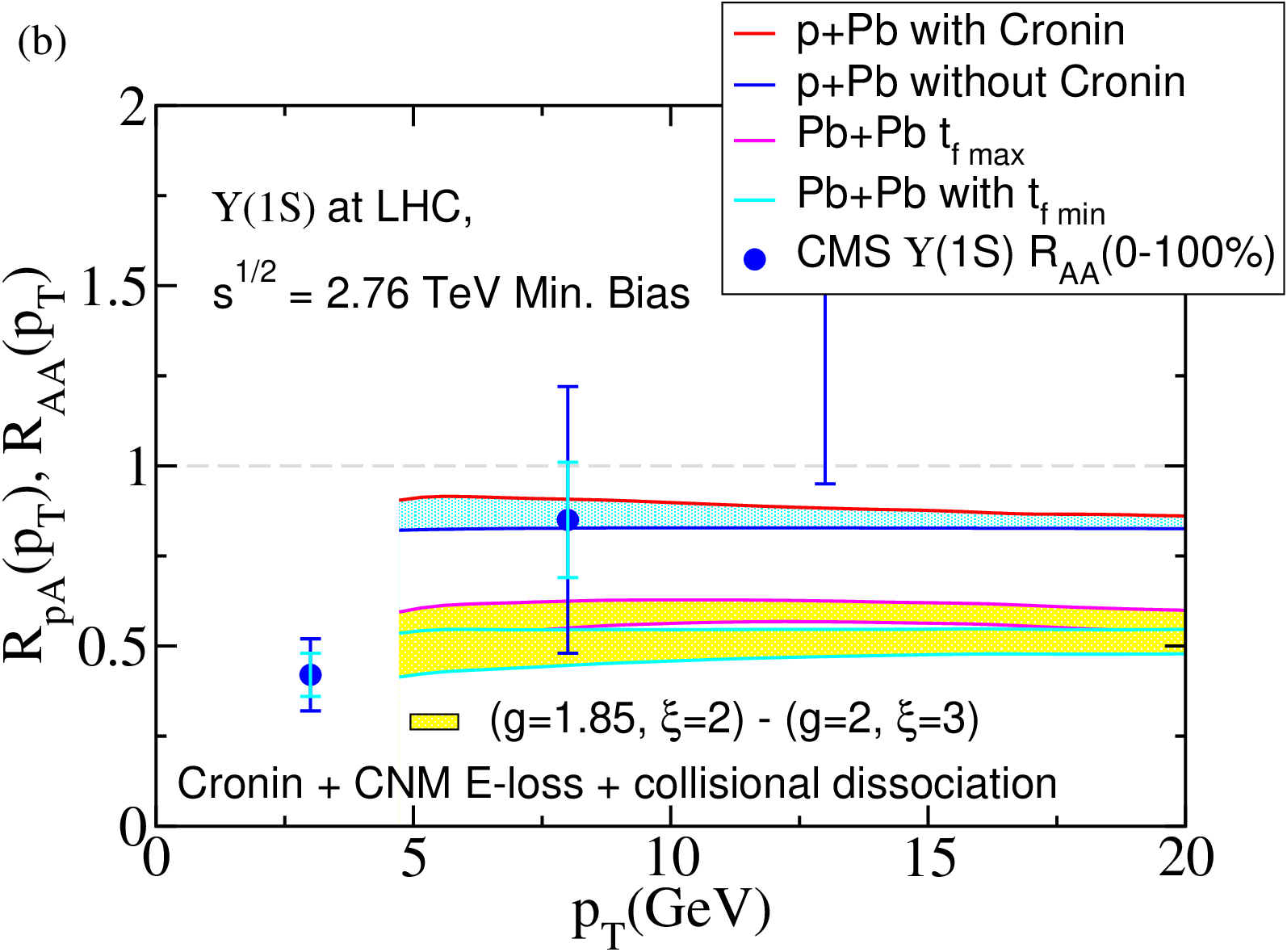} 
\caption{(Color online) Theoretical model predictions for  minimum bias
$R_{pA}(\Upsilon)$ and for $R_{AA}(\Upsilon)$. The left panel is for RHIC p$+$Au and
central ($0-20\%$) Au$+$Au collisions at  $\sqrt{S}=0.2$~TeV.  Data is from
STAR~\cite{Reed:}. The right panel is for LHC p$+$Pb and minimum bias
($N_{\rm{part}}\approx 110$) Pb$+$Pb collisions at $\sqrt{S}=2.76$~TeV.  Data
is from CMS~\cite{Chatrchyan:2012np}.~\label{fig:RpAU1}} 
\end{figure}

The combined results for $R_{AA}$ (0-20\% central) and $R_{pA}$ (minimum bias)
for bottomonia are given in Fig.~\ref{fig:RpAU1}. The effect of transverse
momentum broadening is much smaller for bottomonia when compared to the one for
charmonia. This can be intuitively understood as follows. The mechanism for
Cronin enhancement in this calculation is that initial state scattering
increases the typical transverse momentum carried by the incident partons by a
few GeV. For quarkonia, there is an additional scale $m_H$. For bottomonia the
mass scale is considerably larger than the transverse momentum broadening scale
and few additional GeV do not increase the yields significantly. Preliminary
$\Upsilon $ suppression data are now available at RHIC~\cite{Reed:}. More
differential $p_T$ data will shed light on the similarities and differences in
the CNM effects at RHIC and at the LHC.

\section{Conclusions~\label{section:Conclusions}}

In summary, we carried out a detailed study of high transverse momentum
quarkonium production and modification in heavy ion reactions at RHIC and at
the LHC. We used  a NRQCD approach to calculate the
baseline quarkonium cross sections. We found that for $J/\psi$ mesons the
theoretically computed spectrum is slightly harder than the one observed in the
experiment. For all $\Upsilon$ states ($1S, 2S, 3S$) the agreement is within a
factor of two when we consider both the TeVatron and the LHC data.  In
reactions with heavy nuclei, we presented theoretical model calculations for the
nuclear modification of quarkonium yields at high $p_T$ in minimum bias p(d)$+$A
and  0-20\% central  A$+$A collisions.  We focused on the consistent inclusion of
both cold (CNM) and hot (QGP) nuclear matter effects in different colliding
systems at different center-of-mass energies. We compared our results to
published and preliminary experimental data, where applicable.

In calculating the spectra of quarkonia in heavy ion reactions, 
we included nuclear shadowing (here implemented as coherent power corrections) 
and initial-state energy loss. We also provided, to the best of our knowledge, 
the first implementation of initial-state  transverse momentum broadening  
to study  phenomenologically a possible Cronin-like enhancement for quarkonia. 
All these effects have been well studied for light partons and have been 
recently incorporated in open heavy flavor production. In this paper we extended 
them to $J/\psi$ and $\Upsilon$ mesons. Since the $Q\barQ$ pair is created in the
short-distance hard scattering process and  evolves quickly into a component
of the quarkonium wavefunction, the effects of propagation through the QGP 
were included through (a) quenching of the color octet component (b) collisional 
dissociation model for the formed meson, which was successful 
in describing the attenuation of open heavy flavor, $B\rightarrow \ell+X, 
D\rightarrow \ell+X$, at RHIC. In this paper we restricted our results to  
high transverse momentum and explored the consequences of assuming that
the initial wavefunctions of the quarkonia are well approximated by 
vacuum wavefunctions in the short period before the dissociation.

We found that ignoring the Cronin effect leads to a small overestimate 
of the suppression of $J/\psi$ mesons in the $p_T$ region between 5 GeV and 
10 GeV in central Cu+Cu and Au+Au collisions 
at $\sqrt{S}=0.2$~TeV at RHIC. Including initial-state transverse momentum 
broadening  leads to a somewhat better agreement between theory and the current 
experimental data only for the Cu+Cu reactions. A smaller Cronin enhancement 
will work better. We demonstrated that CNM
effects can be easily constrained in d$+$A reactions at RHIC. For example, 
the d$+$Au calculation that includes power corrections and cold nuclear matter
energy loss predicts ~20\% suppression of the $J/\psi$ cross section. 
Including transverse momentum broadening may lead to as much as 50\% 
enhancement in the region of the Cronin peak. We also found  that 
the Cronin-like modification of the $\Upsilon$ spectrum is much smaller.        
Current data on high-$p_T$ quarkonium production at RHIC does not indicate 
the presence of thermal effects at the level of the quarkonium wavefunction
within our theoretical framework.

The conclusions from our theoretical model comparison to the $\sqrt{S}=2.76$~TeV 
LHC data are not as clear cut. Our calculations underestimated the 
suppression for $J/\psi$ production reported by the ATLAS and CMS experiments 
in the most central Pb+Pb collisions. On the other hand, they agree quite well in 
mid-central and peripheral reactions. We found that the Cronin enhancement at the
LHC is smaller than the one at RHIC due to the harder spectra. However, any Cronin 
enhancement appears incompatible with the experimental results.  
For $\Upsilon$ mesons, $p_T$-differential data in A+A collisions is scarce. 
CMS data for minimum bias $\Upsilon(1S)$ indicate that the low $p_T$ 
suppression may decrease or disappear at high $p_T$. At the same time, at low $p_T$,
where our calculation is not applicable, CMS reported a strong relative 
suppression of $\Upsilon(2S+3S)$ to $\Upsilon(1S)$. If the data is extended 
to high $p_T$ with similar results, it will clearly be incompatible with
our model predictions with quarkonium wavefunctions unaffected by thermal effects.
Together with a more refined $p_T$-differential data on $J/\psi$ suppression at
$\sqrt{S}=2.76$~TeV, this will be a strong indication that thermal QGP effects
may persist for high transverse momentum quarkonia at the LHC in central Pb+Pb
reactions. We plan to address this  possibility in a separate publication.

\section{Acknowledgments}
The authors acknowledge e-mail communication with E. Braaten. The authors also
thank  T. Dahms, A. Dainese, and C. Mironov for useful comments.  This research is
supported by  National Sciences and Engineering Research Council of Canada
(NSERC),  the US Department of Energy, Office of Science, under Contract
No.~DE-AC52-06NA25396 and in part by the JET collaboration.

\begin{appendix}

\section{Fitting color-octet matrix elements for charmonia and a baseline for 
LHC at $\sqrt{S}=2.76$~TeV~\label{appendix:LHCbaseJ}}
\begin{figure}[!ht]
\vspace*{.2in}
\includegraphics[width=4.0in,angle=0]{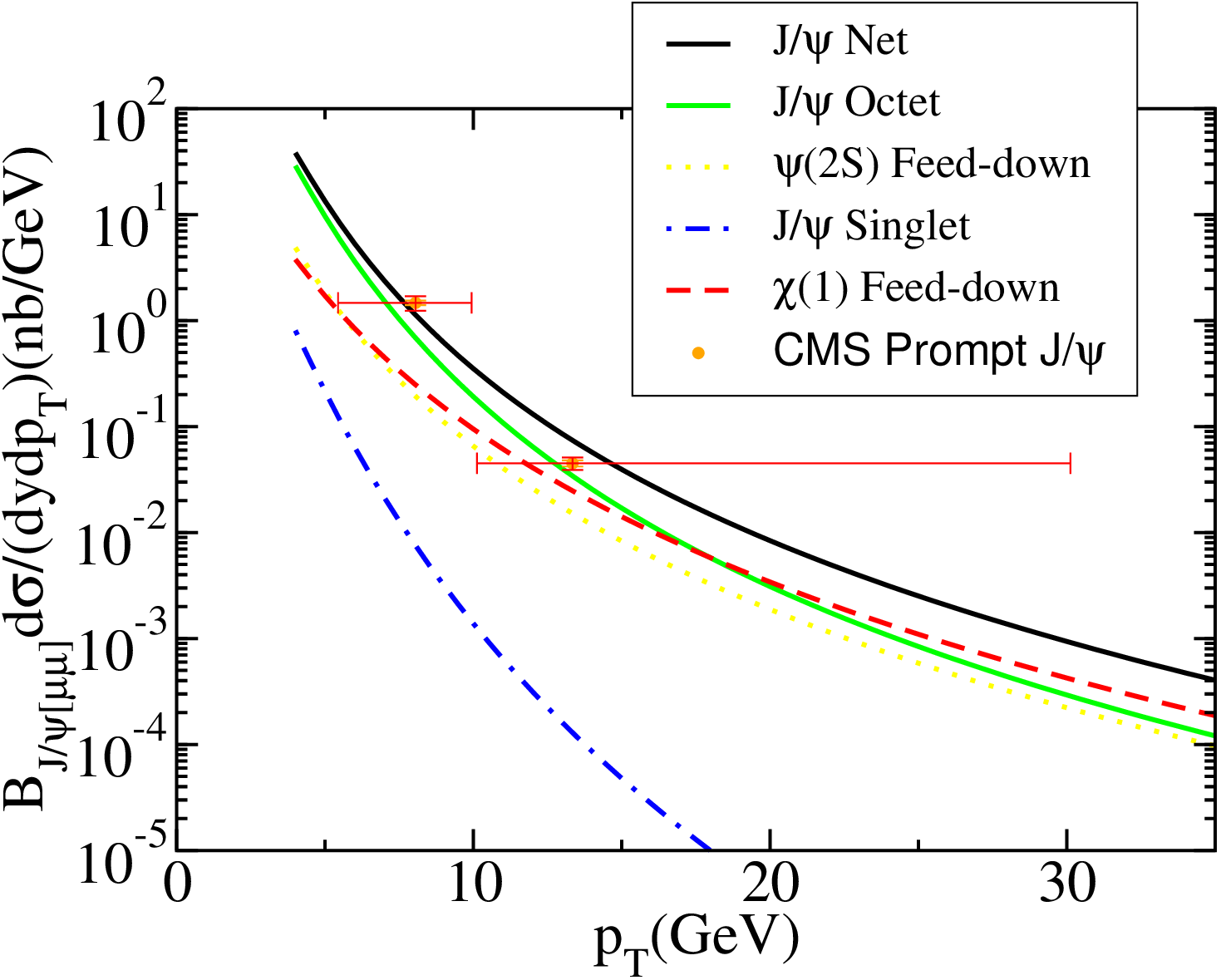} 
\caption{(Color online) The cocktail of contributions that gives the p$+$p
yield for $J/\psi$ production at the LHC at $\sqrt{S}=2.76$~TeV. The uppermost
curve (black online) gives the net yield and gives the p$+$p baseline. We see
that the dominant contribution is the color-octet  contribution. The data is
from CMS~\cite{Chatrchyan:2012np}.~\label{fig:LHC2760Jcocktail}}
\end{figure}

For the net production of $J/\psi$ we consider the direct contribution, and
feed-down contributions from $\chi_{c0}(1P)$, $\chi_{c1}(1P)$,
$\chi_{c2}(1P)$ and $\psi(2S)$. The relevant branching fractions are given
in Table~\ref{table:CharmoniaBFs}~\cite{pdg},
\begin{table*}[h]
\begin{tabular}{c|cccc}
meson from &to $\chi_{c0}$ &to $\chi_{c1}$ &to $\chi_{c2}$ &to $J/\psi$\\ 
\hline
$\psi(2S)$ & 0.0962 & 0.092 & 0.0874 & 0.595   \\
$\chi_{c0}$& &  &  & 0.0116           \\
$\chi_{c1}$& &  &  & 0.344           \\
$\chi_{c2}$& &  &  & 0.195           \\
\end{tabular}
\caption{Relevant branching fractions for charmonia~\cite{pdg}.~\label{table:CharmoniaBFs}}
\end{table*}

$p_T$-differential yields at small rapidity for $J/\psi$ is available from
LHC(\cite{Aad:2011sp,Chatrchyan:2012np}), TeVatron
(\cite{Abe:1997yz,Acosta:2004yw}) and RHIC
(\cite{Adare:2009js,Abelev:2009qaa}). Yields of $\psi(2S)$ have been measured
at TeVatron (\cite{Aaltonen:2009dm,Abe:1997yz,Abe:1997older}) and
LHC(\cite{Chatrchyan:2011kc}). Data for $\chi_{cJ}$ is available from
TeVatron~\cite{Abe:1997yz}.  

 The following color-singlet and color-octet contributions are relevant for our
calculation.
\begin{enumerate}
\item{Direct contributions
\begin{equation}
\begin{split}
{\cal{M}}(\psi[^3S_1]_{1})
 &=\langle \calO(c\barc([^3S_1]_{1})\rightarrow J/\psi)\rangle
 =3N_c\frac{|R_{n=1}(0)|^2}{2\pi}\\
{\cal{M}}(\psi[^3S_1]_{8})
 &=\langle \calO(c\barc([^3S_1]_{8})\rightarrow J/\psi)\rangle \\
{\cal{M}}(\psi[^3P_0]_{1})
 &= \langle \calO(c\barc([^3P_0]_{8})\rightarrow J/\psi)\rangle \\
{\cal{M}}(\psi[^1S_0]_{1})
 &=\langle \calO(c\barc([^1S_0]_{8})\rightarrow J/\psi)\rangle 
~\label{eq:Mpsi1}
\end{split}
\end{equation}
    }
\item{Indirect contribution from $\chi_{cJ}$
\begin{equation}
\begin{split}
{\cal{M}}(\chi[^3P_0]_{1}) 
 &= \langle \calO(Q\barQ([^3P_0]_{1})\rightarrow \chi_{c0})\rangle  = 
3N_c\frac{|R^\prime_{n=1}(0)|^2}{2\pi}\\
{\cal{M}}(\chi[^3S_1]_{8})
 &=\langle \calO(Q\barQ([^3S_1]_{8})\rightarrow \chi_{c0})\rangle
~\label{eq:Mchi}
\end{split}
\end{equation}
    }
\item{Indirect contribution from $\psi(2S)$
\begin{equation}
\begin{split}
{\cal{M}}(\psi[^3S_1]_{1})
 &=\langle \calO(c\barc([^3S_1]_{1})\rightarrow \psi(2S))\rangle
 =3N_c\frac{|R_{n=1}(0)|^2}{2\pi}\\
{\cal{M}}(\psi[^3S_1]_{8})
 &=\langle \calO(c\barc([^3S_1]_{8})\rightarrow \psi(2S))\rangle \\
{\cal{M}}(\psi[^3P_0]_{1})
 &= \langle \calO(c\barc([^3P_0]_{8})\rightarrow \psi(2S))\rangle \\
{\cal{M}}(\psi[^1S_0]_{1})
 &=\langle \calO(c\barc([^1S_0]_{8})\rightarrow \psi(2S))\rangle 
~\label{eq:Mpsi2}
\end{split}
\end{equation}
    }
\end{enumerate}

Hence we have to determine $10$ parameters. The color singlet matrix elements
can be estimated from the wavefunctions of the heavy mesons. We use values from
~\cite{Cho:1995ce,Cho:1995vh,Eichten:1994gt},
\begin{equation}
\begin{split}
{\cal{M}}(J/\psi[^3S_1]_{1}(1S)) &= 1.2 \rGeV^3\\
{\cal{M}}(\chi[^3P_0]_{1}(1P))/m_{\charm}^2 &= 0.054 \rGeV^3\\
{\cal{M}}(\psi[^3S_1]_{1}(2S)) &= 0.76 \rGeV^3\;.
~\label{eq:MCSpsi_val}
\end{split}
\end{equation}

The color octet matrix elements can not be determined from the wavefunction
because it involves the wavefunctional form of a higher Fock state. Therefore,
we fit them to reproduce $p_T$ differential cross sections at the LHC ,
TeVatron and RHIC. We use the following procedure to determine the remaining
$6$ color-octet components. 

CDF~\cite{Abe:1997yz} has measured the feed-down contribution from the
$\chi_{cJ}$ states to $J/\psi$ production. We use this data to fit the octet
matrix element ${\cal{M}}(\chi_{c0}[^3S_1]_{8})$.
\begin{equation}
\begin{split}
{\cal{M}}(\chi_{c0}[^3S_1]_{8}(1P))/m_{\charm}^2 &= (0.00187\pm 0.00025) \rGeV^3\;,
~\label{eq:Mchicn1CO_val}
\end{split}
\end{equation}
where the error includes the change in the matrix elements when we change the
lowest $p_T$ included in the fit by $1$~GeV.  The $\chi^2/\rdof=4.56$ is not
very good because the (dominant) color-octet production is harder than the
experimentally observed spectrum.

Similarly we assume that the measured yields of prompt $\psi(2S)$ is not
substantially contaminated by higher feed-downs and fit the following data
\begin{enumerate}
\item{CDF results at $\sqrt{S}=1.96$~TeV~\cite{Aaltonen:2009dm} and
$\sqrt{S}=1.8$~TeV~\cite{Abe:1997yz,Abe:1997older}}
\item{ATLAS results at $\sqrt{S}=7$~TeV~\cite{Chatrchyan:2011kc}}
\end{enumerate}

We obtain,
\begin{equation}
\begin{split}
{\cal{M}}(\psi[^3S_1]_{8}(2S)) &= (0.0033\pm 0.00021) \rGeV^3\\
{\cal{M}}(\psi[^1S_0]_{8}(2S)) &= (0.0080\pm 0.00067) \rGeV^3=
{\cal{M}}(\psi[^3P_0]_{8}(2S))/m_{\charm}^2\;,
~\label{eq:MJn2CO_val}
\end{split}
\end{equation}
with a $\chi^2/\rdof=5.6$. 

To fit the remaining fit $3$ parameters we use the combined fit for the 
following results for $J/\psi$ (direct+feed-down) yields
\begin{enumerate}
\item{CDF results at $\sqrt{S}=1.96$~TeV~\cite{Acosta:2004yw}}
\item{PHENIX results at $\sqrt{S}=0.2$~TeV~\cite{Adare:2009js}}
\item{ATLAS results at $\sqrt{S}=7$~TeV~\cite{Aad:2011sp}}
\end{enumerate}

Some comments regarding the fits are in order.
\begin{itemize}
\item{Following~\cite{Cho:1995ce,Cho:1995vh}, we do not attempt to fit
${\cal{M}}(\psi[^1S_0]_{8})$ and ${\cal{M}}(\psi[^3P_0]_{8})/m_{\charm}^2$ separately
since the $p_T$ dependence of their short distance coefficients are very
similar. Therefore, we only fit a linear combination of the two parameters (we
take ${\cal{M}}(\psi[^1S_0]_{8})={\cal{M}}(\psi[^3P_0]_{8})/m_{\charm}^2$).}
\item{The results are sensitive to the lower $p_T$ cutoff. We do not include
yields below $p_T=4$~GeV because using a lower cut off gives
a significantly larger $\chi^2/\rdof$. The NRQCD formalism for production is
expected to work well only for large $p_T$.}
\item{The fits are sensitive to the sets of data considered. The ATLAS results
separately prefer to have a larger $^1S_0$ and $^3P_j$ contribution, whereas the
PHENIX results prefer the reverse. The combined fit is most similar to a fit
including only CDF data.}
\item{Additional data from STAR~\cite{Abelev:2009qaa} at $\sqrt{S}=0.2$~TeV and 
CMS~\cite{Chatrchyan:2012np} at $\sqrt{S}=2.76$~TeV is consistent with
our calculations and hence we don't expect them to affect the fits
substantially if we include them.}
\item{The data from TeVatron and LHC is for prompt $J/\psi$. RHIC data includes
feed-down from $B-$mesons. They turn out to be unimportant at RHIC energies.}
\end{itemize}

Using standard fitting techniques we find
\begin{equation}
\begin{split}
{\cal{M}}(J/\psi[^3S_1]_{8}(1S)) &= (0.0013\pm 0.0013) \rGeV^3\\
{\cal{M}}(J/\psi[^1S_0]_{8}(1S)) &= (0.018\pm 0.0087) \rGeV^3=
{\cal{M}}(J/\psi[^3P_0]_{8}(1S))/m_{\charm}^2\;,
~\label{eq:MJn1CO_val}
\end{split}
\end{equation}
with a $\chi^2/\rdof=5.2$. We also note that the uncertainty in the
$[^3S_1]$ matrix element is in particular very large, which is unfortunate
because it is the dominant contribution to the total yield. The $\chi^2/\rdof$
improves if we increase the lowest $p_T$ in the fit.

With all the relevant matrix elements in hand, we can calculate production
yields for all the species which contribute to the production of $J/\psi$,
multiply by the relevant branching fractions (we ignore the small additional
boost in the $J/\psi$), and obtain the prompt yields for $J/\psi$. 

To illustrate the various contributions in a specific case, we show these for
the p$+$p baseline for LHC collisions at $\sqrt{S}=2.76$~TeV. The experimental
data is the prompt $J/\psi$ production yield observed at
CMS~\cite{Chatrchyan:2012np}.

In nuclear collisions, each of the species will undergo a different
modification due to cold nuclear matter and quark gluon plasma effects. To
obtain the p$+$A and A$+$A yields, we calculate the modified yields for each
species and combine them again to obtain the net modification. 

\section{Fitting color-octet matrix elements for bottomonia and a baseline for 
LHC at $\sqrt{S}=2.76$~TeV~\label{appendix:LHCbaseU}}
\begin{figure}[!ht]
\vspace*{.2in}
\includegraphics[width=4.0in,angle=0]{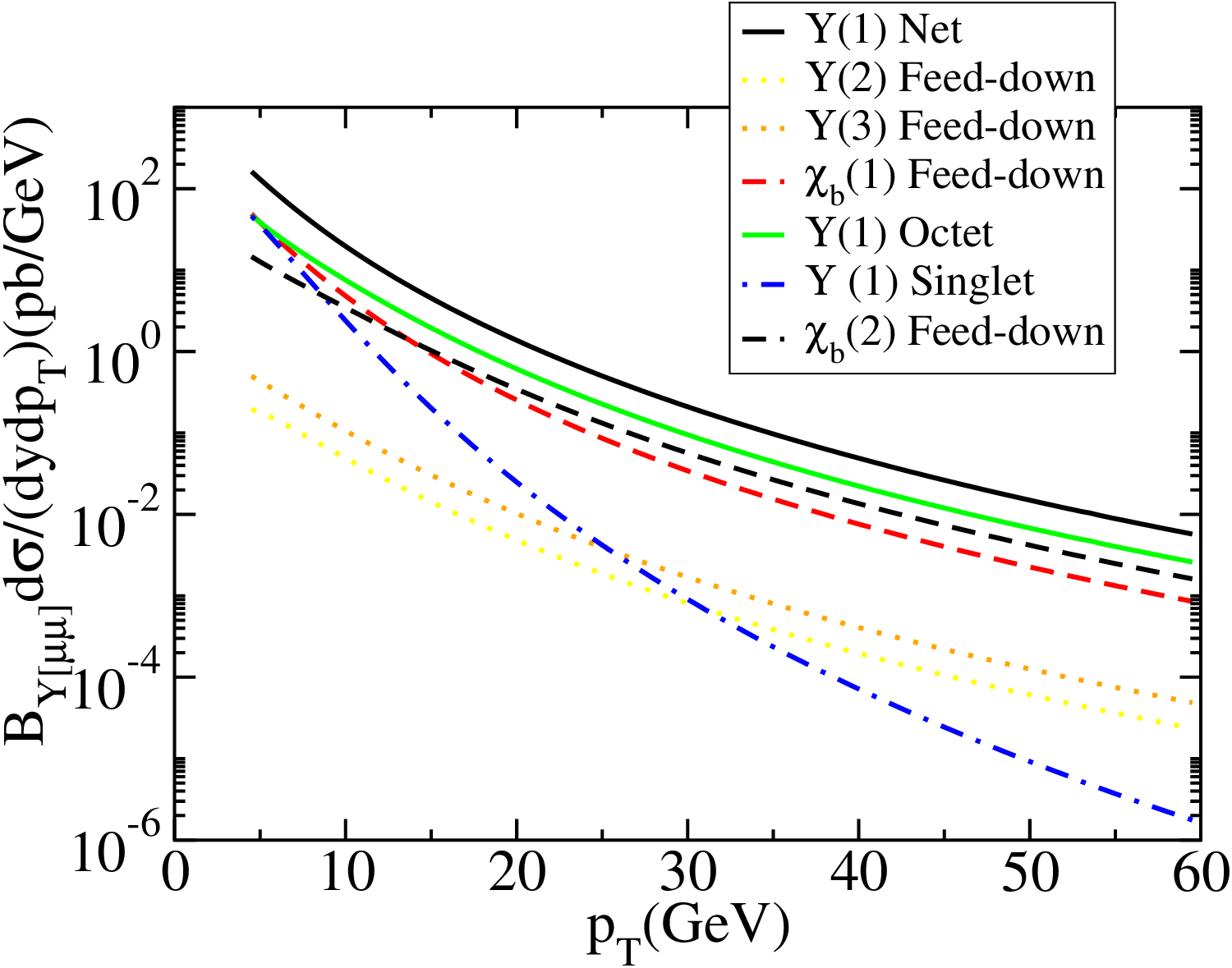} 
\caption{(Color online) The cocktail of contributions that  gives the p$+$p
yield for $\Upsilon(1S)$ production at the LHC at $\sqrt{S}=2.76$~TeV. The uppermost
curve (black online) gives the net yield and gives the p$+$p baseline. ~\label{fig:LHC2760Ucocktail}}
\end{figure}

The analysis of $\Upsilon(1S)$ production and feed-down contributions is very
similar to the analysis for $J/\psi$. Following~\cite{Cho:1995ce,Cho:1995vh}
we consider states up to $n=3$. For the $\Upsilon(3S)$, we consider only the
direct contribution. For $\Upsilon(2S)$, we consider feed-down from
$\Upsilon(2S)$ and $\chi_b(2P)$. For $\Upsilon(1S)$ there are additional
contributions from $\Upsilon(2S)$ and $\chi_b(1P)$. The relevant branching
fractions are given in Table~\ref{table:BottomoniaBFs}.
\begin{table*}[h]
\begin{tabular}{c|cccccccc}
meson from  &to $\chi_{b0}(2)$ &to $\chi_{b1}(2)$ &to $\chi_{b2}(2)$
            &to $\Upsilon(2S)$
            &to $\chi_{b0}(1)$ &to $\chi_{b1}(1)$ &to $\chi_{b2}(1)$ 
            &to $\Upsilon(1S)$\\ 
\hline
$\Upsilon(3S)$ & 0.131 & 0.126 & 0.059 
               & 0.199
               & 0.003 & 0.0017 & 0.019
               & 0.066\\
$\chi_{b0}(2P)$& & &
               & 0.046
               & & &
               & 0.009\\
$\chi_{b1}(2P)$& & &
               & 0.21
               & & &
               & 0.101\\
$\chi_{b2}(2P)$& & &  
               & 0.162
               & & &
               & 0.082\\
$\Upsilon(2S)$ & & &
               &
               & 0.038 & 0.0715 & 0.069 
               & 0.267\\
$\chi_{b0}(1P)$& & &  
               &
               & & &
               & 0.06\\
$\chi_{b1}(1P)$& & &
               &
               & & &
               & 0.35\\
$\chi_{b2}(1P)$& & &
               &
               & & &
               & 0.22\\
\end{tabular}
\caption{Relevant branching fractions for bottomonia~\cite{pdg}.~\label{table:BottomoniaBFs}}
\end{table*}

The multitude of feed-down contributions makes the fitting of the matrix
elements much more subtle than for $J/\psi$. We use data from
CDF~\cite{Acosta:2001gv} and CMS~\cite{Khachatryan:2010zg} to determine the
matrix elements.

As for $J/\psi$, the color singlet matrix elements are estimated using the
radial wavefunctions of the mesons.
\begin{equation}
\begin{split}
{\cal{M}}(\Upsilon[^3S_1]_{1}(1S)) &= 10.9 \rGeV^3\\
{\cal{M}}(\chi_b[^3P_0]_{1}(1P))/m_{\bottom}^2 &= 0.100 \rGeV^3\\
{\cal{M}}(\Upsilon[^3S_1]_{1}(2S)) &= 4.5 \rGeV^3\\
{\cal{M}}(\chi_b[^3P_0]_{1}(2P))/m_{\bottom}^2 &= 0.036 \rGeV^3\\
{\cal{M}}(\Upsilon[^3S_1]_{1}(3S)) &= 4.3 \rGeV^3
\end{split}
\end{equation}

For the color-octets, we use the following procedure. We first fit the highest
state, $\Upsilon(3S)$. 
\begin{equation}
\begin{split}
{\cal{M}}(\Upsilon[^3S_1]_{8}(3S)) &= (0.0513\pm 0.0085) \rGeV^3\\
{\cal{M}}(\Upsilon[^1S_0]_{8}(3S)) &= (0.0002\pm 0.0062) \rGeV^3=
{\cal{M}}(\Upsilon[^3P_0]_{8}(3S))/(5m_{\bottom}^2)\;,
~\label{eq:MUn3CO_val}
\end{split}
\end{equation}
The fit of the to parameters is quite good and gives a $\chi^2/\rdof=1.33$. 

Having fixed the $\Upsilon(3S)$ yields, we now consider $\Upsilon(2S)$ which has
feed-down from $\Upsilon(3S)$ and from $\chi_b(2)$, as well as a direct
production contribution. There are $4$ new parameters for $n=2$ (the $^3P_0$
and $^1S_0$ are not treated as independent as in the $J/\psi$). An unconstrained fit for these $4$
parameters converges to an unphysical point where the direct contribution is negative and
is canceled by a positive contribution from the feed-down contributions.
Considering the CDF data separately does not help, and a problem of similar
nature appears in fitting the CMS data, albeit with the $\chi_b$ contribution.
As expected, letting the $^1S_0$ and $^3P_0$ matrix elements free to float does
not help with the fitting because the shapes of the two contributions is very
similar.

To resolve this issue, we limit the variation in the $^3S_1$ contribution by 
assuming that it roughly scales with mass in going from $J/\psi$ to $\Upsilon(2S)$.
\begin{equation}
\begin{split}
{\cal{M}}(\Upsilon[^3S_1]_{8}(2S)) 
&\in [\frac{1}{10}, 10]\times{\cal{M}}(J/\psi[^3S_1]_{8}(1S))
\bigl(\frac{m_{\Upsilon(2S)}}{m_{\psi(1S)}}\bigr)^3\\
{\cal{M}}(\chi_b[^3S_1]_{8}(2P)) 
&\in [\frac{1}{10}, 10]\times{\cal{M}}(\chi_c[^3S_1]_{8}(1P))
\bigl(\frac{m_{\chi_b(2P)}}{m_{\chi_c(1P)}}\bigr)^3
~\label{eq:MUn2CO_scaling}
\end{split}
\end{equation}
With these assumptions, the $5$ matrix elements can be fit while satisfying the
constraint that production cross sections for all the particles are positive.
To obtain an error estimate, we fit the LHC and CDF data separately, and the
difference gives an estimate of the variation in the parameters. The results obtained are
as follows,
\begin{equation}
\begin{split}
{\cal{M}}(\Upsilon[^3S_1]_{8}(2S)) 
&= (0.0224\pm 0.0196) \rGeV^3\\ 
{\cal{M}}(\Upsilon[^1S_0]_{8}(2S)) &= (-0.0067\pm0.0084) \rGeV^3=
{\cal{M}}(\Upsilon[^3P_0]_{8}(2S))/(5m_{\bottom}^2)\\
{\cal{M}}(\chi_b[^3S_1]_{8}(2P)) &= (0.0324) \rGeV^3\;.
~\label{eq:MUn2CO_val}
\end{split}
\end{equation}
Note that having $^1S_0$ contributions to be negative is not a problem. The
main physical requirement is that the net cross sections should be positive for
all $p_T$.  The quality of the fit is worse than the fit for
$\Upsilon(3S)$, with a $\chi^2/\rdof=3.5$. In particular the yields at CDF are
underpredicted. Perhaps including higher order corrections in $\alpha_s$, and
more data can improve the fit in the future. We also comment on the fact that 
we don't quote an error estimate for the $\chi_b(2P)$ matrix element in
Eq.~\ref{eq:MUn2CO_val}. In our fitting procedure, the ${\cal{M}}(\chi_b[^3S_1]_{8}(2P))$
coverges to the boundary of the constrained region in 
Eq.~\ref{eq:MUn2CO_scaling} for both LHC and CDF data. Hence a trustworthy
error estimate can not be obtained in this case. A similar situation arises for
the fitting of the $\Upsilon(3S)$ matrix elements.

Next we consider the production of $\Upsilon(1S)$. We handle the $^3S_1$
components in the same way as for $\Upsilon(2S)$. The results are given below,
\begin{equation}
\begin{split}
{\cal{M}}(\Upsilon[^3S_1]_{8}(1S)) 
&= (0.0477\pm0.0334) \rGeV^3\\ 
{\cal{M}}(\Upsilon[^1S_0]_{8}(1S)) &= (0.0121\pm0.040) \rGeV^3=
{\cal{M}}(\Upsilon[^3P_0]_{8}(1S))/(5m_{\bottom}^2)\\
{\cal{M}}(\chi_b[^3S_1]_{8}(1P)) &= (0.1008) \rGeV^3\;.
~\label{eq:MUn1CO_val}
\end{split}
\end{equation}
The $\chi^2/\rdof=3.8$.

Some comments regarding the fits are in order.
\begin{itemize}
\item{Following~\cite{Cho:1995ce,Cho:1995vh}, we do not attempt to fit
${\cal{M}}(\Upsilon[^1S_0]_{8}(nS))$ and
${\cal{M}}(\Upsilon[^3P_0]_{8}(nS))/m_{\bottom}^2$
separately since the forms of their short distance pieces are very similar.
Therefore, we only fit a linear combination of the two parameters (we take
${\cal{M}}(\Upsilon[^1S_0]_{8}(nS))={\cal{M}}(\Upsilon[^3P_0]_{8}(nS))/(5m_{\bottom}^2)$ which
is the relative scale of the two contributions).}
\item{The results are sensitive to the lower $p_T$ cutoff. We do not include
yields below $p_T=5$~GeV.}
\end{itemize}

\section{$B$ feed-down contribution~\label{appendix:Bfeed}}

\begin{figure}[!ht]
\vspace*{.2in} 
\includegraphics[width=3.38in,angle=0]{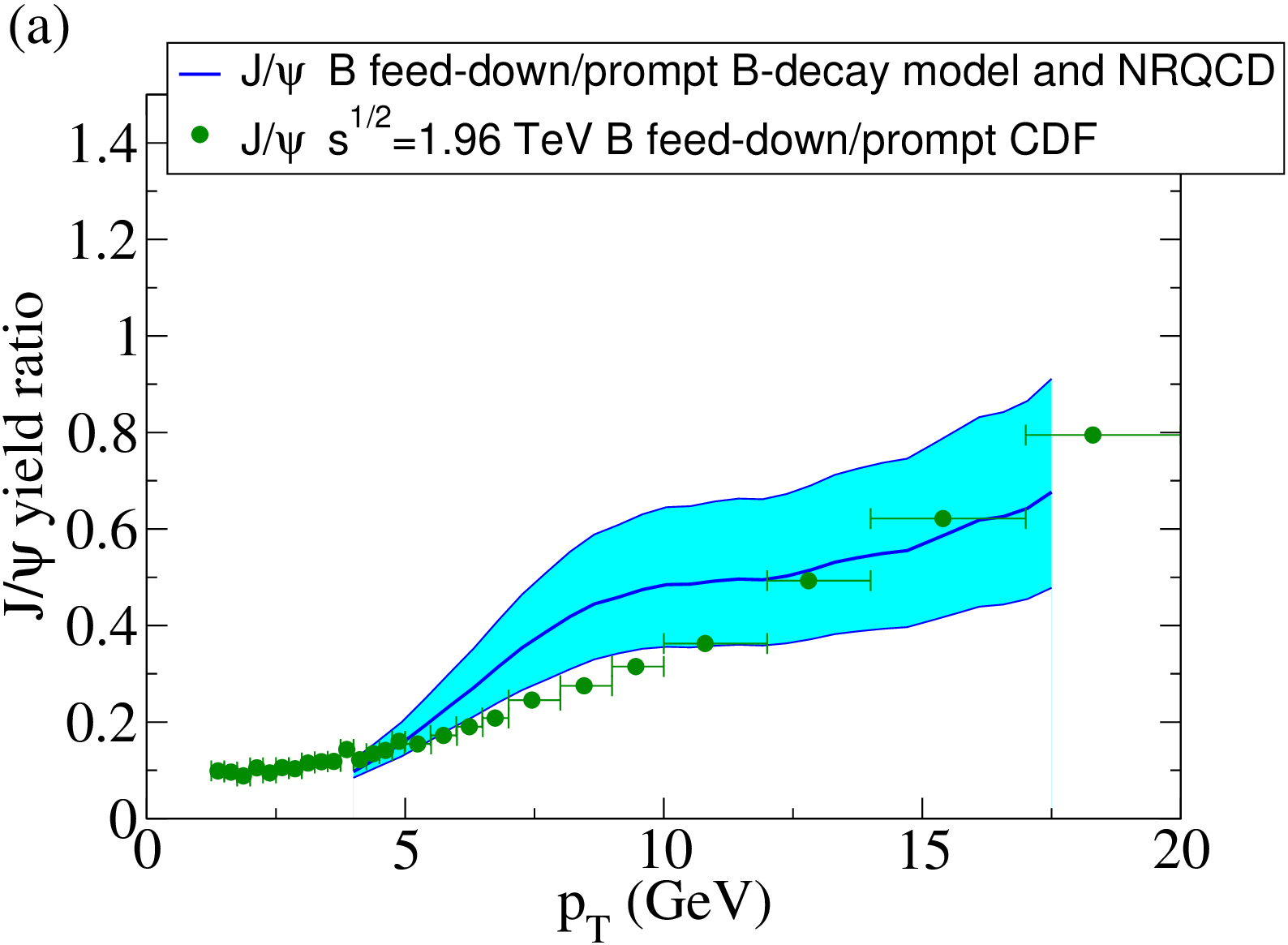} 
\includegraphics[width=3.38in,angle=0]{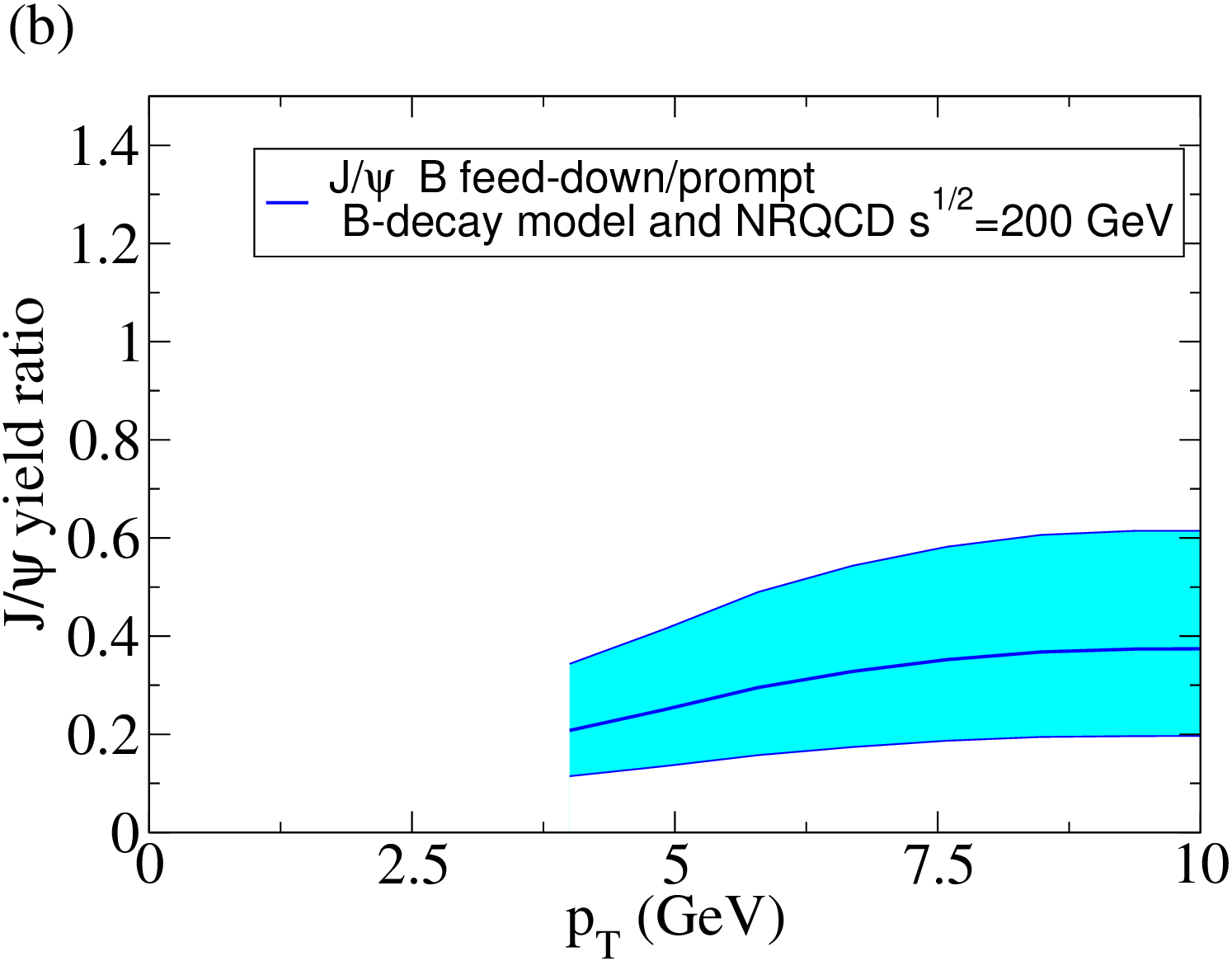} 
\includegraphics[width=3.38in,angle=0]{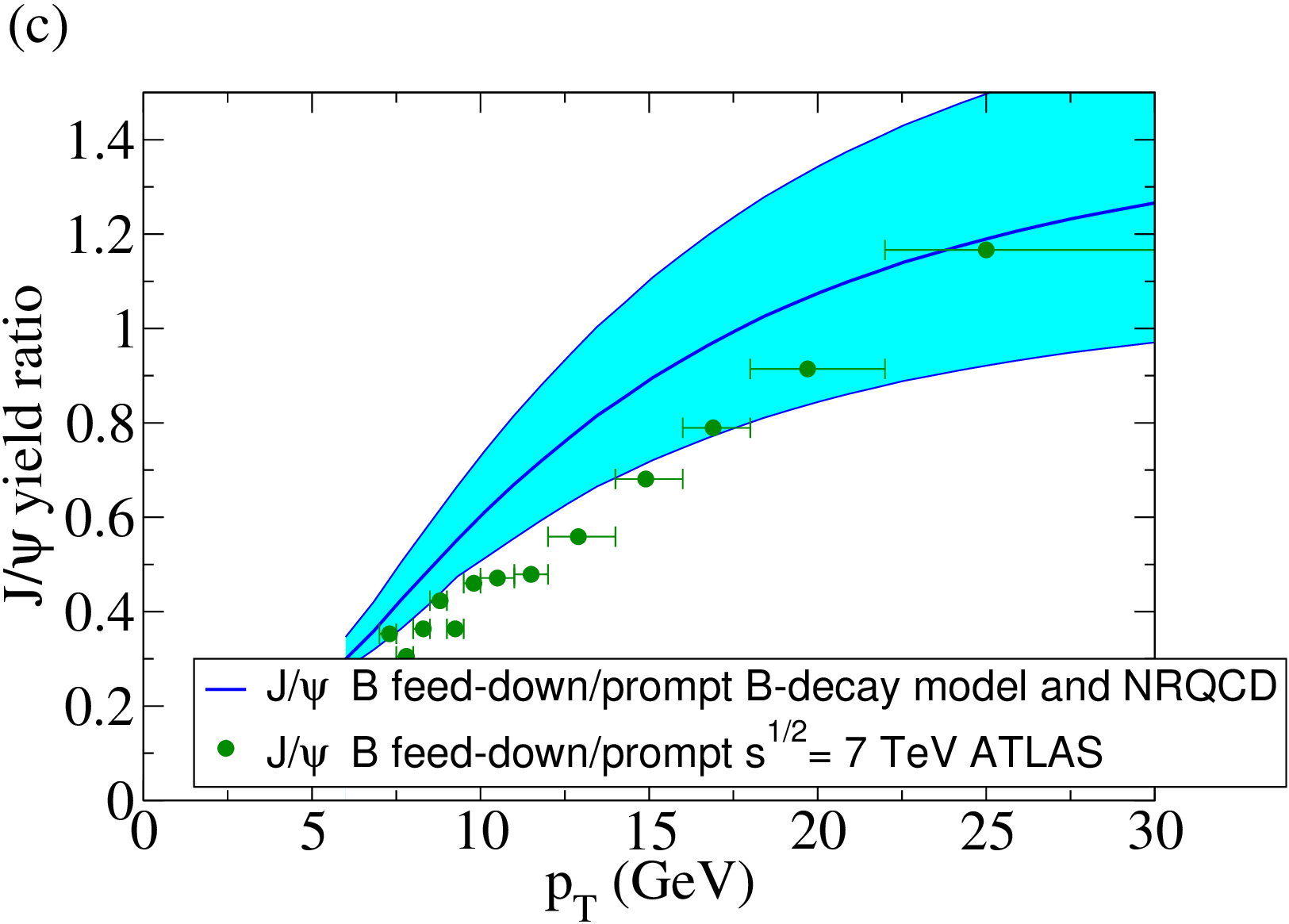} 
\caption{(Color online) Ratio of the $J/\psi$ yield from $B$ feed-down to the
prompt yield.  The upper left panel corresponds to data for the yield of
$J/\psi$ from CDF at $\sqrt{S}=1.96$~TeV~\cite{Acosta:2004yw}. The upper right
panel is for RHIC at $\sqrt{S}=0.2$~TeV.  The bottom panel has data from the
LHC at $\sqrt{S}=7$~TeV from the ATLAS collaboration~\cite{Aad:2011sp} The
darker colored error bars (dark green online) show the experimental results for
the ratio and the error bar represents the width of the larger energy bin of
the prompt and $B$ feed-down yield. The solid line (blue online) is the
theoretically calculated result. The uncertainty band is associated with the
scale variation in the production of $J/\psi$, as in Fig.~\ref{fig:Jyields}.
Uncertainties in the $B$-feeddown are not displayed. \label{fig:Jyieldsratio}}
\end{figure} 

Using previous results for the production of $B-$hadrons~\cite{hep-ph/0611109}
and calculating their decay to $J/\psi$, we have estimated the feed-down
contribution from $B-$hadrons to the $J/\psi$ yields. Our model for $B-$ decay
gives results that give the right trend when compared with the observations
from CDF and ATLAS. At RHIC we find that at least up to $p_T\sim20$~GeV, the
feed-down contribution is smaller than the prompt production but is not
insignificant. We attribute this fact partly to the hardness of the NRQCD
spectrum.

\section{Results for $J/\psi$ production  at forward rapidity $y=3$
~\label{appendix:y3}}

While this paper focuses on results for quarkonium production and modification  
relevant to the central rapidity region at the LHC, it is useful to investigate the 
implications of our theoretical model for forward rapidity.  In this example, we select 
$y=3$ and evaluate the  $J/\psi$ production cross section in p+p, p+Pb and Pb+Pb reactions.
We are motivated by the possibility that the ALICE experiment~\cite{Abelev:2012rv} will 
soon measure such cross sections up to a high transverse momentum. 

Due to the large center-of-mass energy per nucleon pair and the wide rapidity range at the
LHC, the steepness of the underlying quarkonium production cross sections does  
not differ considerably between $y=0$ and $y=3$. Cold nuclear matter effects are expected 
to be qualitatively similar for these 2 rapidities. Furthermore, the charged particle
rapidity density, which is proportional to the QGP density in the Bjorken expansion
scenario, is expected to vary by less than 20\% over the rapidity range of interest.
In the co-moving plasma frame (in our case this means boosting to $y=3$), quarkonia 
propagate strictly in the transverse direction. Thus, the final-state dynamics at forward 
rapidity in the Bjorken expansion scenario is described by the same set of equations 
that we used to obtain the mid-rapidity results as long as one is careful to use
transverse sizes and transverse momenta.
 
\begin{figure}[!ht]
\vspace*{.2in}
\includegraphics[width=3.38in,angle=0]{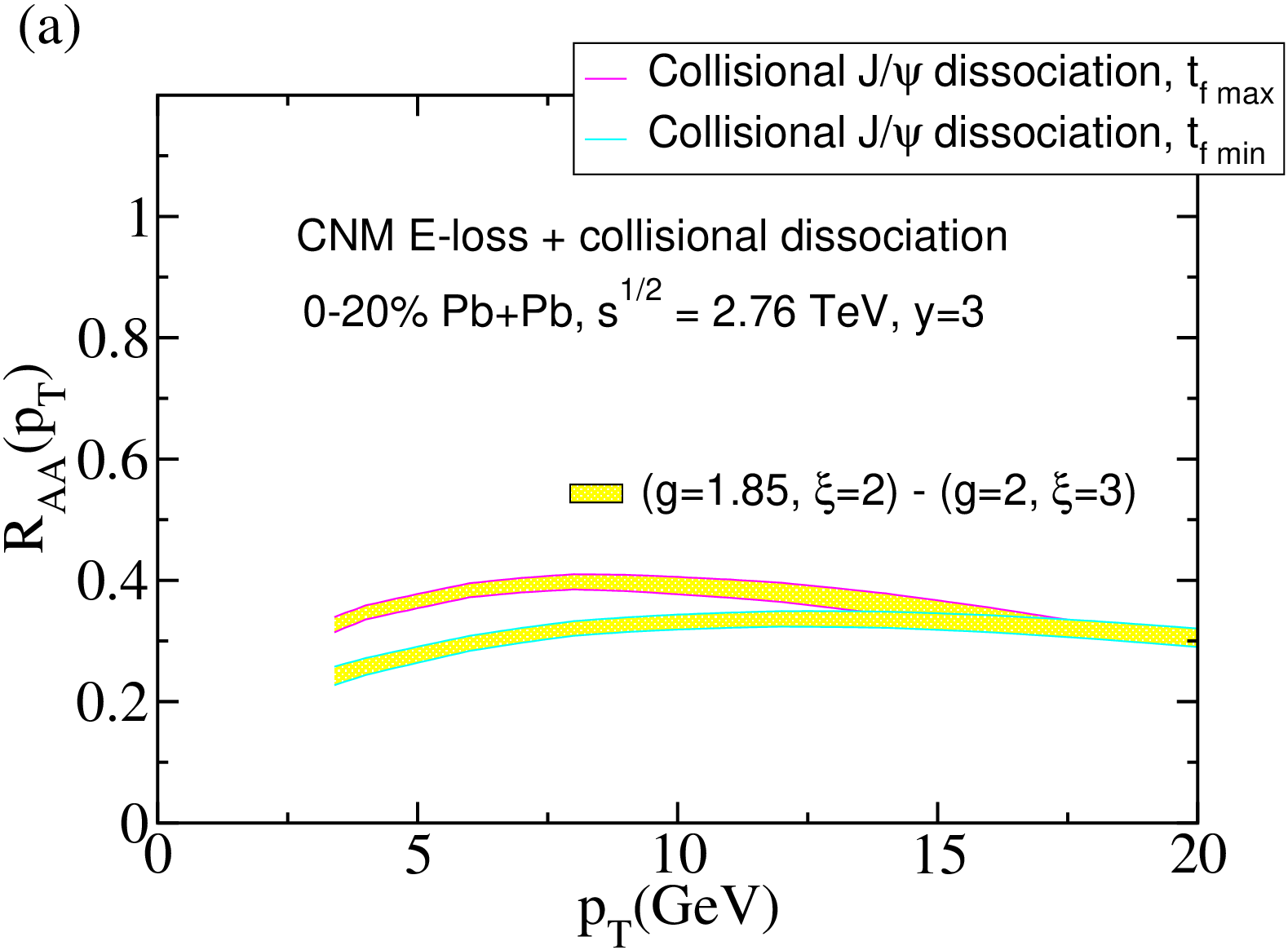}
\includegraphics[width=3.38in,angle=0]{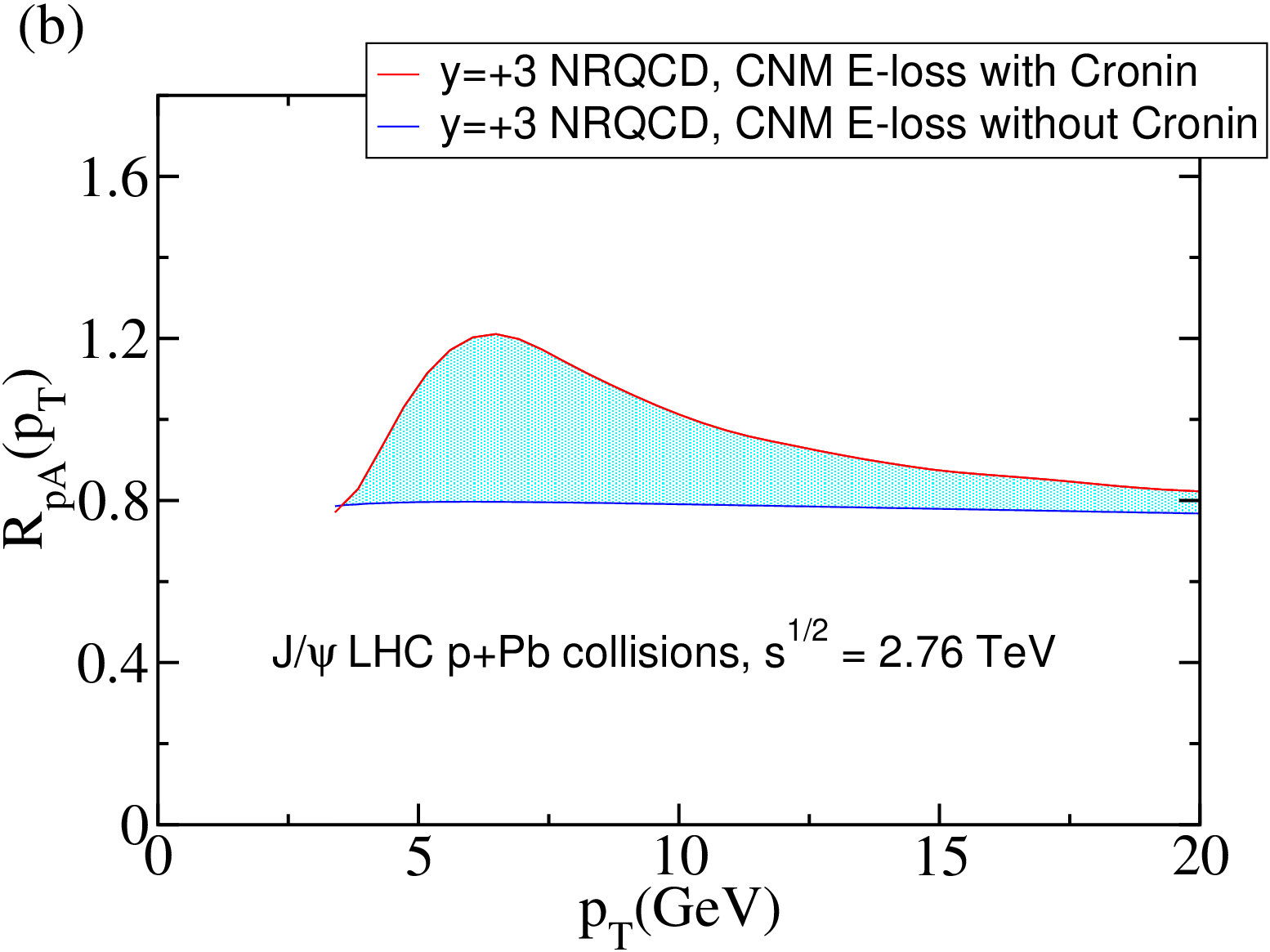} 
\caption{(Color online) Results for $R_{AA}$ (left panel) and $R_{pA}$ right panel at $y=3$
where positive rapidity corresponds to motion parallel to the proton in p+A
collisions.  We only show results for $R_{AA}$ ignoring the Cronin effect. One
can constrain the Cronin effect from experiments by comparing with $R_{pA}$ for
$y=3$.~\label{fig:y3}}
\end{figure}

Our results of the high-$p_T$  $J/\psi$ suppression  at $y=3$ are presented 
in Fig.~\ref{fig:y3}. The conventions in the plots are the same as in
Section~\ref{section:Results}. 

\end{appendix}

\end{document}